\documentclass[english]{aastex}
\usepackage[T1]{fontenc}
\usepackage{float}
\usepackage{amstext}
\usepackage{amssymb}
\usepackage{graphicx}

\makeatletter
\slugcomment{}
\shorttitle{}
\shortauthors{}

\makeatother

\usepackage{babel}
\begin{document}

\title{The Impact of Helium-Burning Reaction Rates on Massive Star Evolution
and Nucleosynthesis}

\author{Christopher West$^{\text{1,2}}$}

\email{west0482@umn.edu}

\and{}

\author{Alexander Heger$^{\text{1,2},3}$}

\email{alexander.heger@monash.edu}

\and{}

\author{Sam M. Austin$^{\text{2},4}$}

\email{austin@nscl.msu.edu}

\affil{$^{\text{1}}$Minnesota Institute for Astrophysics}

\affil{School of Physics and Astronomy, University of Minnesota }

\affil{$^{\text{2}}$Joint Institute for Nuclear Astrophysics}

\affil{$^{\text{3}}$Monash Centre for Astrophysics }

\affil{School of Mathematical Sciences, Monash University}

\affil{$^{\text{4}}$National Superconducting Cyclotron Laboratory }

\affil{640 South Shaw Lane, E. Lansing, MI 48824, Michigan State University}
\begin{abstract}
We study the sensitivity of presupernova evolution and supernova nucleosynthesis
yields of massive stars to variations of the helium-burning reaction
rates within the range of their uncertainties. The current solar abundances
from Lodders (2009) are used for the initial stellar composition.
We compute a grid of 12 initial stellar masses and 176 models per
stellar mass to explore the effects of independently varying the $^{\text{12}}$C($\alpha,\gamma$)$^{\text{16}}$O
and 3$\alpha$ reaction rates, denoted $R_{\alpha,12}$ and $R_{3\alpha}$,
respectively. The production factors of both the intermediate-mass
elements (A=16-40) and the \emph{s}-only isotopes along the weak \emph{s}-process
path\textbf{ }($^{70}$Ge, $^{76}$Se, $^{80}$Kr, $^{82}$Kr, $^{86}$Sr,
and $^{87}$Sr) were found to be in reasonable agreement with predictions
for variations of \textbf{$R_{3\alpha}$} and \textbf{$R_{\alpha,12}$}
of $\pm25\,\%$; the s-only isotopes, however, tend to favor higher
values of \textbf{$R_{3\alpha}$} than the intermediate-mass isotopes.
The experimental uncertainty (one standard deviation) in \textbf{$R_{3\alpha}$}(\textbf{$R_{\alpha,12}$})
is approximately $\pm10\,\%$($\pm25\,\%$). The results show that
a more accurate measurement of one of these rates would decrease the
uncertainty in the other as inferred from the present calculations.
We also observe sharp changes in production factors and standard deviations
for small changes in the reaction rates, due to differences in the
convection structure of the star.\textbf{ }The compactness parameter
was used to assess which models would likely explode as successful
supernovae, and hence contribute explosive nucleosynthesis yields.
We also provide the approximate remnant masses for each model and
the carbon mass fractions at the end of core-helium burning as a key
parameter for later evolution stages.
\end{abstract}


\section{Introduction}

Massive stars are responsible for the production of most intermediate-mass
($\mathrm{A}=16-40$) isotopes through hydrostatic burning phases
and subsequent supernovae \citep{B2FH,Woosley2002}. During core-He
burning the $^{\text{12}}$C($\alpha,\gamma$)$^{\text{16}}$O and
3$\alpha$ reactions compete to determine the relative abundances
of oxygen and carbon prior to core-C burning. Changes in these abundances
have significant effects on subsequent stellar evolution and structure
and on the resulting nucleosynthesis. The carbon abundance influences
subsequent shell burning episodes and affects whether core-C burning
will be radiative or convective. There is also a non-monotonic relation
between the carbon abundance and the resulting remnant mass \citep{Woosley2003},
so that these reactions are important for understanding the populations
of neutron stars (NS) and black holes (BH). 

These rates can also affect weak \emph{s}-process yields. The weak
\emph{s}-process is a slow neutron capture process occurring at the
end of convective core-He burning and during shell carbon burning
\citep{Pignatari2010}, and contributes to isotopic production along
the \emph{s}-process path above iron and up to a mass number of A$\approx$100
\citep{Raiteri1993}. A change in the helium-burning rates can induce
a corresponding change in temperature to keep the star at constant
luminosity, and the reaction for the neutron source for the weak \emph{s}-process,
$^{\text{22}}$Ne($\alpha,\mathrm{n}$)$^{\text{25}}$Mg, is highly
temperature-dependent. Additionally, the amounts of neutron poisons
have been shown to vary with these rates \citep{Rayet2000,Tur2009}. 

Although not discussed in this paper the production of the important
radioactive nuclei $^{26}$Al, $^{44}$Ti, and $^{60}$Fe \citep{Tur2010}
is also sensitive to these rates. Gamma rays from these nuclei provide
observational information that may help to test models of massive
star internal structure and nucleosynthesis through the constraints
imposed by the abundance ratios of $\mathrm{^{44}\mathrm{Ti}/^{56}Co}$
and $\mathrm{^{26}Al/^{60}Fe}$ \citep{Diehl2006,Leising2009}.

This work is an extension of the study by Tur et al. (\citeyear{Tur2007},
\citeyear{Tur2009}), who calculated a limited subset of models of
our full 2D parameter space. They concluded that across the 2$\sigma$
uncertainty range root-mean-square (rms) deviations for the production
factors sometimes vary non-monotonically with the rates, as do deviations
in the remnant mass, indicating that both helium burning reactions
are independently important \citep{Tur2007}. We extend upon their
study by performing a much finer sampling of the 2$\sigma$ uncertainty
range to assess the effect of changing the rates \emph{independently},
in order to map the whole 2D landscape to determine the true non-monotonic
behavior\textbf{ }with a higher resolution grid. The purpose is to:
i) examine the effect of varying the helium burning reaction rates
on the production factors of intermediate-mass isotopes, ii) examine
the effect on the production factors of the six \emph{s}-only isotopes
along the weak \emph{s}-process path, iii) assess the impact on i)
and ii) of including only models that are likely to explode as successful
supernovae, and iv) examine the effect on the remnant mass. We do
not address the effect of varying the solar abundance set, which indeed
has been shown to impact the final nucleosynthesis, and is studied
in \citet{Tur2007} for the intermediate-mass isotopes and in \citet{Tur2009}
for the weak \emph{s}-process isotopes. We use an updated solar abundance
set \citep{Lodders2009} that was not yet available for the previous
studies. It was corrected for the weak \emph{s}-only isotopes by subtracting
estimated main \emph{s}-process contributions, as described in Section
2. We note that recent 3-body calculations of the 3$\alpha$ reaction
show an increase in this rate at temperatures below $\sim0.07\,\mathrm{GK}$
\citep{Nguyen2012}. This will not impact the current study as He
ignition in massive stars occurs beyond this threshold, at $T_{3\alpha}\sim0.1\,\mathrm{GK}$.

This paper has the following outline: in Section 2 we describe the
stellar models and methods used in the analysis. In Section 3 we compare
the intermediate-mass and weak \emph{s}-process isotopes across all
models, and for the subset of models likely to explode as supernovae
rather than collapse to black holes (we ignore hypernovae and gamma
ray bursts). This subset is chosen using a compactness parameter filter.
We also discuss the remnant mass and carbon mass fractions at the
end of core-He burning for the different stellar masses. Our conclusions
are given in Section 4.

\section{Stellar Models and Analysis}

All models were computed using KEPLER, a time-implicit one-dimensional
hydrodynamics package for stellar evolution \citep{Weaver1978,Rauscher2002}.
A grid of 12 initial stellar masses M/$\mathrm{M}_{\text{\ensuremath{\odot}}}$=12,
13, 14, 15, 16, 17, 18, 20, 22, 25, 27, and 30 was used, with the
revised pre-solar abundances from \citet{Lodders2009} for the initial
composition. For each stellar mass, 176 models were computed to scan
at least a 2$\sigma$ uncertainty range for both the $^{\text{12}}$C($\alpha,\gamma$)$^{\text{16}}$O
and 3$\alpha$ reaction rates, denoted $R_{\alpha,12}$ and $R_{3\alpha}$,
respectively. This range was parametrized as a multiplier on the centroid
values for the rates, as done by \citet{Tur2007}. The centroid values
used in KEPLER are taken from \citet{Caughlan1988} for $R_{3\alpha}$,
and 1.2 times the rate recommended by \citet{Buchmann2000} for $R_{\alpha,12}$.
The range for the $R_{\alpha,12}$ multipliers was $\left(0.5,2.0\right)$
with a resolution of $\Delta=0.1$, and the range for the $R_{3\alpha}$
multipliers was $\left(0.75,1.25\right)$ with a resolution of $\Delta=0.05$.
Commonly accepted uncertainties for $R_{3\alpha}$ and $R_{\alpha,12}$
are $\pm10\,\%$ \citep{Chernykh2010} and $\pm25\,\%$. Our total
range was, conservatively, slightly more than $\pm2\sigma$. 

In Fig.\,\ref{fig1} we show the grid of models computed both by
\citet{Tur2007} and in the present work.

\includegraphics[scale=0.5,angle=90]{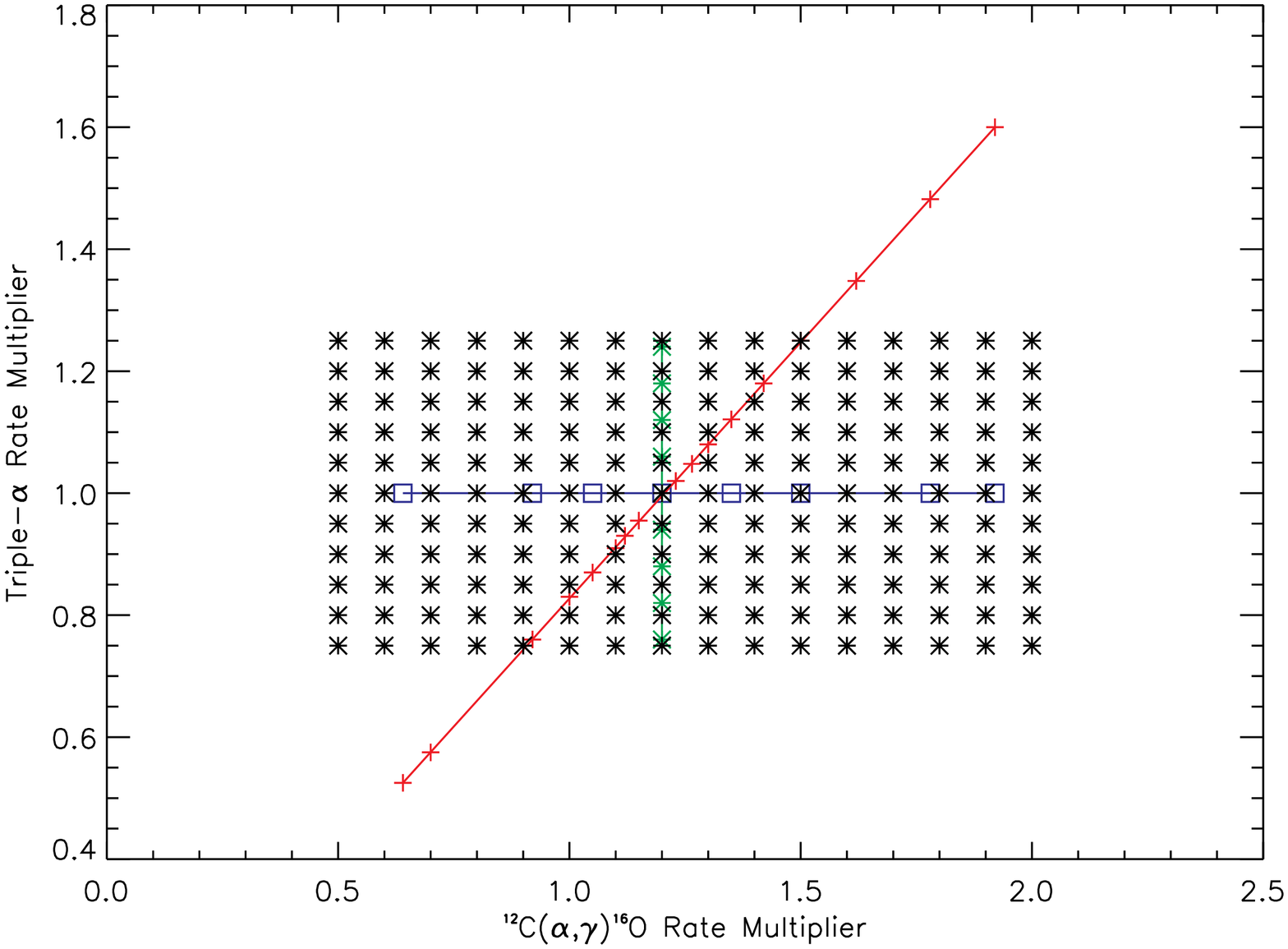}

\figcaption{\label{fig1} The $R_{\alpha,12}$ and $R_{3\alpha}$ reaction rate
multiplier values used in the stellar models. The models using the
reaction rate multiplier pairs given in red plusses, green asterisks,
and blue squares were performed by \citet{Tur2007}. The black asterisks
show the models computed in the present work.}

All stellar models were first evolved through hydrostatic burning
until the Fe core collapsed, and an inward velocity of $10^{8}\,\mathrm{cm\, s^{-1}}$
was reached. The explosion mechanism for the resulting supernova was
modeled as a mechanical piston that imparted an acceleration at constant
Lagrangian mass coordinate to provide the desired total kinetic energy
of the ejecta, taken in these models to be 1.2\,B ($\mathrm{1\, B}=10^{51}\,\mathrm{erg}$)
at 1 year after the explosion. For details on the parametrization
of the explosion used in KEPLER see \citet{Woosley2007}, and references
therein. For details on the treatment of convection and mixing see
\citet{Woosley1998} and \citet{Woosley2002}, and a discussion of
the mass cut is given in \citet{Tur2007} and \citet{Heger2010}.
Effects of rotation and magnetic fields are ignored. The final supernova
yields for all models were then averaged by integrating over the Salpeter
initial mass function (IMF) for each reaction rate multiplier pair
in Fig.\,\ref{fig1}. The yields from stellar winds were included. 

\begin{equation}
Y_{i}^{*}=\sum_{j}\intop_{m_{j}}^{m_{j+1}}\xi\left(m\right)\cdot\left(s_{i,j}\cdot\left[m-m_{j}\right]+Y_{i,j}\right)\cdot dm\label{eq:1}
\end{equation}

\begin{equation}
P_{i}=\frac{Y_{i}^{*}}{X_{i,\odot}\cdot\sum_{k}Y_{k}^{*}}\label{eq:2}
\end{equation}

In Equation\,\ref{eq:1} the IMF interpolated yield mass for isotope
\emph{i }is given by $Y_{i}^{*}$, and the Salpeter mass spectrum
is\emph{ $\xi\left(m\right)=C\cdot m^{-2.35}$}, where $C$ is the
proportionality constant\emph{ .} The mass grid used for the integrations
are the \emph{ejected} masses, defined as the baryonic remnant masses
subtracted from the initial stellar mass grid. Yields are linearly
interpolated between adjacent masses in the integral, with the slope
defined by, \emph{$s_{i,j}=\left(Y_{i,j+1}-Y_{i,j}\right)/\left(m_{j+1}-m_{j}\right)$},
where $Y_{i,j}$ is the yield mass of isotope \emph{i} from a model
with initial mass $m_{j}$. In Equation\,\ref{eq:2} the production
factor for isotope \emph{i }is given by $P_{i}$, the sum in the denominator
runs over all isotopes, and the solar mass fraction of isotope \emph{i}
is given by $X_{i,\odot}$. 

Massive stars are responsible for producing nearly the entire solar
abundance of a subset of intermediate-mass isotopes, namely $^{16,18}$O,
$^{20}$Ne, $^{23}$Na, $^{24}$Mg, $^{27}$Al, $^{28}$Si, $^{32}$S,
$^{36}$Ar, and $^{40}$Ca. Hence, in order to make the solar abundance
pattern, massive near-solar metallicity stars need to produce these
intermediate-mass isotopes in solar ratios. An analysis of the standard
deviations of the production factors for this set of isotopes is used
in the present work to identify helium reaction rate values that agree
with solar observations. This agreement is an approximation that the
above isotopes owe their entire solar abundance to massive, near-solar
metallicity stars, and relies on sufficient sampling of the initial
mass function (IMF) and understanding of the initial compositions.
As mentioned, the impact of uncertainties in the initial composition
is not addressed in this work, but an analysis of the effect of different
compositions can be found in \citet{Tur2007}.

For the weak \emph{s}-process isotopes a correction is necessary,
since the solar abundances of the six \emph{s}-only isotopes along
the weak \emph{s}-process path ($^{70}$Ge, $^{76}$Se, $^{80}$Kr,
$^{82}$Kr, $^{86}$Sr, and $^{87}$Sr) have additional contributions
from the main \emph{s}-process, which occurs in asymptotic giant branch
(AGB) stars, not massive stars. Hence what is needed are the production
factors relative to the contributions from the weak \emph{s}-process
only, not relative to the entire solar abundance. To achieve this,
the solar abundance decomposition from \citet{West2012} was used,
which gives, in part, the approximate solar contributions for the
six weak \emph{s}-only isotopes. This modifies Equation\,\ref{eq:1}.
with the substitution of $X_{i,w}$ for $X_{i,\odot}$, where $X_{i,w}$
denotes the contribution to the solar mass fraction of isotope \emph{i
}from the weak \emph{s}-process.

\section{Results and Discussion}

\subsection{Comparison of $^{\text{12}}$C(\textmd{$\alpha,\gamma$})$^{\text{16}}$O
and 3\textmd{$\alpha$} Reaction Rates}

\begin{figure}[H]
\begin{minipage}[t]{0.5\columnwidth}%
\includegraphics[angle=90,scale=0.3]{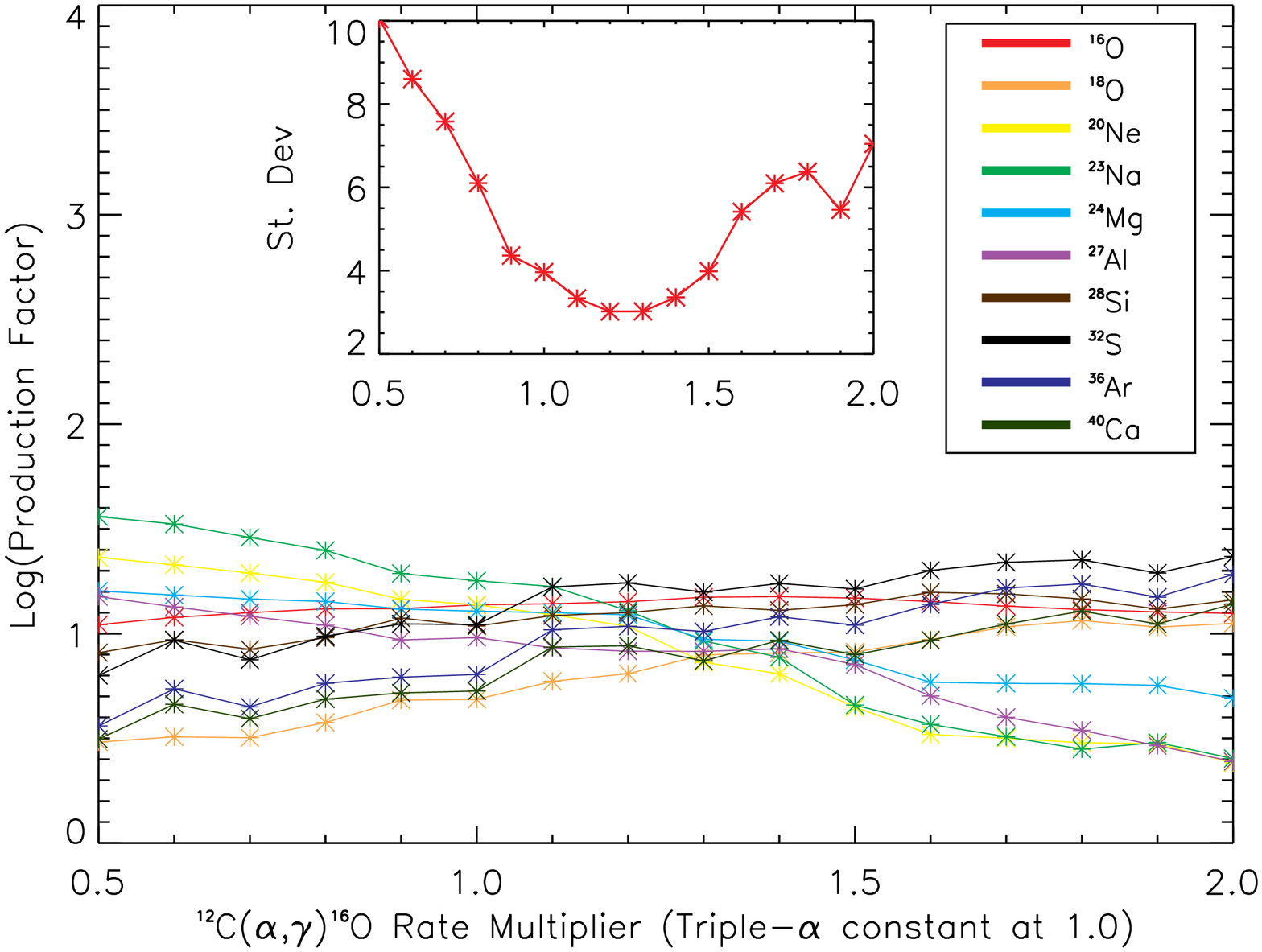}%
\end{minipage}%
\begin{minipage}[t]{0.5\columnwidth}%
\includegraphics[angle=90,scale=0.3]{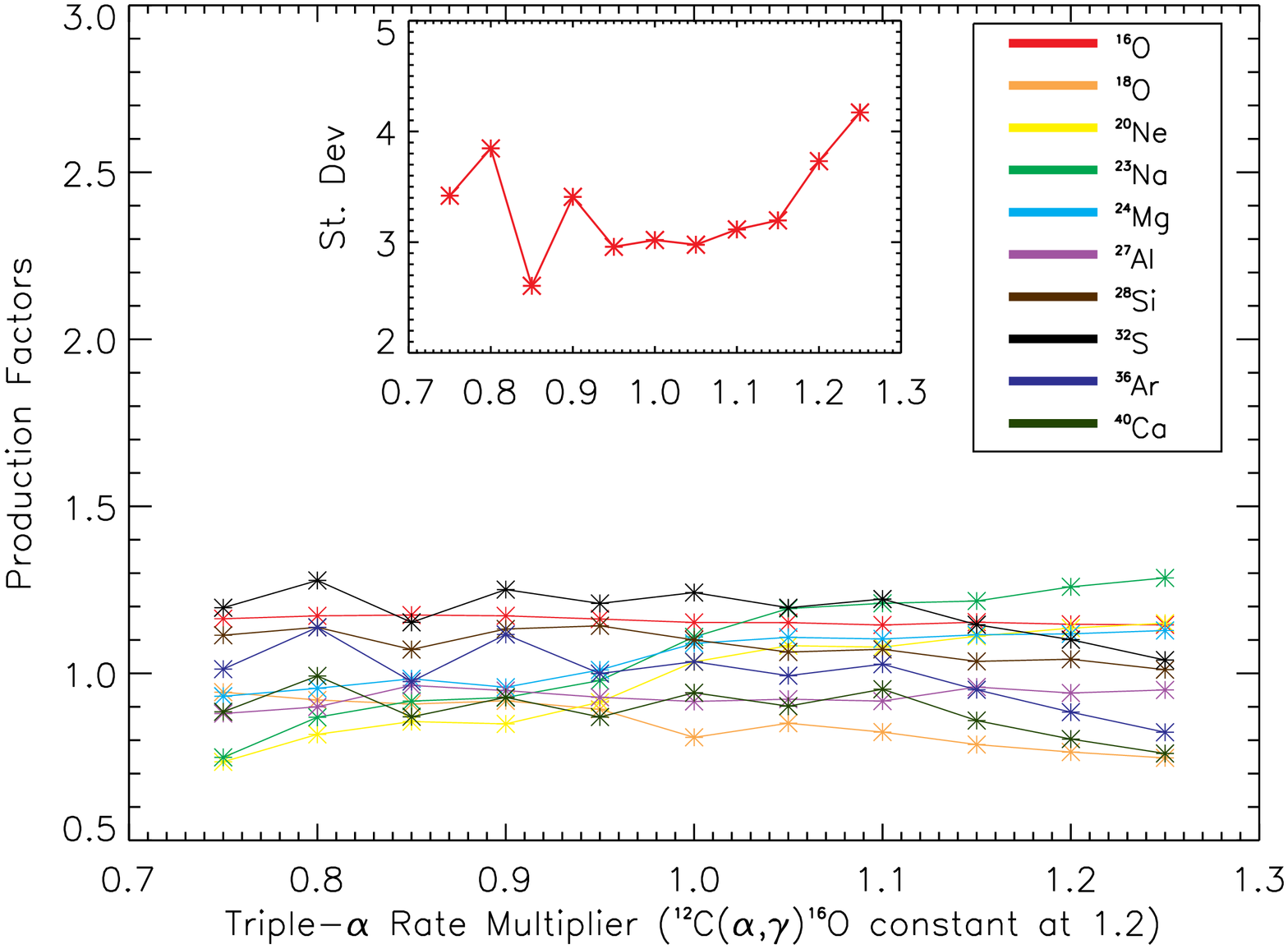}%
\end{minipage}

\caption{\emph{Left: }The production factors for the intermediate-mass isotopes
averaged over the IMF as a function of the $R_{\alpha,12}$ multiplier,
at a constant $R_{3\alpha}$ multiplier of 1.0. \emph{Right: }The
production factors for the intermediate-mass isotopes averaged over
the IMF as a function of the $R_{3\alpha}$ multiplier, at a constant
$R_{\alpha,12}$ multiplier of 1.2. The inset graph shows the standard
deviations.\label{fig2} }

\end{figure}

The production factors for the models were computed using Equations\,\ref{eq:1}
and \ref{eq:2}, and standard deviations were calculated for each
model, using the intermediate-mass isotope set (in Section 4.1) and
the six \emph{s}-only isotopes along the weak \emph{s}-process path
(in Section 4.2), 

\begin{equation}
\sigma_{P}=\sqrt{\frac{\sum_{k=1}^{n}P{}_{k}^{2}}{n-1}-\left(\frac{\sum_{k=1}^{n}P{}_{k}}{n-1}\right)^{2}}
\end{equation}
where $P{}_{k}$ are the production factors computed in Equation\,\ref{eq:2},
and $n$ is the number of entries in the isotope list: 10 for the
intermediate-mass isotopes and 6 for the \emph{s}-only isotopes along
the weak \emph{s}-process path. We distinguish $\sigma_{P}$ (the
standard deviation of the production factors) from $\sigma$ (the
uncertainty in the rates). Since massive stars contribute to most
of the solar abundances for the intermediate-mass isotopes considered
(or a fraction of them in the case of the \emph{s}-only isotopes),
low standard deviations should indicate combinations of $R_{\alpha,12}$
and $R_{3\alpha}$ that agree with observations. We first follow the
type of analysis performed by \citet{Tur2007}, for constant $R_{\alpha,12}$
and $R_{3\alpha}$ multipliers, shown in Fig.\,\ref{fig2}.

For a constant $R_{3\alpha}$ multiplier of 1.0 (Fig.\,\ref{fig2},
\emph{left}), values for the $R_{\alpha,12}$ multiplier are favored
within about $\pm$25\,\% of the centroid multiplier of 1.2. For
a constant $R_{\alpha,12}$ multiplier of 1.2 (Fig.\,\ref{fig2},
\emph{right}), values for the $R_{3\alpha}$ multiplier have a minimum
standard deviation at 0.85, and vary across a similar range of standard
deviation values for a $\pm25\,\%$ change in $R_{3\alpha}$. Since
$R_{3\alpha}$ is better experimentally determined, however, the extremes
of this range are less likely than for $R_{\alpha,12}$. The results
in Fig.\,\ref{fig2} show approximate qualitative agreement with
the findings of \citet{Tur2007}, but care must be taken in a comparison
as they do not use the \citet{Lodders2009} abundances, and they have
demonstrated that there is non-trivial variation among different solar
abundance sets. 

\begin{figure}[H]
\begin{minipage}[t]{0.5\columnwidth}%
\includegraphics[angle=90,scale=0.3]{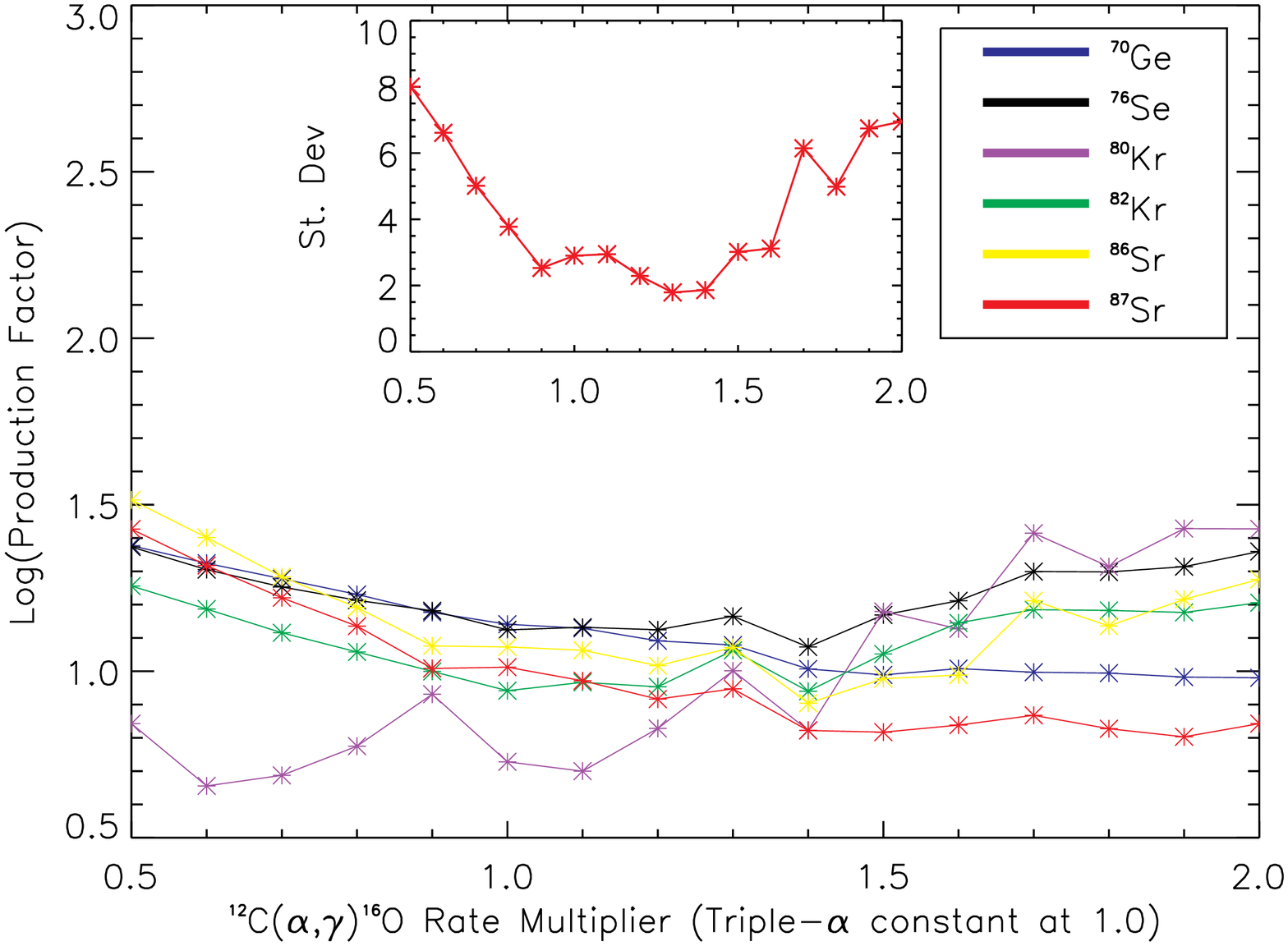}%
\end{minipage}%
\begin{minipage}[t]{0.5\columnwidth}%
\includegraphics[angle=90,scale=0.3]{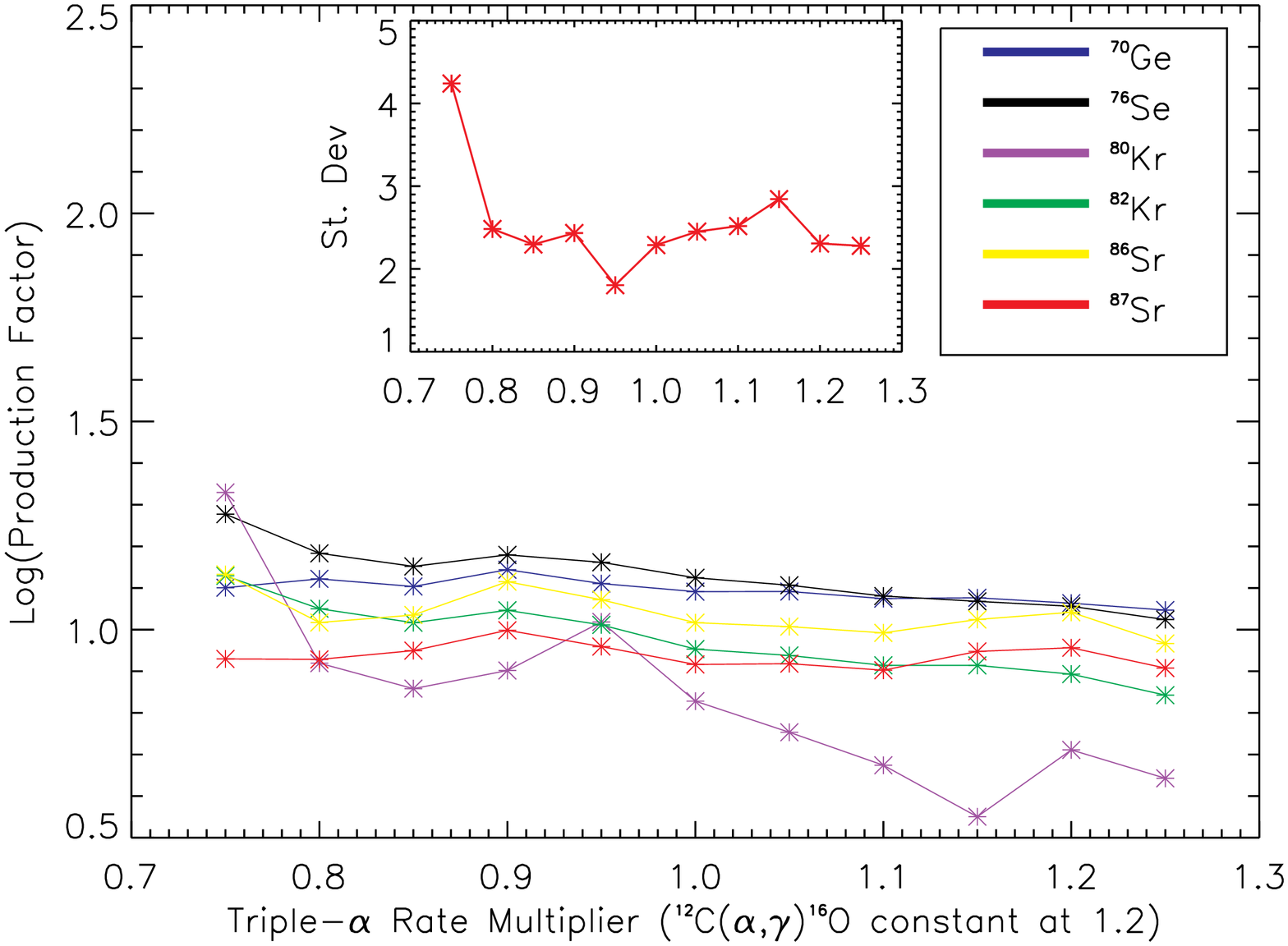}%
\end{minipage}

\caption{\emph{Left: }The production factors for the weak \emph{s}-only isotopes
averaged over the IMF as a function of the $R_{\alpha,12}$ multiplier,
at a constant $R_{3\alpha}$ multiplier of 1.0. \emph{Right: }The
production factors for the weak \emph{s}-only isotopes averaged over
the IMF as a function of the $R_{3\alpha}$ multiplier, at a constant
$R_{\alpha,12}$ multiplier of 1.2. The inset graph shows the standard
deviations.\label{fig3}}
\end{figure}

The corresponding plot for the weak \emph{s}-only isotopes is given
in Fig.\,\ref{fig3}.

For a constant $R_{3\alpha}$ multiplier of 1.0 (Fig.\,\ref{fig3},
\emph{left}), values for the $R_{\alpha,12}$ multiplier have a minimum
standard deviation at 1.3. For a constant $R_{\alpha,12}$ multiplier
of 1.2 (Fig.\,\ref{fig3}, \emph{right}), the standard deviation
has a minimum at the $R_{3\alpha}$ multiplier value of 0.95. Significant
variations in the production factors exists across both multiplier
ranges.

\subsection{Intermediate-Mass Isotopes}

To address how changing the rate multipliers independently affects
the nucleosynthesis, the entire set of models in $\left(R_{\alpha,12},R_{3\alpha}\right)$
space was mapped in a 2D grid, rather than restricting ourselves to
1D slices. Results are first given for just the 25\,$\mathrm{M}_{\text{\ensuremath{\odot}}}$
models (Fig.\,\ref{fig4}). The corresponding plots for all models
can be found in the Appendix.

\includegraphics[scale=0.5,angle=90]{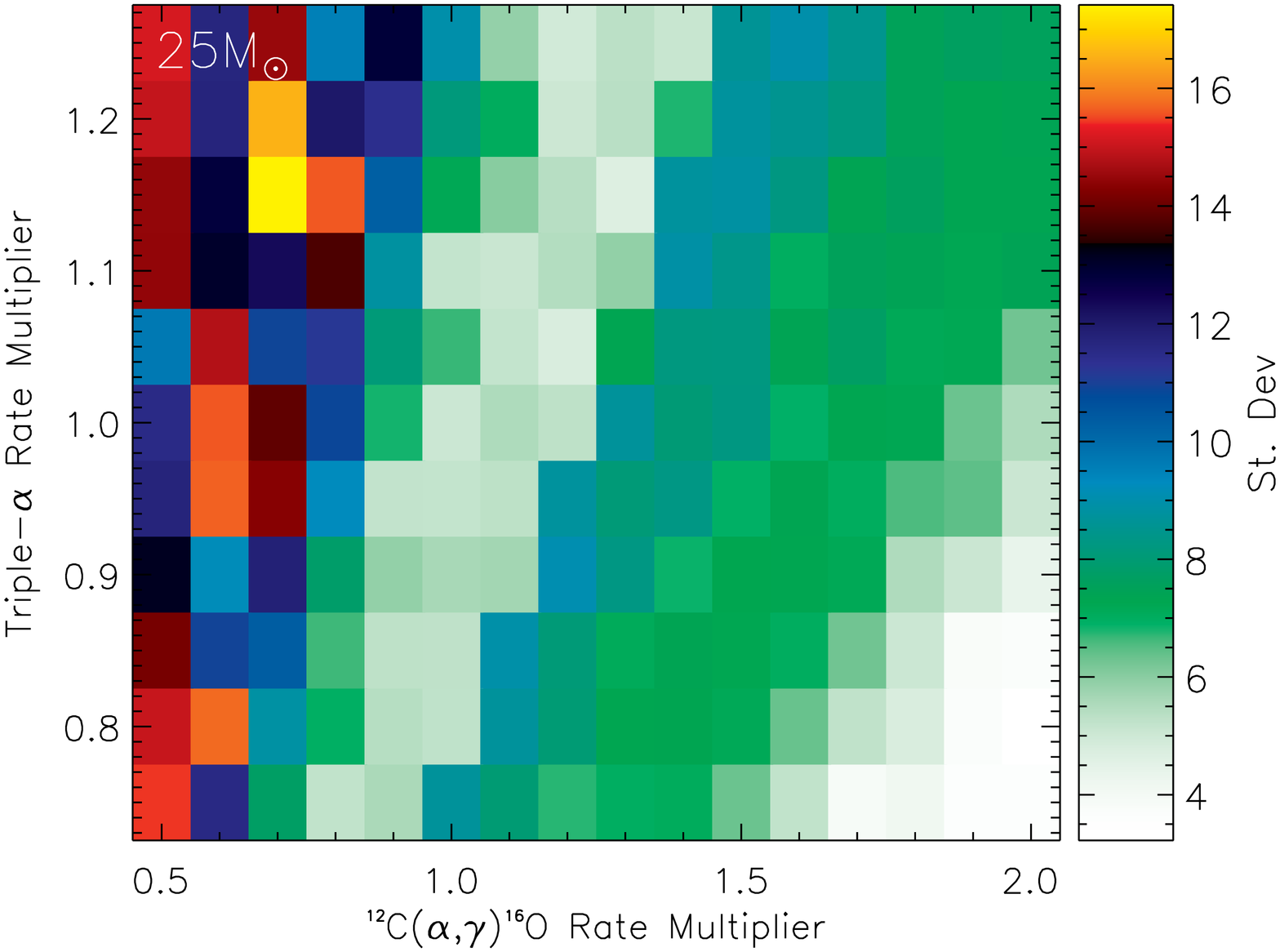}

\figcaption{\label{fig4} Standard deviations of the production factors as a
function of the $R_{\alpha,12}$ and $R_{3\alpha}$ reaction rate
multipliers for the 25\,$\mathrm{M}_{\text{\ensuremath{\odot}}}$
models. Each model is given by the $R_{\alpha,12}$ and $R_{3\alpha}$
reaction rate multiplier pair used for the helium rates. }

The best fit $R_{\alpha,12}$ and $R_{3\alpha}$ values occupy a region
in the lower right-hand corner, and a strip running approximately
though the centroid value for each rate. It is interesting that some
adjacent models display significant differences in their nucleosynthesis
despite small change in the reaction rate multipliers, for example
$\left(R_{\alpha,12},R_{3\alpha}\right)=\left(0.5,0.9\right)$,$\left(0.6,0.9\right)$
or $\left(R_{\alpha,12},R_{3\alpha}\right)=\left(0.5,1.05\right)$,$\left(0.6,1.05\right)$.
In some cases adjacent models can evolve with quite different shell
burning episodes, whereas in other cases can be very similar, such
as for $\left(R_{\alpha,12},R_{3\alpha}\right)=\left(0.9,0.9\right)$,$\left(1.0,0.9\right)$
or $\left(R_{\alpha,12},R_{3\alpha}\right)=\left(1.0,1.1\right)$,$\left(1.1,1.1\right)$.
How this occurs can be understood by considering the convective history
of adjacent models. First, the $\left(R_{\alpha,12},R_{3\alpha}\right)=\left(0.5,0.9\right)$
and $\left(R_{\alpha,12},R_{3\alpha}\right)=\left(0.6,0.9\right)$
models are given in Fig.\,\ref{fig5}, which have a difference in
$\sigma_{P}$ of 3 to 4 (Fig.\,\ref{fig4}). 

\begin{figure}[H]
\begin{minipage}[t]{1\columnwidth}%
\begin{center}
\includegraphics[width=7cm,height=12cm,angle=90]{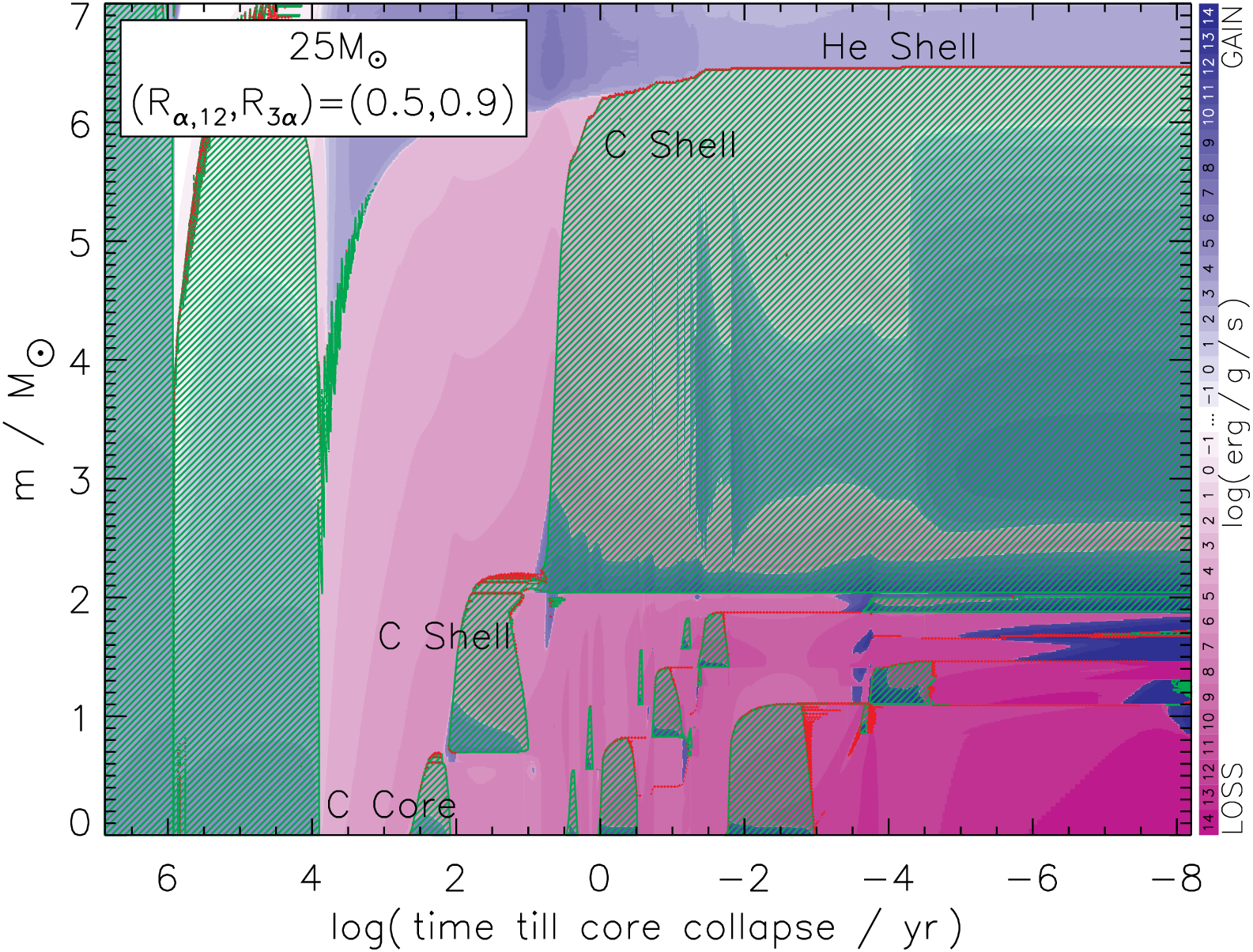}
\par\end{center}%
\end{minipage}

\medskip{}
\begin{minipage}[t]{1\columnwidth}%
\begin{center}
\includegraphics[width=7cm,height=12cm,angle=90]{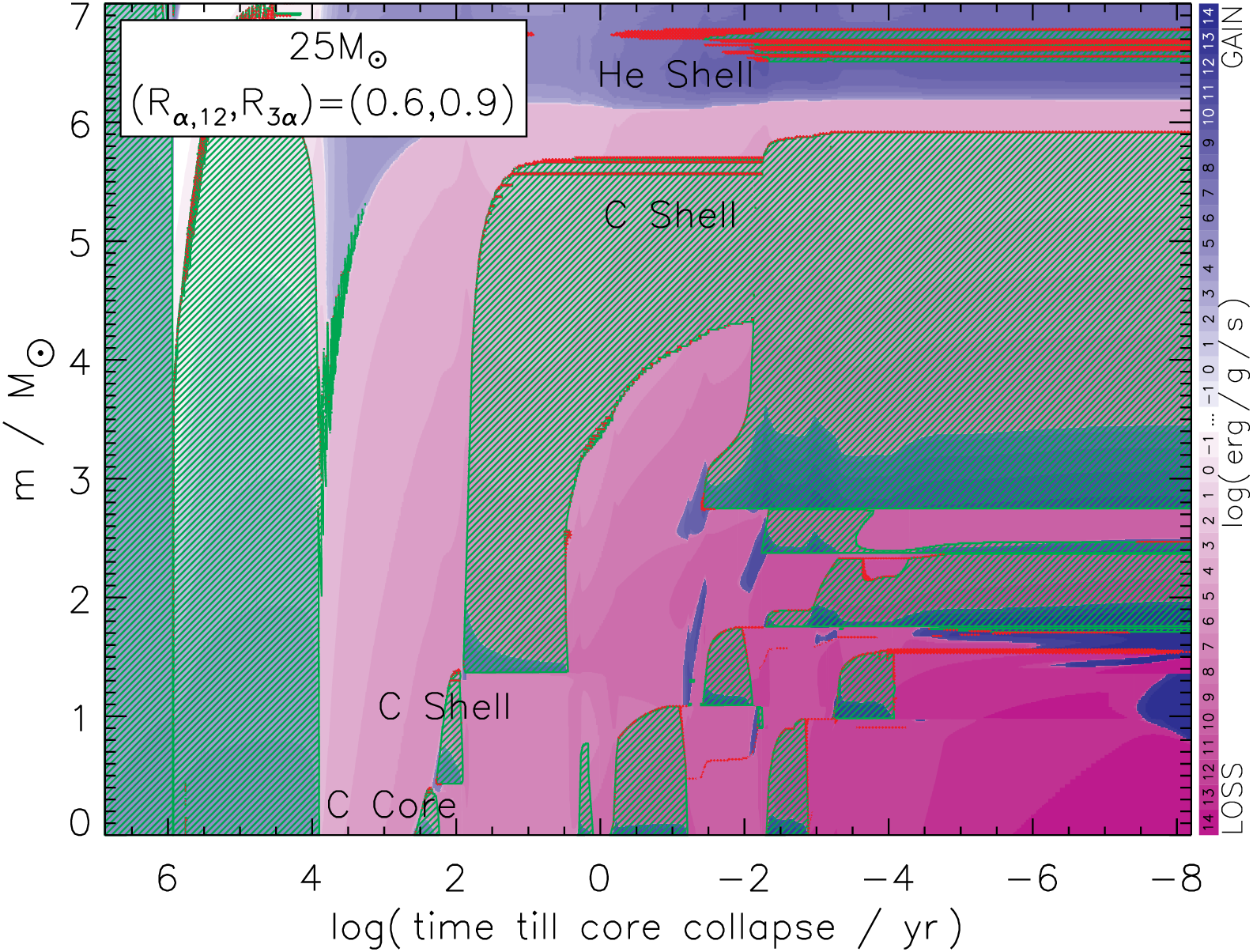}
\par\end{center}%
\end{minipage}
\end{figure}

\figcaption{\label{fig5} \emph{Top}: The convection plot for the inner 7$\,\mathrm{M}_{\odot}$
of the $\left(R_{\alpha,12},R_{3\alpha}\right)=\left(0.5,0.9\right)$
model. \emph{Bottom}: The convection plot for the $\left(R_{\alpha,12},R_{3\alpha}\right)=\left(0.6,0.9\right)$
model. Shown are convective regions (green hatch-marks), semi-convective
(red cross-hatching), energy generation from nucleosynthesis (blue),
and radiative/neutrino cooling (pink). The entire evolution from the
main sequence to onset of core collapse is shown. }

In the $\left(0.5,0.9\right)$ model we observe a convective region
that extends past the C shell and into the above He layer, with subsequent
He ingestion into the shell burning with C into O. In the $\left(0.6,0.9\right)$
model the convective layer terminates before the He shell, and the
mixing and subsequent nucleosynthesis seen in the $\left(0.5,0.9\right)$
model does not occur.

In contrast, the $\left(R_{\alpha,12},R_{3\alpha}\right)=\left(0.9,0.9\right)$
and $\left(R_{\alpha,12},R_{3\alpha}\right)=\left(1.0,0.9\right)$
models are given in Fig.\,\ref{fig6}, which show a difference in
$\sigma_{P}$ of $\leq1$. 

\begin{figure}[H]
\begin{minipage}[t]{1\columnwidth}%
\begin{center}
\includegraphics[width=7cm,height=12cm,angle=90]{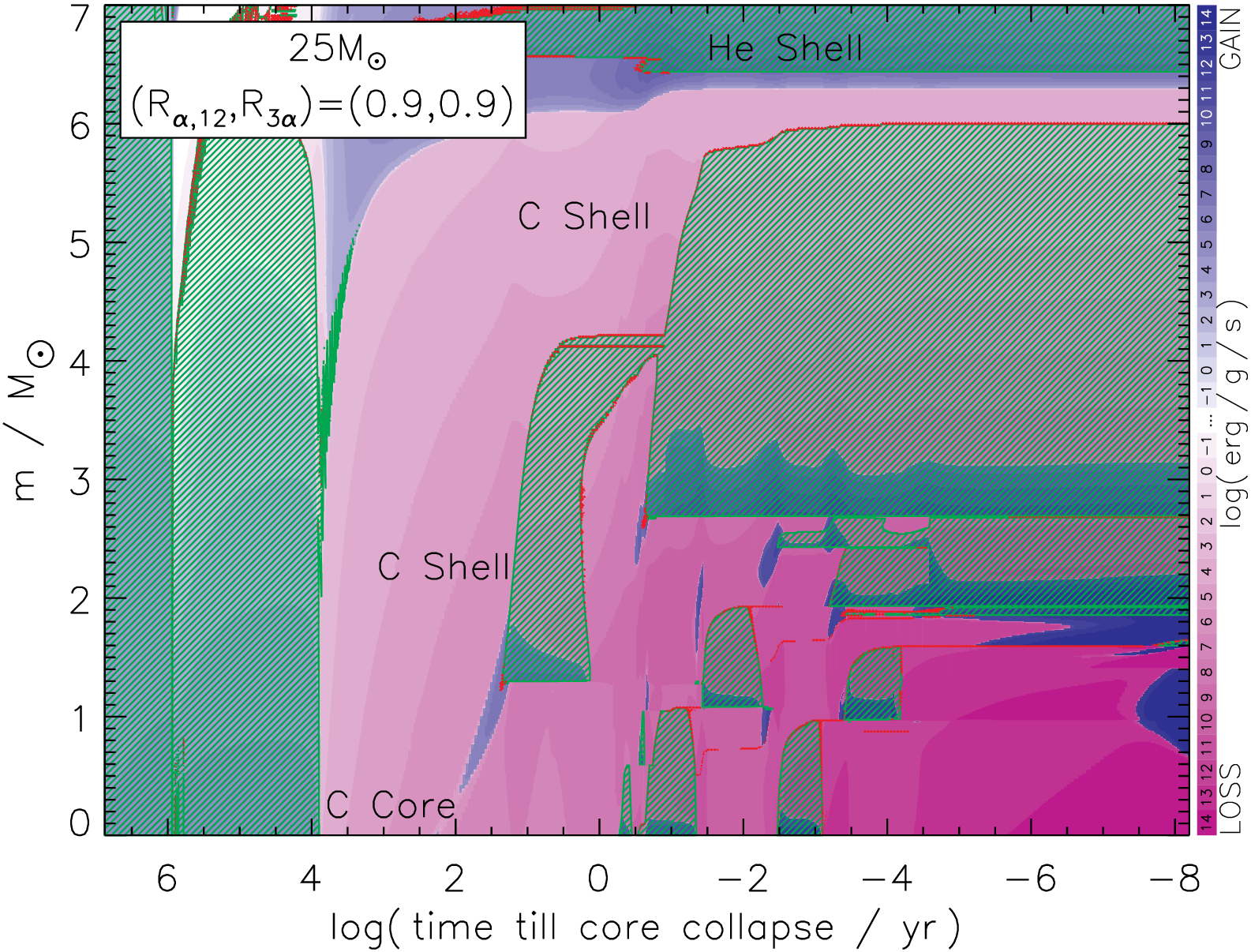}
\par\end{center}%
\end{minipage}

\medskip{}
\begin{minipage}[t]{1\columnwidth}%
\begin{center}
\includegraphics[width=7cm,height=12cm,angle=90]{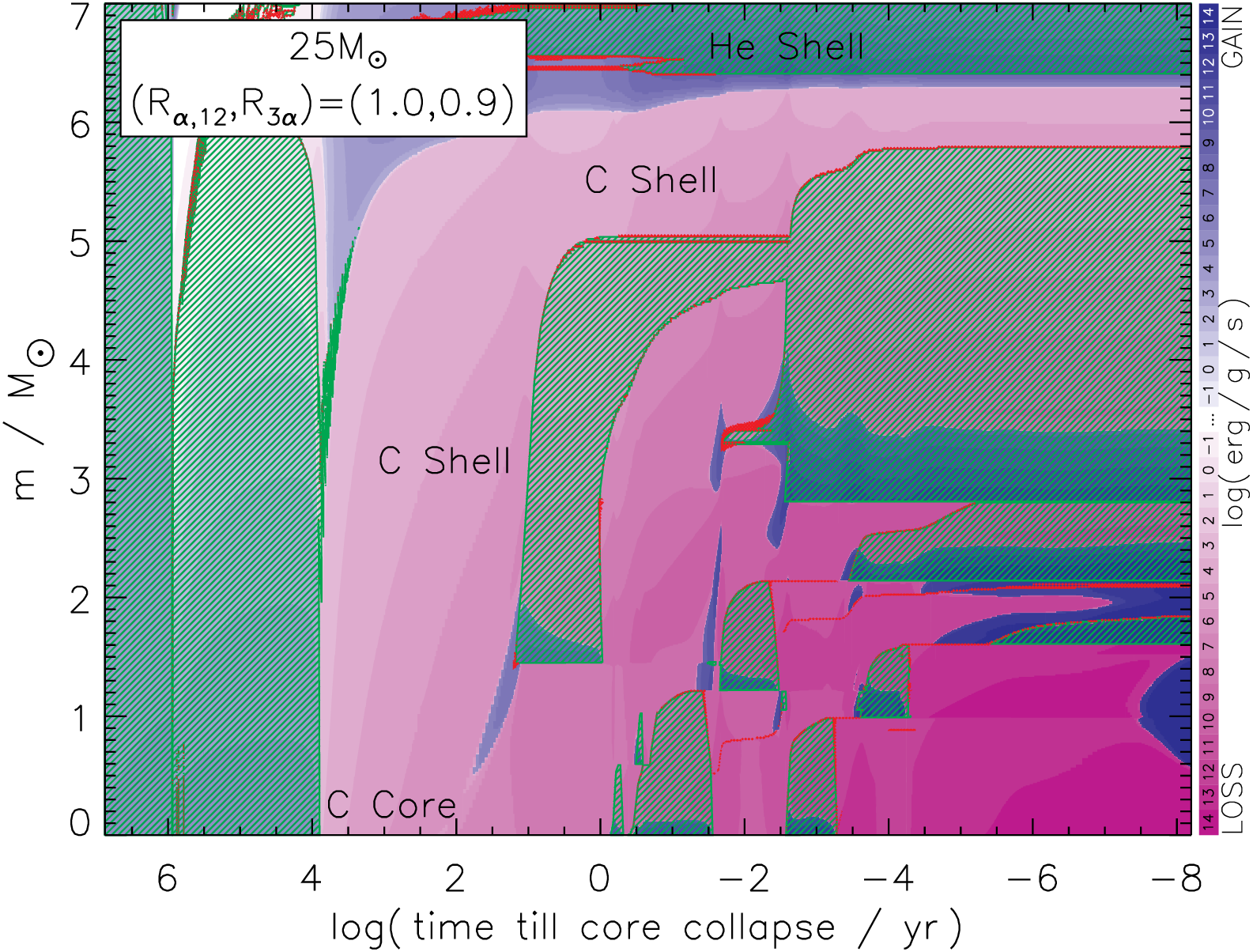}
\par\end{center}%
\end{minipage}
\end{figure}

\figcaption{\label{fig6} \emph{Top}: The convection plot for the inner 7$\,\mathrm{M}_{\odot}$
of the $\left(R_{\alpha,12},R_{3\alpha}\right)=\left(0.9,0.9\right)$
model. \emph{Bottom}: The convection plot for the $\left(R_{\alpha,12},R_{3\alpha}\right)=\left(1.0,0.9\right)$
model. See Fig.\,\ref{fig5} for a detailed description.}

In both Fig.\,\ref{fig5} and Fig.\,\ref{fig6} the $\Delta R_{3\alpha}$
is the same; however, in the latter the $R_{\alpha,12}$ values are
sufficiently large to result in radiative core-C burning, and C shell
burning episodes in the $\left(R_{\alpha,12},R_{3\alpha}\right)=\left(0.9,0.9\right)$
and $\left(1.0,0.9\right)$ models (Fig.\,\ref{fig6}) that do not
interact convectively with the He layer. Generally, larger carbon
abundances at the end of core-He burning can support longer and more
energetic carbon shell burning episodes, which can result in He ingestion
leading to different nucleosynthesis.

The differences of adjacent models depend on the initial stellar mass,
and it is expected that regions in the $\left(R_{\alpha,12},R_{3\alpha}\right)$
parameter space that have different standard deviations at one mass
may not at another. To average these effects across the IMF, Equations\,\ref{eq:1}
and \ref{eq:2} were used for the entire set of models for all masses,
and the standard deviations for the production factors are given Fig.\,\ref{fig7}.

Changes in the \textbf{$^{12}\mathrm{C}+{}^{12}\mathrm{C}$} reaction
rate may also affect nucleosynthesis; rate uncertainties at low temperatures
were thought to be large, perhaps orders of magnitude. A heavy ion
fusion study by \textbf{\citet{Jiang2007}} reported a rate decrease
at low energy. On the other hand, \textbf{\citet{Spillane2007}} found
a strong increase in rates due to a low energy resonance. A still
lower energy resonance at lower energy had weak experimental support.
Possible effects have been studied by \citet{Pignatari2013}\textbf{
}for a 25\,$\mathrm{M}_{\odot}$ half-solar metallicity star. They
found significant changes in the production of \emph{s}-process elements.
However, their calculations were not carried through to solar collapse
and an explosion, which affects the production of these isotopes \citep{Tur2009}.
In addition, recent measurements by \citet{Zickefoose2010}, reported
by \citet{Notani2012} give an S-factor 50 times smaller than the
previously reported value. For these reasons it is not clear how much
the nucleosynthesis considered in this paper would be affected.

\includegraphics[angle=90,scale=0.5]{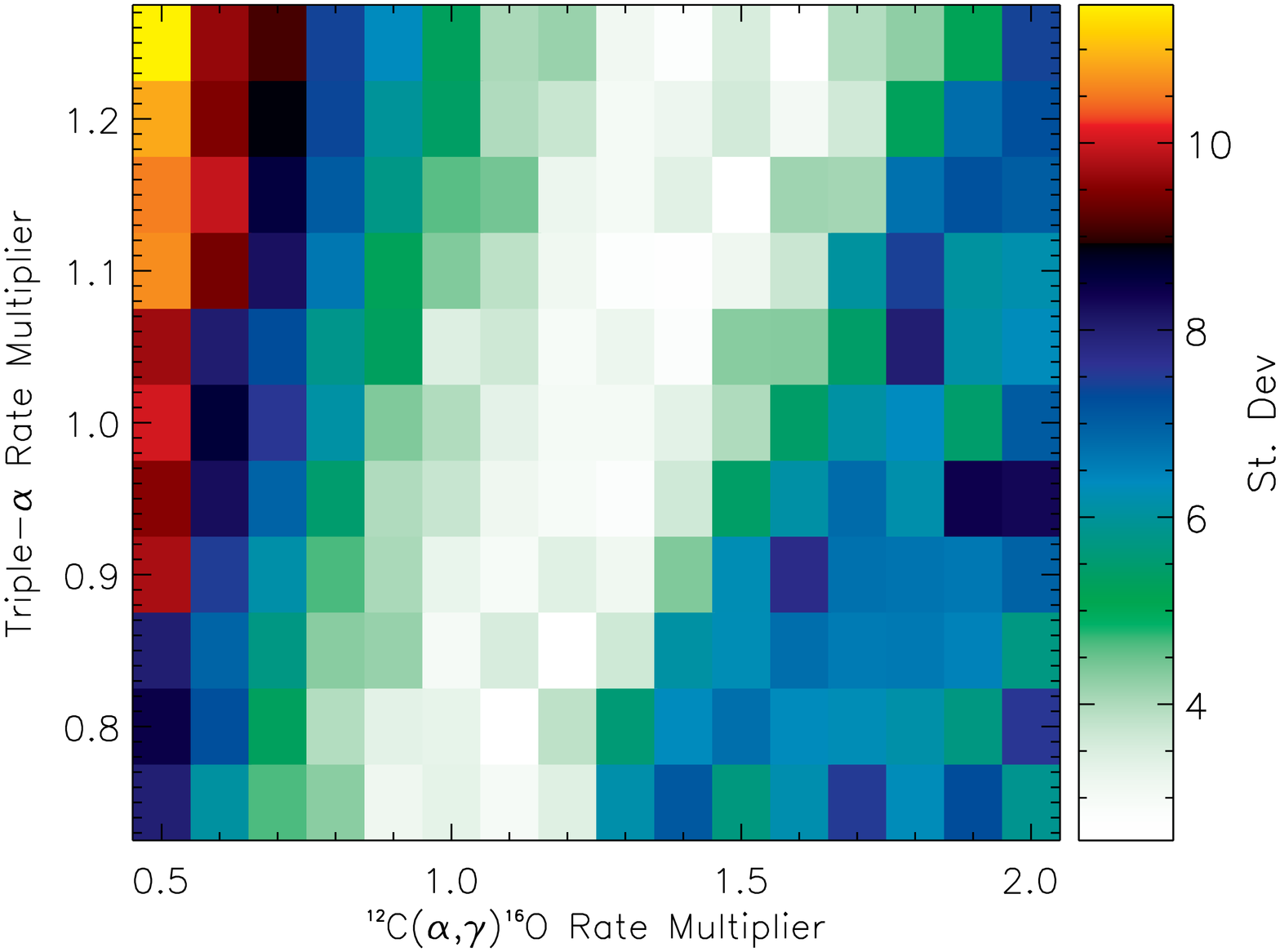}

\figcaption{\label{fig7} Standard deviations of the IMF averaged production
factors for the intermediate-mass isotopes. The entire grid of initial
masses was used.}

The results in Fig.\,\ref{fig7} show a region of small standard
deviation $\left(\sigma_{P}\lesssim4\right)$ that extends across
models within $\left(R_{\alpha,12},R_{3\alpha}\right)=\left(1.0,0.75\right)$
to $\left(1.6,1.25\right)$ and defined with a slope (in rate multiplier
ratios) close to unity with a spread of $\approx\pm0.2$ in $R_{\alpha,12}$.
The IMF averaged production factors for the weak \emph{s}-process
isotopes are shown in Fig.\,\ref{fig8}. The results for the individual
masses can be found in the appendix for the weak \emph{s}-only isotopes.

\includegraphics[angle=90,scale=0.5]{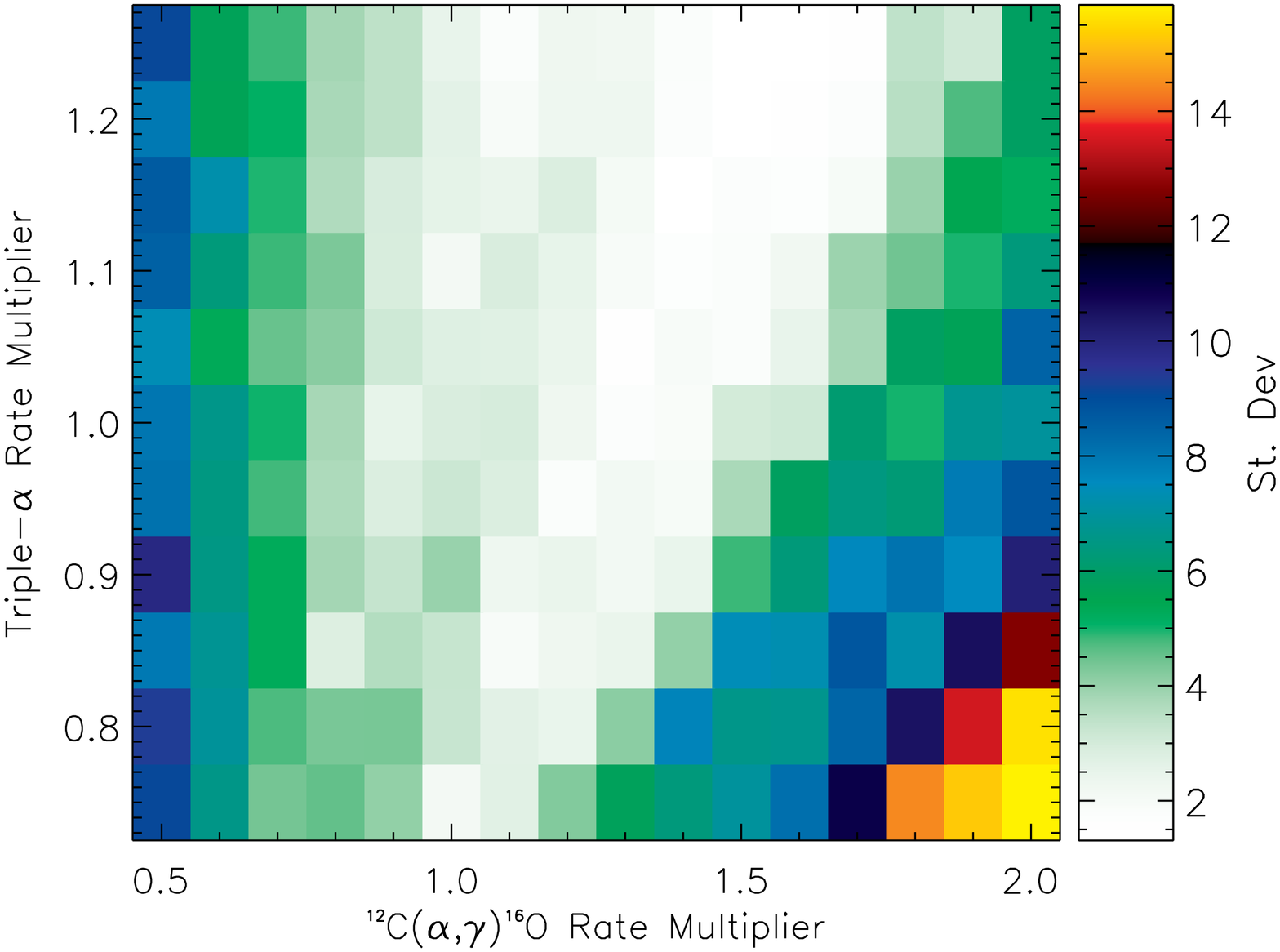}

\figcaption{\label{fig8}Standard deviations for the IMF averaged production
factors for the weak \emph{s}-only isotopes, using the entire grid
of initial masses.}

The results in Fig.\,\ref{fig8} show a region of small standard
deviation $\left(\sigma_{P}\lesssim4\right)$ that extends across
models within $\left(R_{\alpha,12},R_{3\alpha}\right)=\left(1.3,1.0\right)$
to $\left(1.5,1.25\right)$ and defined with a slope close to unity
with a spread of $\approx\pm0.1$ in $R_{\alpha,12}$. The production
factors for all isotopes for the $\left(R_{\alpha,12},R_{3\alpha}\right)=\left(1.3,1.0\right)$
model are given in Fig.\,\ref{fig9}. This model lies within the
region of minimum standard deviation for both the intermediate-mass
and weak \emph{s}-only isotopes. The neutrino-process isotopes, $\mathrm{^{7}Li}$,
$\mathrm{^{11}B}$, $^{19}\mathrm{F}$, $^{138}\mathrm{La}$, and
$^{180}\mathrm{Ta}$ all show over-productions. Specifically, $\mathrm{^{11}B}$
shows a production factor very close to $^{16}\mathrm{O}$ $\left(P_{11\mathrm{B}}/P_{16\mathrm{O}}=0.97\right)$,
which agrees with \citet{Austin2011}; these authors also show that
this ratio varies by more than a factor of 2 at different values of
$R_{\alpha,12}$. The low values for most of the heavy nuclei in Fig.\,\ref{fig9}
are expected, as we did not include the \emph{r}-process or \emph{s}-process
contributions from AGB stars. 

\includegraphics[angle=90,scale=0.5]{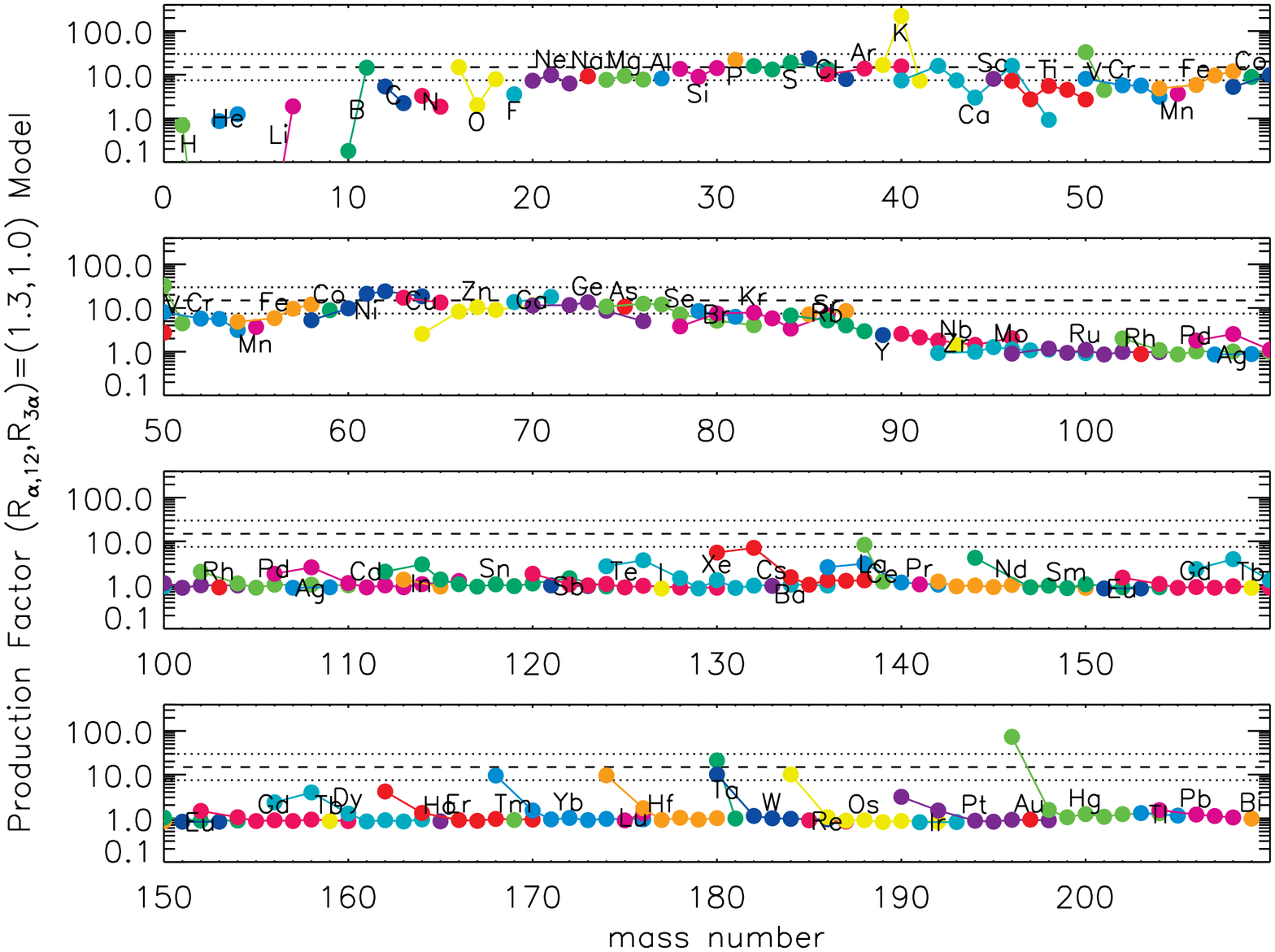}

\figcaption{\label{fig9}Production factors for all isotopes for the $\left(R_{\alpha,12},R_{3\alpha}\right)=\left(1.3,1.0\right)$
model. Shown are the $^{16}\mathrm{O}$ production factor (dashed
line) with $\pm2$ ranges (dotted lines).}

\subsection{Implications for Stellar Remnants }

The analysis above assumed a perfect supernova success rate; i.e.,
abundances from all models were included even though some would collapse
to a black hole without enriching the ISM with a SN event. We thus
performed an additional analysis that removed the models that may
result in possible ``failed'' supernovae, prior to IMF averaging.
Black hole formation following core collapse has been investigated
recently by \citet{OConnor2011}, who identified a single parameter
that can be used to roughly infer the fate of the core collapse event,
using 62 progenitors. This compactness parameter is defined generally
as,

\begin{equation}
\xi_{M}=\frac{M/\mathrm{M}_{\odot}}{R\left(M\right)/1000\,\mathrm{km}},
\end{equation}
where $M$ is the baryonic mass, and $R\left(M\right)$ is the radial
coordinate that encloses $M$ at the time of core bounce. The relevant
specific $\xi_{M}$ for black hole formation is at a mass of $M=2.5\,\mathrm{M}_{\odot}$,
and $\xi_{2.5}$ is used by \citet{OConnor2011} to distinguish a
possible boundary between successful and failed supernova explosions
at the value of $\xi_{2.5}=0.45$. Their models with $\xi_{2.5}<0.45$
were concluded to be likely successful supernova, using considerations
of the time-averaged neutrino heating efficiency and subject (albeit
mildly) to the equation of state (EOS) employed.

A more recent analysis of over 100 supernova simulations by \citet{Ugliano2012}
have resultant NS and BH mass ranges that are compatible with a possible
paucity of low mass BHs, which may imply a lower $\xi_{2.5}$ value
than used by \citet{OConner2011}. A more refined boundary of $\xi_{2.5}\thickapprox0.25$
has been proposed by \citet{Woosley2012}, and will be adopted in
the present work. 

In our analysis all models are assumed to have the same final kinetic
energy for the ejecta (1.2\,B), however, the explosion energies can
vary. A larger explosion energy would cause successful SNe above $\xi_{2.5}=0.25$,
since now larger densities would be required to overcome this larger
energy and prevent a successful SN explosion. It would also cause
more material above the Fe core to escape the gravitational potential,
resulting in a smaller mass cut and remnant mass for models already
below this limit. 

The $\xi_{2.5}$ values were computed for each model. The distribution
of $\xi_{2.5}$ values for the 25\,$\mathrm{M}_{\text{\ensuremath{\odot}}}$
models is given as an example in Fig.\,\ref{fig10}. Figures for
the $\xi_{2.5}$ values of the other masses can be found in the Appendix.

\includegraphics[scale=0.5,angle=90]{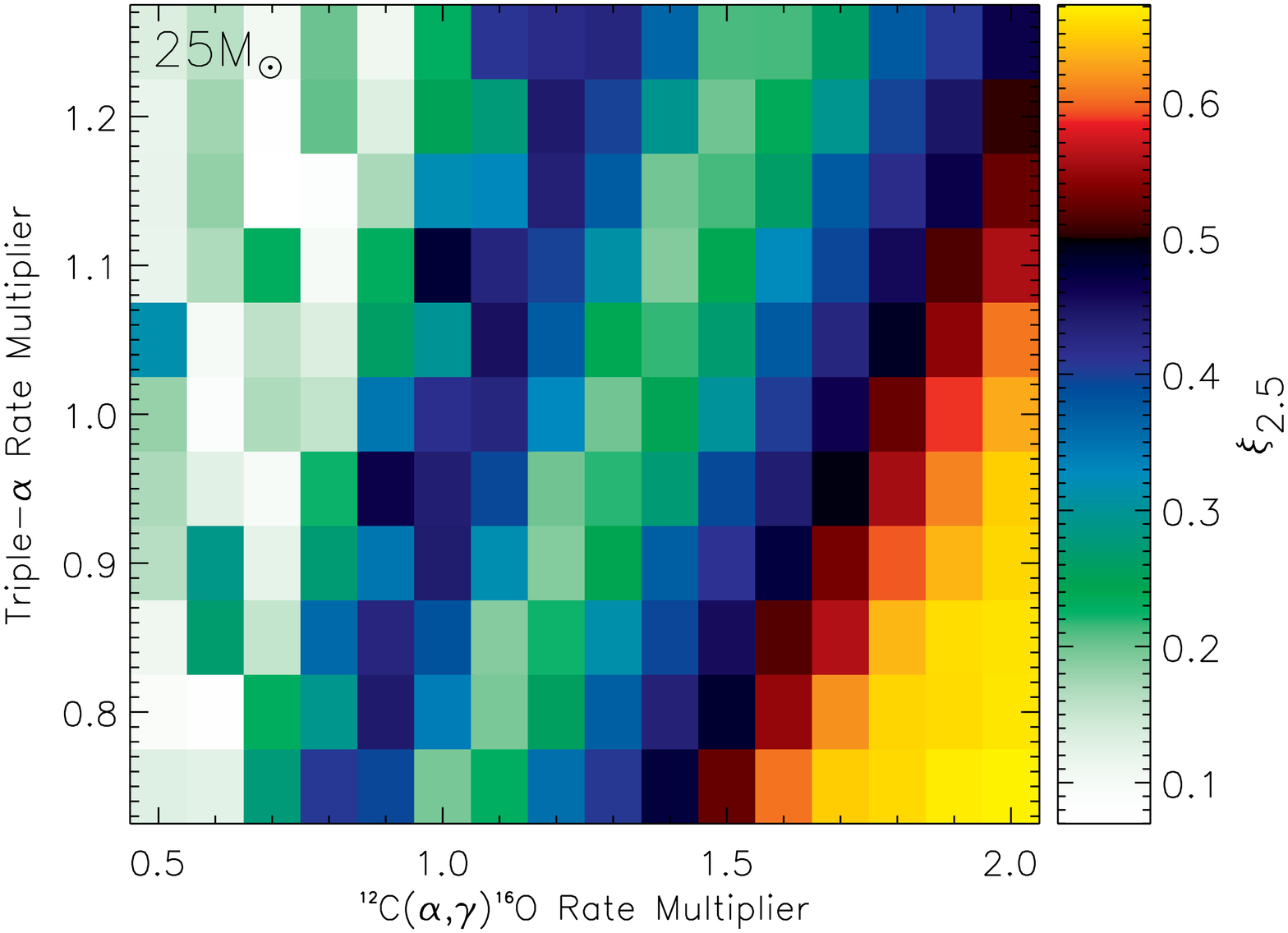}

\figcaption{\label{fig10} The distribution of $\xi_{2.5}$ values for the 25\,$\mathrm{M}_{\text{\ensuremath{\odot}}}$
models.}

For the 25\,$\mathrm{M}_{\text{\ensuremath{\odot}}}$ models, those
with low $R_{\alpha,12}$ values are favored for successful SNe events,
in addition to a local minimum in $\xi_{2.5}$ defined by a narrow
strip close to the centroid value for this rate. In comparison, the
12\,$\mathrm{M}_{\text{\ensuremath{\odot}}}$, 13\,$\mathrm{M}_{\text{\ensuremath{\odot}}}$,
14\,$\mathrm{M}_{\text{\ensuremath{\odot}}}$, 15\,$\mathrm{M}_{\text{\ensuremath{\odot}}}$,
and 16\,$\mathrm{M}_{\text{\ensuremath{\odot}}}$ models \emph{all}
explode as successful SNe, whereas the 17\,$\mathrm{M}_{\text{\ensuremath{\odot}}}$
models have only 6 of 176 that fail. The 18\,$\mathrm{M}_{\text{\ensuremath{\odot}}}$
models have 16 failed SNe, all above $R_{\alpha,12}\geq0.9$ and spanning
the whole range of $R_{3\alpha}$. Among our models of 20\,$\mathrm{M}_{\text{\ensuremath{\odot}}}$,
several fail towards high $R_{\alpha,12}$ and low $R_{3\alpha}$
values, with a fairly delineated boundary between successful and failed
SNe beginning at a multiplier pair value of $\left(R_{\alpha,12},R_{3\alpha}\right)=\left(1.2,0.75\right)$,
and ending at $\left(1.8,1.25\right)$. The 22\,$\mathrm{M}_{\text{\ensuremath{\odot}}}$
and 27\,$\mathrm{M}_{\text{\ensuremath{\odot}}}$ models show the
same behavior as the 20\,$\mathrm{M}_{\text{\ensuremath{\odot}}}$
models but with a boundary favoring more BHs, delineated by a slope
defined by $\left(1.0,0.75\right)$ and $\left(1.4,1.25\right)$ for
the former, and $\left(0.5,0.75\right)$ and $\left(0.8,1.25\right)$
for the latter. Among our models of 30\,$\mathrm{M}_{\text{\ensuremath{\odot}}}$,
there are 10 that explode at $R_{\alpha,12}=0.5$ along with a strip
of successful SNe from $\left(0.8,0.75\right)$ to $\left(1.3,1.25\right)$.

The yields from the stellar winds of all models, as well as the explosive
yields of the models that satisfy the condition $\xi_{2.5}<0.25$
were then averaged over an IMF. The result is given in Fig.\,\ref{fig11}
for the intermediate-mass isotopes and in Fig.\,\ref{fig12} for
the weak \emph{s}-only isotopes.

\includegraphics[angle=90,scale=0.5]{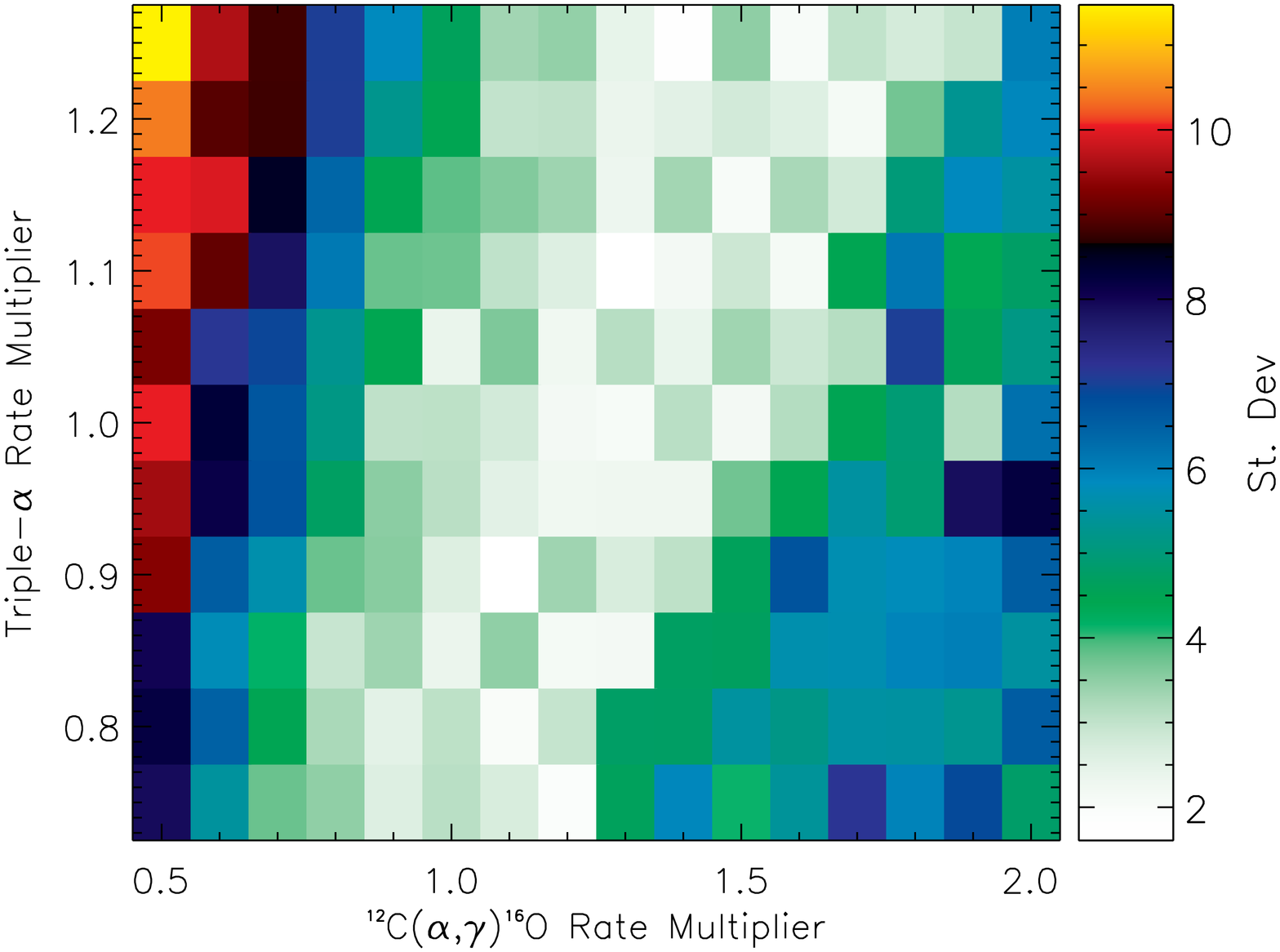}

\figcaption{\label{fig11} Standard deviations for the IMF-averaged production
factors for the intermediate-mass isotopes. The yields from all stellar
winds and the explosive yields from models that satisfy the condition
$\xi_{2.5}<0.25$ were used in the averaging.}

\includegraphics[angle=90,scale=0.5]{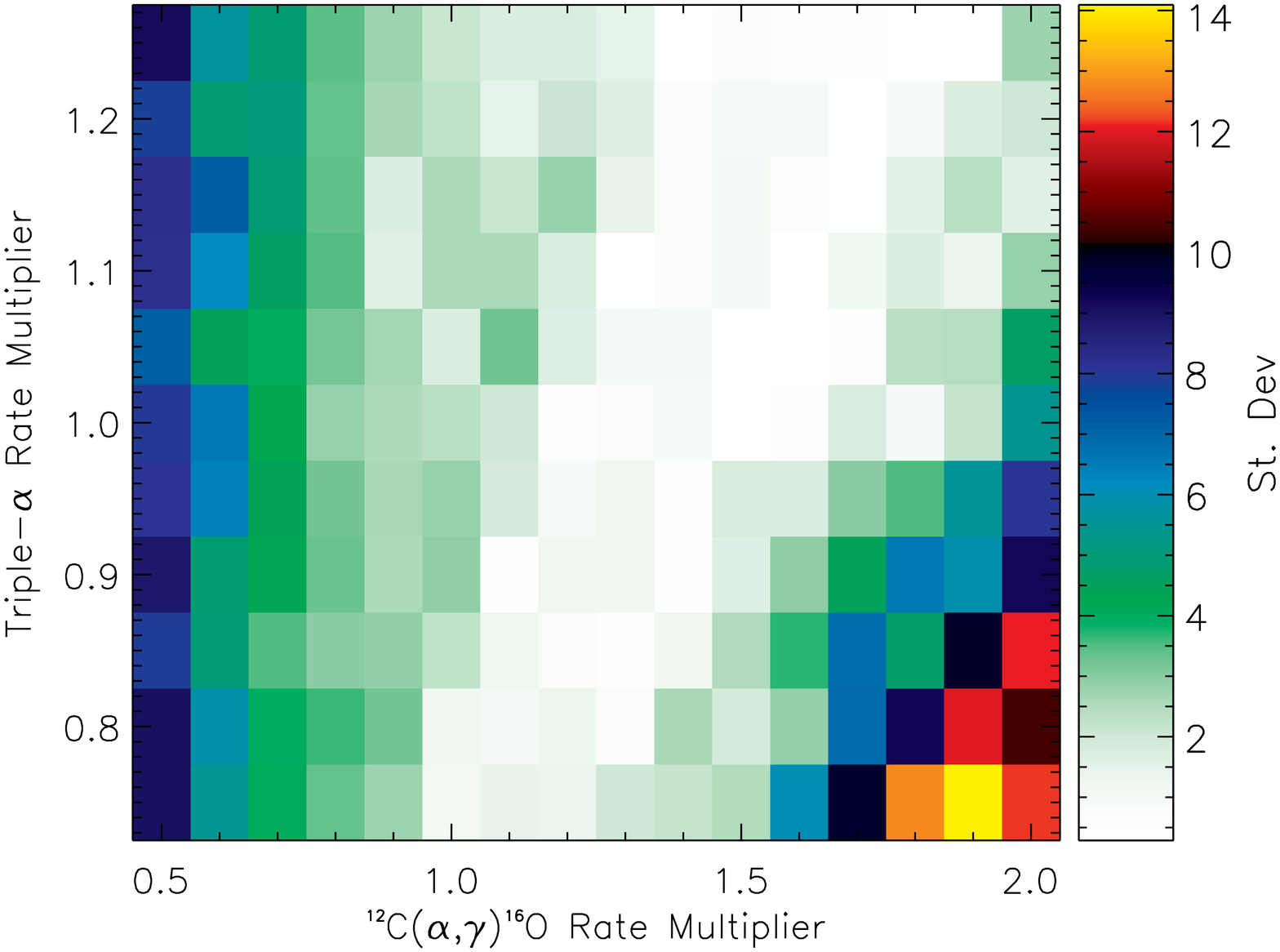}

\figcaption{\label{fig12} Standard deviations for the IMF-averaged production
factors for the weak \emph{s}-only isotopes. The yields from stellar
winds and from the explosive yields of models that satisfy the condition
$\xi_{2.5}<0.25$ were used in the averaging.}

As stated, Fig.\,\ref{fig7} showed a small standard deviation region
$\left(\sigma_{P}\lesssim4\right)$. This region extended across models
within $\left(R_{\alpha,12},R_{3\alpha}\right)=\left(1.0,0.75\right)$
to $\left(1.5,1.25\right)$ and was defined with a slope close to
unity with a spread of $\approx\pm0.2$ in $R_{\alpha,12}$. The impact
of the compactness parameter for the intermediate-mass isotopes is
that the $\left(\sigma_{P}\lesssim4\right)$ region from Fig.\,\ref{fig7}
is still observed in Fig.\,\ref{fig11}, but the latter has models
in this region whose standard deviation has $\sigma_{P}$ larger by
1, although this effect may be less if an even finer mass grid was
used. We conclude that the current $25\,\%$ uncertainty range of
$R_{\alpha,12}$ agrees with observations. The region of small standard
deviations for Fig.\,\ref{fig11} has a slope defined approximately
by $\left(R_{\alpha,12},R_{3\alpha}\right)=\left(1.1,0.75\right)$
to $\left(1.6,1.25\right)$ with a spread of $\approx\pm0.2$ in $R_{\alpha,12}$
for $R_{3\alpha}\lesssim0.95$, and spread of $\approx\pm(0.3-0.4)$
in $R_{\alpha,12}$ for $R_{3\alpha}\gtrsim0.95$. It is then possible
to define the relation, 

\begin{equation}
R_{\alpha,12}=1.0R_{3\alpha}+(0.25\pm0.3).\label{eq:5}
\end{equation}
For a chosen $R_{3\alpha}$ value in the range 0.75 to 1.25, Equation\,\ref{eq:5}
gives a range of $R_{\alpha,12}$ values that agree with observations,
taking into account only the intermediate-mass isotopes.

The impact of the compactness parameter on the weak \emph{s}-only
isotopes is that the $\left(\sigma_{P}\lesssim4\right)$ region from
Fig.\,\ref{fig8} becomes somewhat \emph{more} sharply defined, in
that the models above $R_{3\alpha}\approx0.95$, from $R_{\alpha,12}\approx1.0$
to $1.3$ have increased standard deviations by $\approx1-2$. In
Fig.\,\ref{fig12} there is a region of small $\sigma_{P}$ with
a slope defined by $\left(1.1,0.75\right)$ to $\left(1.6,1.25\right)$,
with an average spread of $\approx\pm0.2$ in $R_{\alpha,12}$. This
result should be interpreted with care, however, since the IMF averaging
is subject to the $X_{i,w}$ values, taken from an approximate analysis
of weak \emph{s}-process contributions to the solar abundances \citep{West2012}
which uses only the relevant nuclear physics and does not employ stellar
modeling. Furthermore, the decomposition of the weak \emph{s}-process
contributions does not address recent works that indicate possible
evidence for an increase of the \emph{s}-elements in the Galactic
disk \citep{Mashonkina2007,Maiorca2011,Maiorca2012,Jacobson2012}.
It is unclear what impact this would have on our analysis, however,
since the weak s-isotopic abundances do not dominate their respective
elemental abundances. Despite these issues, the weak \emph{s}-process
analysis in the present work is a step in the right direction; one
simply cannot compare the weak \emph{s}-process yields directly to
the solar abundances (which contain main \emph{s}-process abundances
also). We thus caution the reader that whereas our weak \emph{s}-process
analysis is an improvement, it is also weakly constrained. 

The best values for the helium rates from our analysis should agree
with observations for \emph{both} the intermediate-mass and weak \emph{s}-only
isotope sets. We thus computed the average of the standard deviation
values for both sets (Fig.\,\ref{fig13}). Whereas the optimal values
for the helium burning rates from our analysis should reproduce the
observed abundance ratios for both the intermediate-mass and weak
\emph{s}-only isotope sets, the two sets will not be at the same level
of production factor, because they have different astrophysical natures
and galactic chemical evolution histories. For example, the weak \emph{s}-process,
being of secondary nature, is overproduced by a factor of $\approx2.0$
at solar metallicity and less is made at lower metallicities, yet
their \emph{relative} abundance ratio should be about solar - and
this is what we match. In contrast, the intermediate-mass isotopes
are primary and should be produced at a solar level, and again within
this subgroup the isotope ratios should be at solar level. Hence,
they were analyzed separately before the results were combined, instead
of computing the standard deviations for all isotopes as a single
group. The ratio of weak \emph{s}-process to intermediate-mass isotopes
is not the solar abundance ratio and not expected to be, so fitting
both at once would be wrong.

\includegraphics[angle=90,scale=0.5]{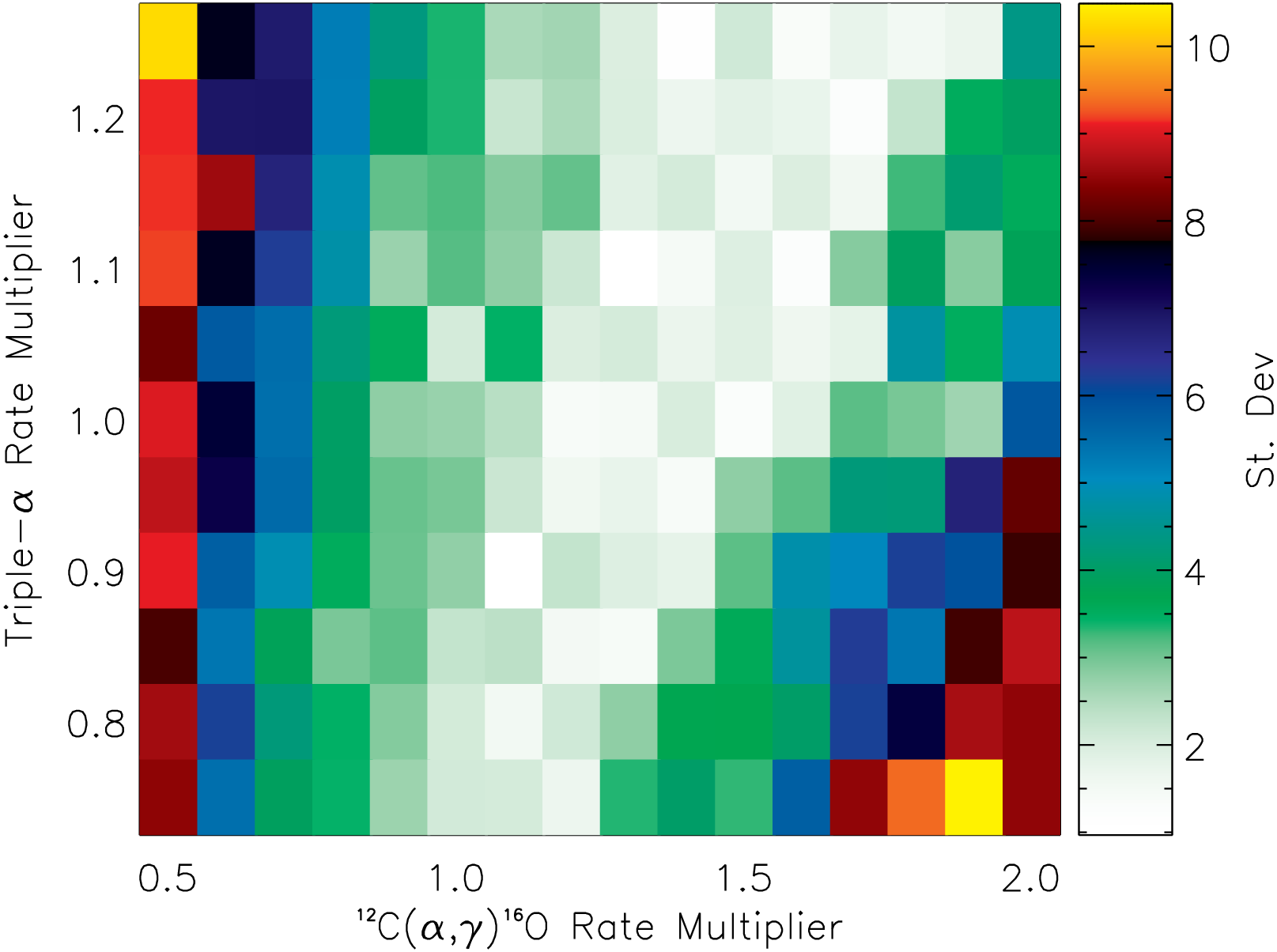}

\figcaption{\label{fig13} The average of the standard deviations for the production
factors of the intermediate-mass (Fig.\,\ref{fig11}) and weak \emph{s}-only
isotopes (Fig.\,\ref{fig12}). }

The region of small standard deviations for Fig.\,\ref{fig13} has
a slope defined approximately by $\left(R_{\alpha,12},R_{3\alpha}\right)=\left(1.1,0.75\right)$
to $\left(1.6,1.25\right)$ with a spread of $\approx\pm0.1$ in $R_{\alpha,12}$
for $R_{3\alpha}\lesssim0.95$, and spread of $\approx\pm(0.2-0.3)$
in $R_{\alpha,12}$ for $R_{3\alpha}\gtrsim0.95$. We then define
the relation, 

\begin{equation}
R_{\alpha,12}=1.0R_{3\alpha}+(0.35\pm0.2).\label{eq:6}
\end{equation}
For a chosen $R_{3\alpha}$ value in the range 0.75 to 1.25, Equation\,\ref{eq:6}
gives a range of $R_{\alpha,12}$ values that agree with observations,
taking into account both the intermediate-mass and weak \emph{s}-only
isotopes. Note that this relation does not accurately define the regions
of small standard deviation for either isotope set individually (see
Fig.\,\ref{fig11} and Fig.\,\ref{fig12}). Note further that the
region of small $\sigma_{P}$ in Fig.\,\ref{fig13} is ``noisy,''
and has values along the line of best fit that are not minima of this
region. We must re-emphasize that the analysis used is approximate,
and the region contains values that may change if we change the parameters
of the model or improve the completeness of the models included. Hence,
whereas we can define this region, the analysis is likely insufficient
to reliably distinguish between neighboring values within it. Subject
to the approximations employed in the analysis, coordinate pairs for
the reaction rate multipliers that satisfy this relation results in
nucleosynthesis that equally agrees with current observations for
both the intermediate and weak \emph{s}-only isotopes, as far as the
present study can determine. 

We also explored the dependence of our results on the IMF used in
Equation\,\ref{eq:1}. We computed IMF averaged production factors
for both the intermediate and weak \emph{s }isotope list using two
modified Salpeter IMFs ($+0.3$ and $-0.3$ added to the exponent).
For the weak\emph{ s}-isotope list, both modified IMFs resulted in
a difference in IMF-averaged production factors by $\leq6\,\%$ in
the region of best fit rates identified by Equation\,\ref{eq:6}.
For the intermediate-mass isotope list, both modified IMFs resulted
in a difference in IMF averaged production factors by $\leq4\,\%$
in the region of best fit rates. We thus believe the choice of IMF
has only a small impact on the results, provided a reasonable IMF
is chosen.

The stellar model KEPLER, however, is only approximate: convection
is treated using mixing length theory, effects of rotation, binary
star evolution, and magnetic fields are ignored, many reaction rates
and mass loss rates are not well enough known, the opacities have
uncertainties, etc. Some of these effects have been investigated.
For example, \citet{Chieffi2012} studied the impact of rotation on
solar metallicity massive stars in the range $13-120\,\mathrm{M}_{\odot}$,
and found over-productions of F and slight over-productions of weak
\emph{s}-process isotopes. Another study by \citet{Iliadis2011} found
several reaction rate uncertainties that influence massive star Al
production, and found a range of $^{26}\mathrm{Al}$ larger than those
found by \citet{Tur2007}. Although it was not the purpose of the
present work to address the effects of the approximations in the stellar
models, it is important to note that they can have a non-negligible
impact on the results in some cases.\textbf{ }Additionally, we interpolated
yields across a finite IMF sampling and only for solar metallicity
CCSNe stars instead of a full galactic chemical evolution model from
big bang nucleosynthesis to the present Galaxy including all nucleosynthesis
sources. For the isotopes we compare in this work, however, the assumption
that solar composition stars should produce about their solar ratios
is reasonable.

\subsection{Variations in Carbon Mass Fractions and Remnant Mass}

The baryonic mass of the progenitor of the remnant depends on the
central carbon mass at the end of core-He burning%
\footnote{The gravitational mass of the remnant also depends on type, formation
scenario \citep{Zhang2008}, and equation of state \citep{Lattimer2001}.%
}. An increase in $R_{3\alpha}$ or decrease in $R_{\alpha,12}$ results
in an increase in the carbon abundance. An example of the resulting
trend in baryonic mass is given in Fig.\,\ref{fig14}, for the 25\,$\mathrm{M}_{\text{\ensuremath{\odot}}}$
models. Figures for the baryonic masses of the other models can be
found in the Appendix.

\includegraphics[angle=90,scale=0.5]{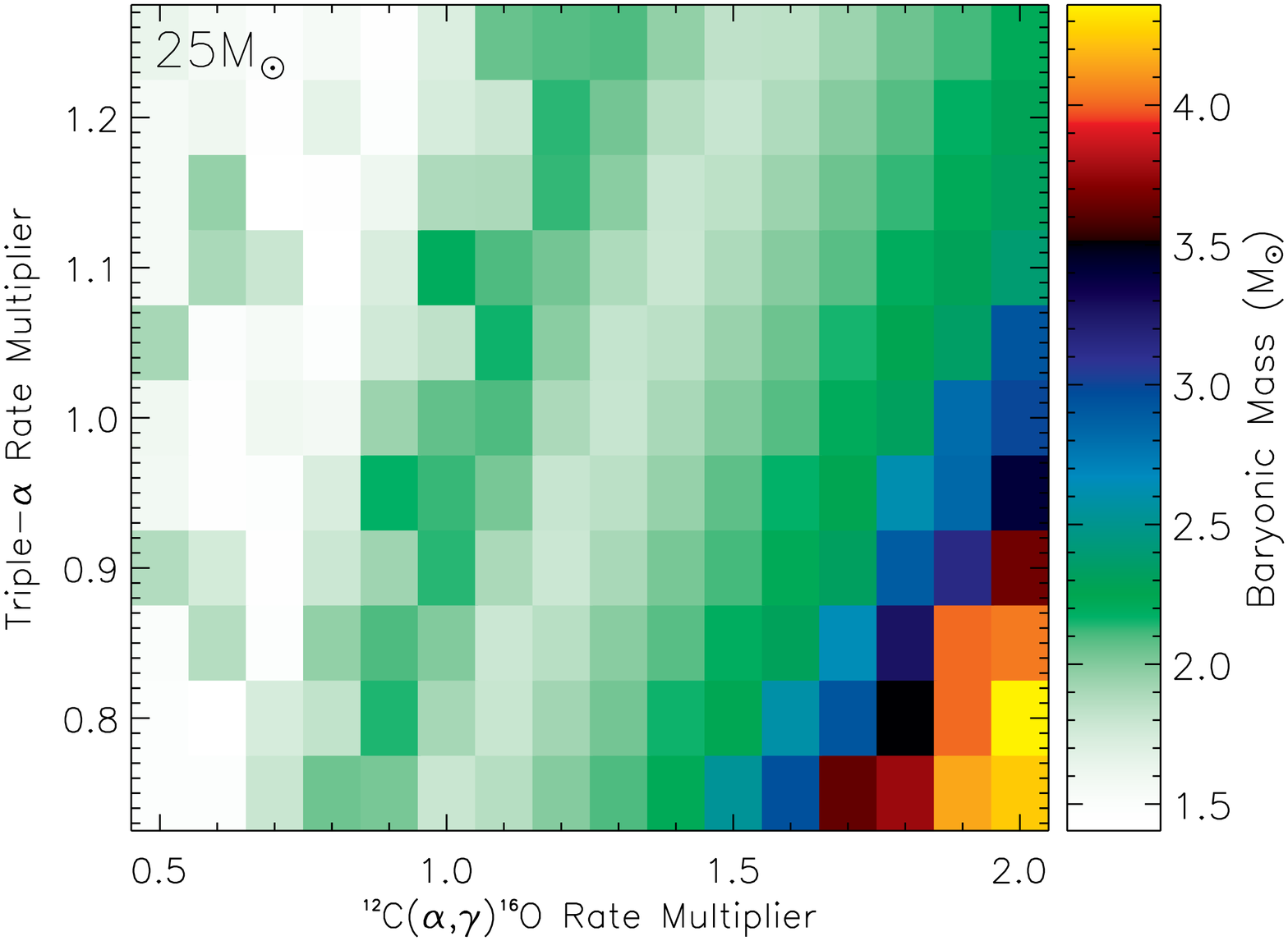}

\figcaption{\label{fig14} The baryonic mass of the progenitor of the remnant
for the 25\,$\mathrm{M}_{\text{\ensuremath{\odot}}}$ models as a
function of the $R_{3\alpha}$ and $R_{\alpha,12}$ multipliers.}

Fig.\,\ref{fig14} shows that there is a decrease in baryonic remnant
mass for increasing $R_{3\alpha}$ and decreasing $R_{\alpha,12}$.
This is because a larger carbon abundance at the end of core-He burning
can support longer and more energetic carbon shell burning episodes,
which allows the core to cool to lower entropy, yielding a smaller
progenitor \citep{Woosley2003,Tur2007}. Note the apparent local maximum
beginning at a $R_{\alpha,12}$ multiplier value of $\sim0.7$ and
extending to $\sim1.2$, which coincides with the $\xi_{2.5}$ local
maximum in Fig.\,\ref{fig10}. This may be caused by non-convective
core-C burning that results in a more compact star and more massive
baryonic remnant \citep{Heger2001}. This non-monotonicity has also
been observed by \citet{Tur2007}.

The cut-off mass for the remnant becoming a BH versus a NS is difficult
to assess. A fraction of the NS progenitor is radiated away by neutrinos,
which is dependent on the EOS \citep{Lattimer2001}. A larger maximum
for NS masses result from the ``stiffest'' EOS (those with largest
pressures for a given density), and an upper limit of $M_{high}=2.9\,\mathrm{M}_{\odot}$
has been calculated by \citet{Tolos2012}. Observational evidence
places this limit between $2.0\,\mathrm{M}_{\odot}\lesssim M_{\mathrm{high}}\lesssim2.5\,\mathrm{M}_{\odot}$,
although this may only be an indication of the limit to the mass that
can possibly be accreted in a binary system \citep{Lattimer2010}. 

An example of the correlation between baryonic mass and central carbon
mass at the end of core-He burning can be seen by comparing Fig.\,\ref{fig14}
with Fig.\,\ref{fig15} for the 25\,$\mathrm{M}_{\text{\ensuremath{\odot}}}$
models. Figures for the carbon mass fractions of the other models
can be found in the Appendix.

\includegraphics[scale=0.5,angle=90]{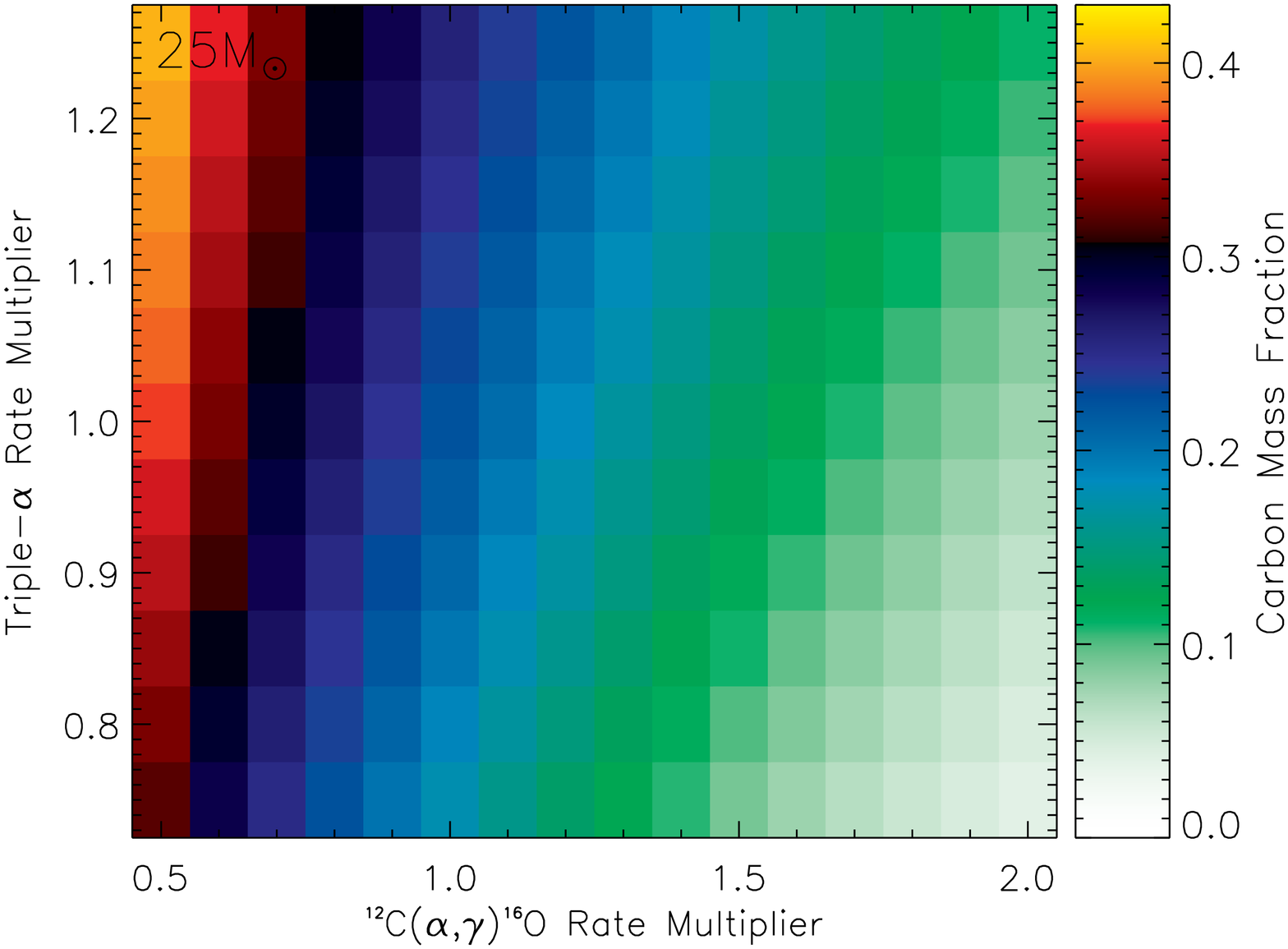}

\figcaption{\label{fig15} Central carbon mass fraction at the end of core-He
burning for the 25\,$\mathrm{M}_{\text{\ensuremath{\odot}}}$ models
as a function of the $R_{3\alpha}$ and $R_{\alpha,12}$ multipliers. }

As shown in Fig.\,\ref{fig15}, high $R_{3\alpha}$ and low $R_{\alpha,12}$
multipliers result in higher carbon mass fractions as expected, whereas
low $R_{3\alpha}$ and high $R_{\alpha,12}$ multipliers result in
lower carbon mass fractions. Comparing Fig.\,\ref{fig14} and Fig.\,\ref{fig15}
shows an overall inverse relation between the carbon mass fraction
and the remnant mass (see also Fig.\,\ref{fig:baryonics} and Fig.\,\ref{fig:carbons}).

\section{Conclusions}

This paper studies the effect of changing the helium burning rates
$^{\text{12}}$C($\alpha,\gamma$)$^{\text{16}}$O and $^{\text{4}}$He($2\alpha,\gamma$)$^{\text{12}}$C
independently to map the effects on stellar evolution and nucleosynthesis.
We follow the entire evolution from hydrogen burning through the SN
explosion, including wind and SN yields, and considering fallback,
mixing, and which stars make SNe ($\xi_{2.5}<0.25$) or collapse to
BHs without SN, and finally integrating the yields over an IMF. In
total, we calculated nucleosynthesis for a grid of 176 models for
each of 12 stellar masses from $\mathrm{12\, M}_{\text{\ensuremath{\odot}}}$
to $\mathrm{30\, M}_{\text{\ensuremath{\odot}}}$ (2112 models). This
is by far the most extensive investigation on the effects of rate
variations for the helium burning reactions to date, and the first
to use updated solar abundances \citep{Lodders2009}. 

Combining constraints on intermediate-mass and weak \emph{s}-only
isotopes, we find a best fit for rate multipliers $\left(R_{\alpha,12},R_{3\alpha}\right)=\left(1.1,0.75\right)$
to $\left(1.6,1.25\right)$ with an average spread of $\approx\pm0.2$
in $R_{\alpha,12}$. More generally, we find a relation between $R_{\alpha,12}$
and $R_{3\alpha}$ for good fits to nucleosynthesis given by $R_{\alpha,12}=1.0R_{3\alpha}+(0.35\pm0.2)$
in the range $R_{3\alpha}$ = 0.75 to 1.25. We also provide the line
of best fit using only the intermediate-mass isotope list, given by
\textbf{$R_{\alpha,12}=1.0R_{3\alpha}+(0.25\pm0.3)$.}

In this analysis, all models are assumed to have a final kinetic energy
of the ejecta of 1.2\,B. Real supernovae have a range of explosion
energies. A larger explosion energy would allow successful SNe above
the $\xi_{2.5}=0.25$ limit, since now larger densities would be required
to overcome this larger energy to prevent a successful SN explosion.
It would also cause more material above the Fe core to escape the
gravitational potential, resulting in a smaller mass cut and remnant
mass for models already below $\xi_{2.5}=0.25$. 

Small changes in the reaction rates can result in significant differences
in the convection structure of the star, which is not just a numerical
artifact but due to the physics of shell burning. This introduces
``noise'' into the comparisons, and is a warning for calculations
done for specific cases; the general conclusion may be influenced
by isolated processes. Given an astrophysical model that includes
all the important physics with perfectly known physical input parameters,
the values for the standard deviations shown in the figures should
have a minimum region reflecting only the uncertainties in isotopic
abundances. We know, however, that the stellar model used is only
approximate and can affect the nucleosynthesis, as discussed in Section
3.3. For the isotopes we compare in this work, however, the assumption
that solar composition stars should produce about their solar ratios
is reasonable.

If we change the parameters of the models or improve the completeness
of the models, we expect the best fit rates to change also. Hence
it is not necessarily the case that the best fit rate for an observable
(in our case the abundances) coincides with the true rate. This suggests
that the best reaction rates we obtain in our analyses are at some
level effective rates. Analogous procedures have been used in many
areas of physics. For example, in the shell model of nuclear physics,
an effective nucleon-nucleon interaction (close to but not identical
to the true interaction) is chosen to fit the low-lying spectra of
many nuclei, and this reaction is very successfully used to predict
other observables. Similarly, we think that the present procedure
can provide better predictions of other astrophysical quantities,
remnant masses or neutrinos synthesis of certain isotopes, for example. 

Whereas our comparisons do not have the power to accurately determine
the helium burning reaction rates, they do show that the experimental
values are not too far from the truth and that changes needed to compensate
(in an effective interaction sense) for model uncertainties are not
large. Thus, similar calculations necessary to assess the situation,
as overall model uncertainties decrease, will be less demanding; we
have shown that a significant part of the uncertainty space is irrelevant.
It is also clear that if one of the two helium burning reactions is
much better determined, the effective rate for the other will be much
better determined. For example, if it is later determined that $R_{3\alpha}$
is near 1.25 times the present experimental centroid value, then the
best fit $R_{\alpha,12}$ will be 33\,\% larger than the experimental
centroid value. It may also happen that if both reactions become well
determined, they would not agree with the effective interaction that
best reproduces the abundances. In such an event, this work may still
serve to provide an evaluation of other model uncertainties and point
the way to improvements.

\acknowledgements{}

This research was supported by the US DOE Program for Scientific Discovery
through Advanced Computing (SciDAC; DE-FC02-09ER41618), by the US
Department of Energy under grant DE-FG02-87ER40328, by the Joint Institute
for Nuclear Astrophysics (JINA; NSF grant PHY08-22648 and PHY110-2511).
AH acknowledges support by a Future Fellowship from the ARC and from
Monash University through a Larkins Fellowship.

\appendix{}

\section{Appendix}

\begin{figure}[H]
\begin{minipage}[t]{0.32\columnwidth}%
\includegraphics[angle=90,scale=0.2]{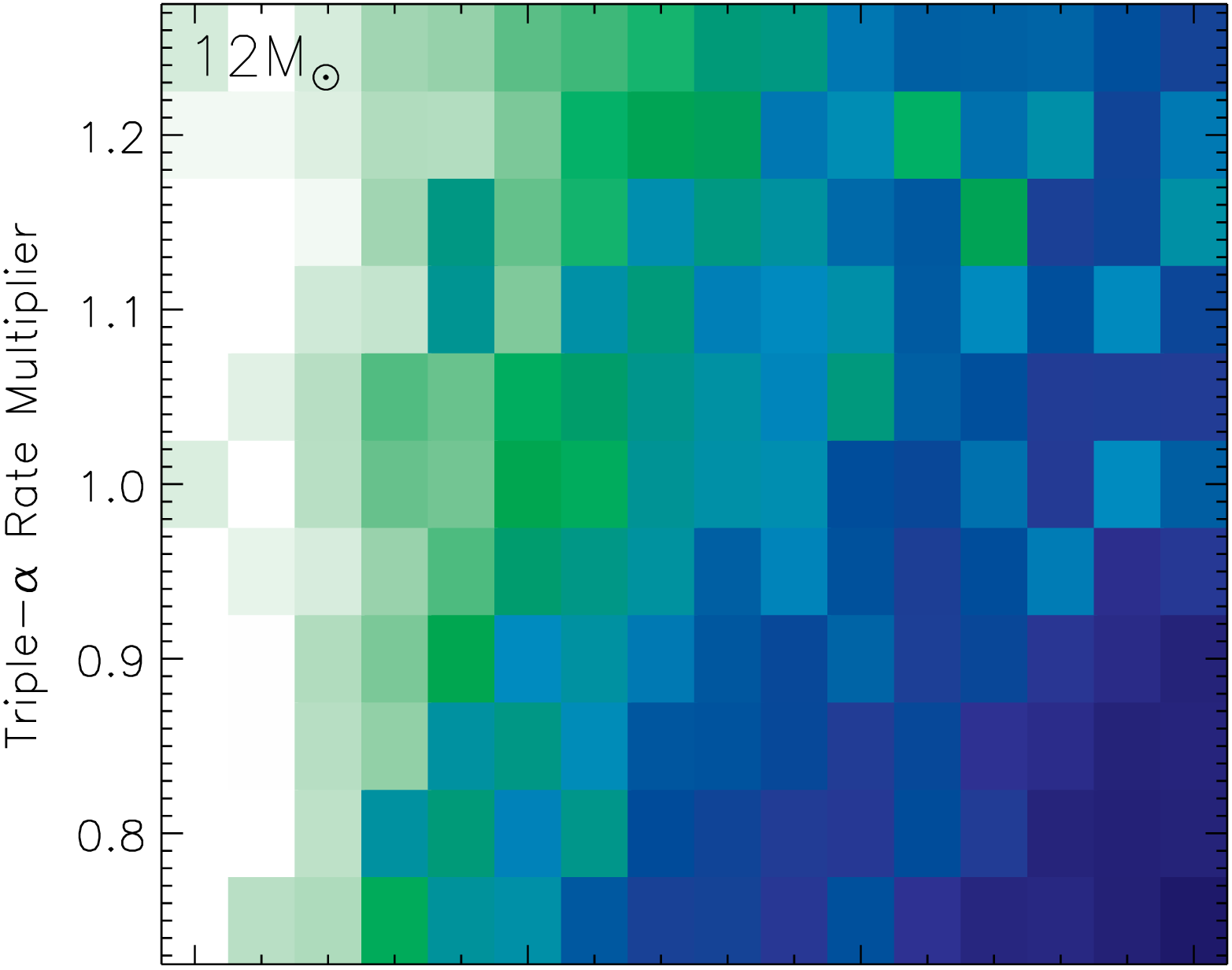}%
\end{minipage}\hspace{-1.345cm}%
\begin{minipage}[t]{0.32\columnwidth}%
\includegraphics[angle=90,scale=0.2]{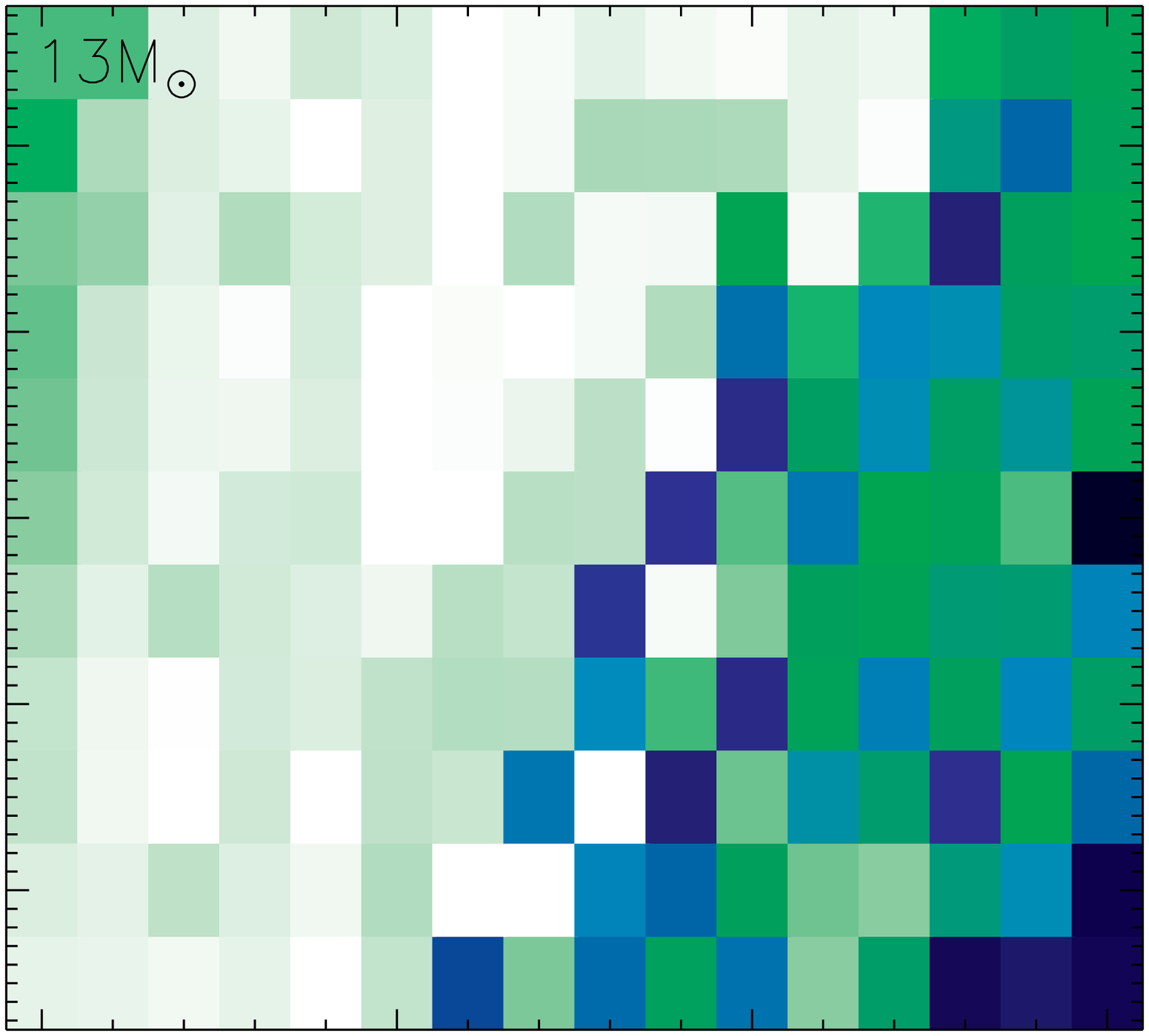}%
\end{minipage}\hspace{-1.345cm}%
\begin{minipage}[t]{0.32\columnwidth}%
\includegraphics[angle=90,scale=0.2]{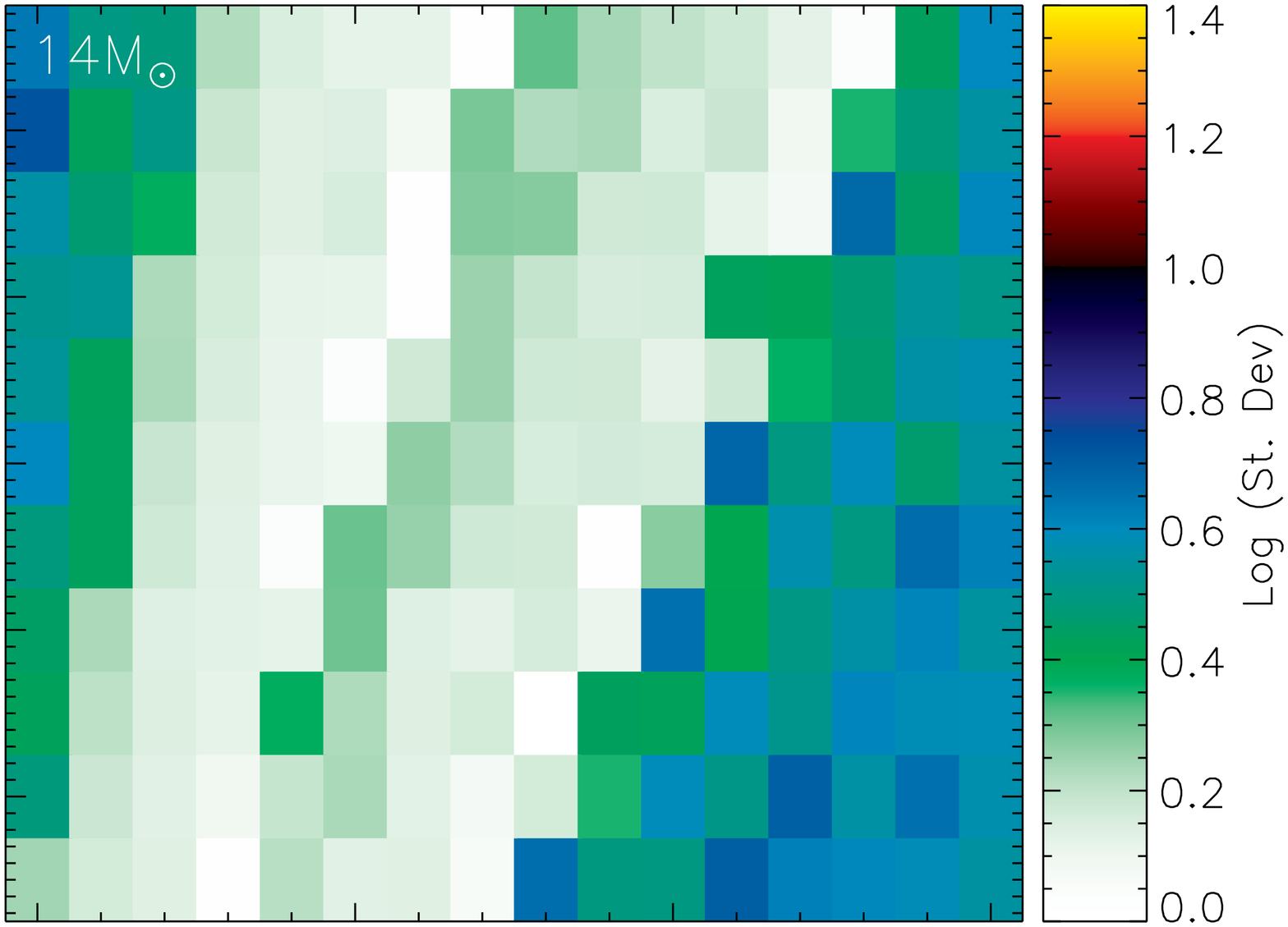}%
\end{minipage}

\vspace{-0.385cm}

\begin{minipage}[t]{0.32\columnwidth}%
\includegraphics[angle=90,scale=0.2]{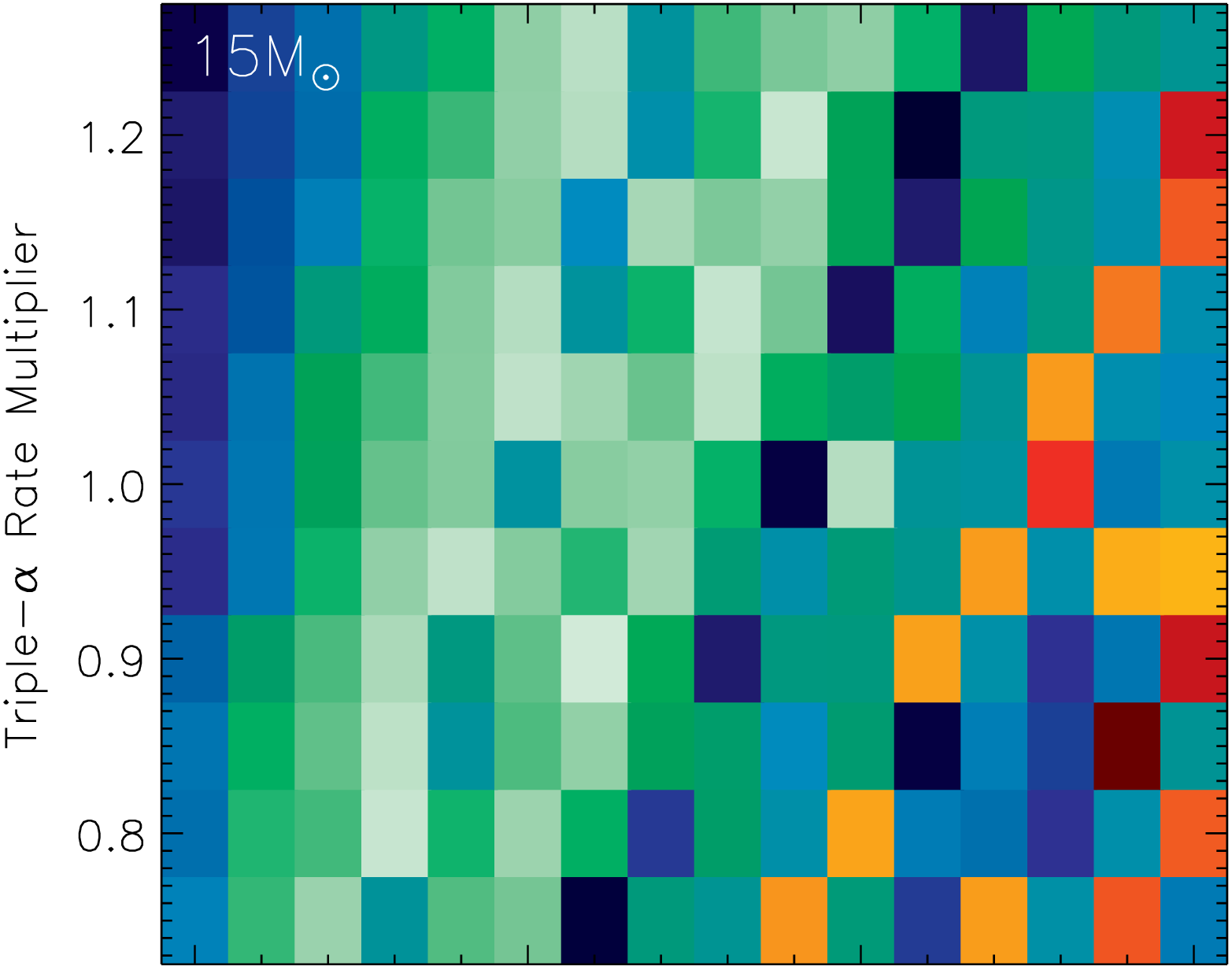}%
\end{minipage}\hspace{-1.345cm}%
\begin{minipage}[t]{0.32\columnwidth}%
\includegraphics[angle=90,scale=0.2]{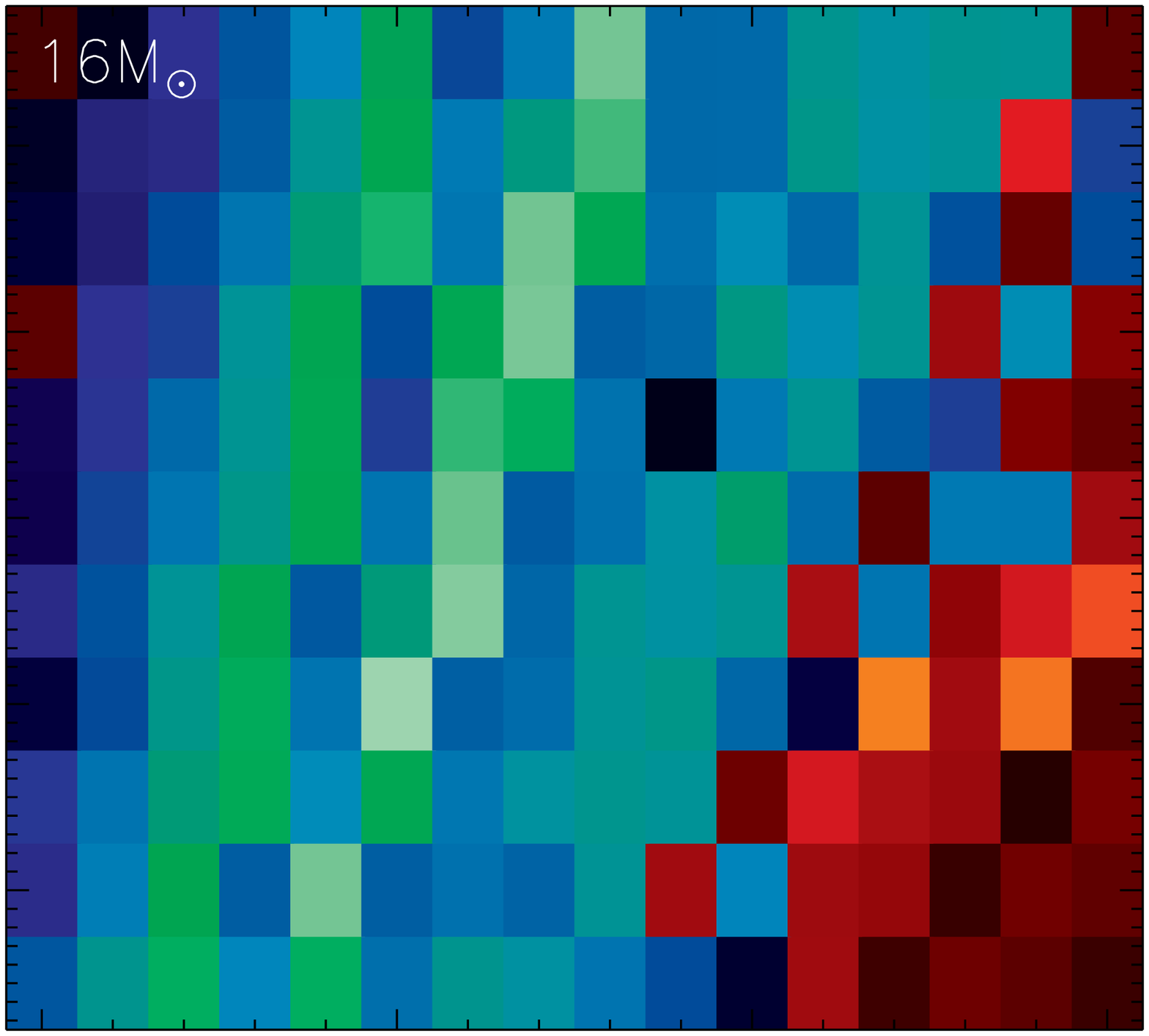}%
\end{minipage}\hspace{-1.345cm}%
\begin{minipage}[t]{0.32\columnwidth}%
\includegraphics[angle=90,scale=0.2]{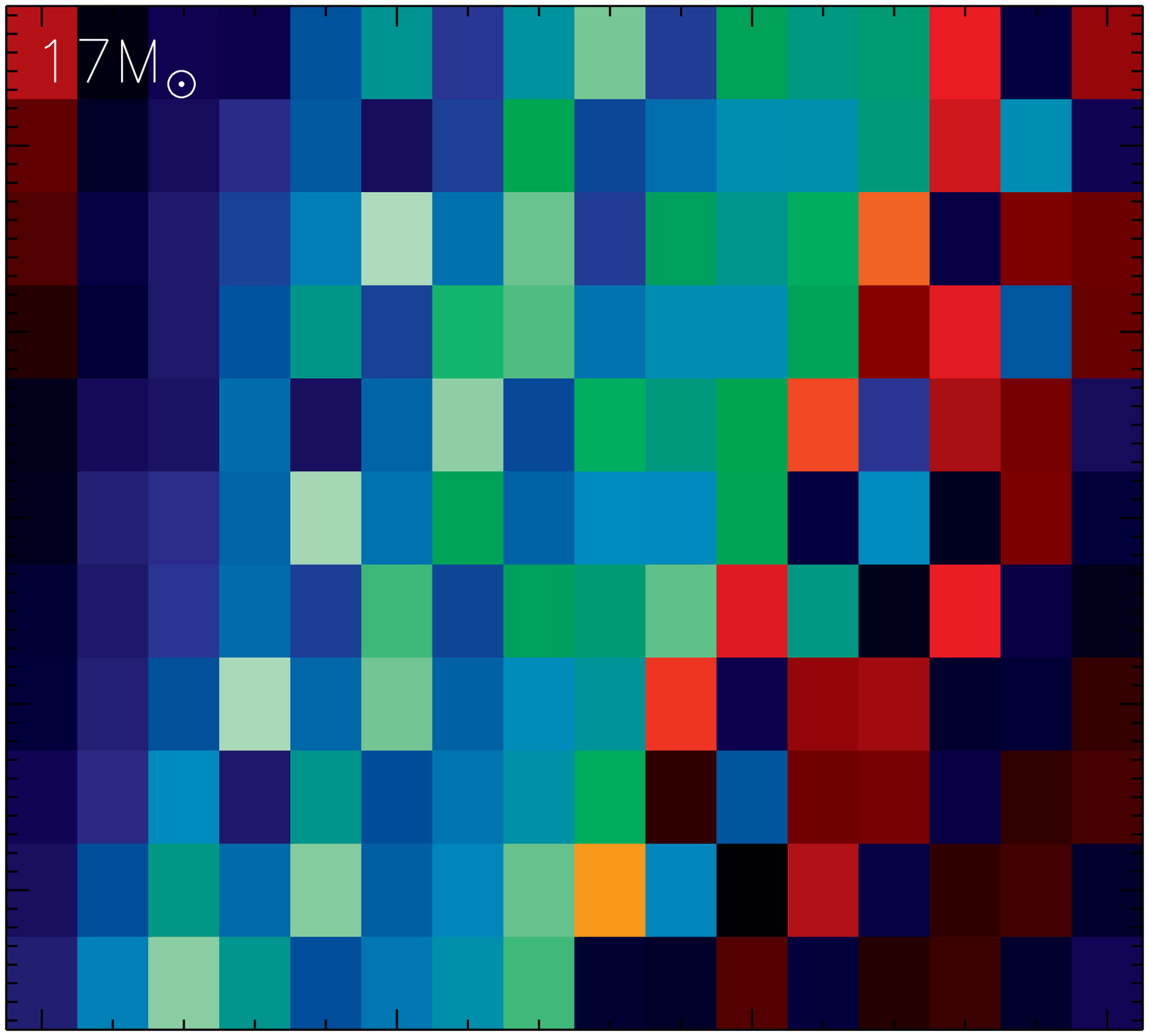}%
\end{minipage}

\vspace{-0.385cm}

\begin{minipage}[t]{0.32\columnwidth}%
\includegraphics[angle=90,scale=0.2]{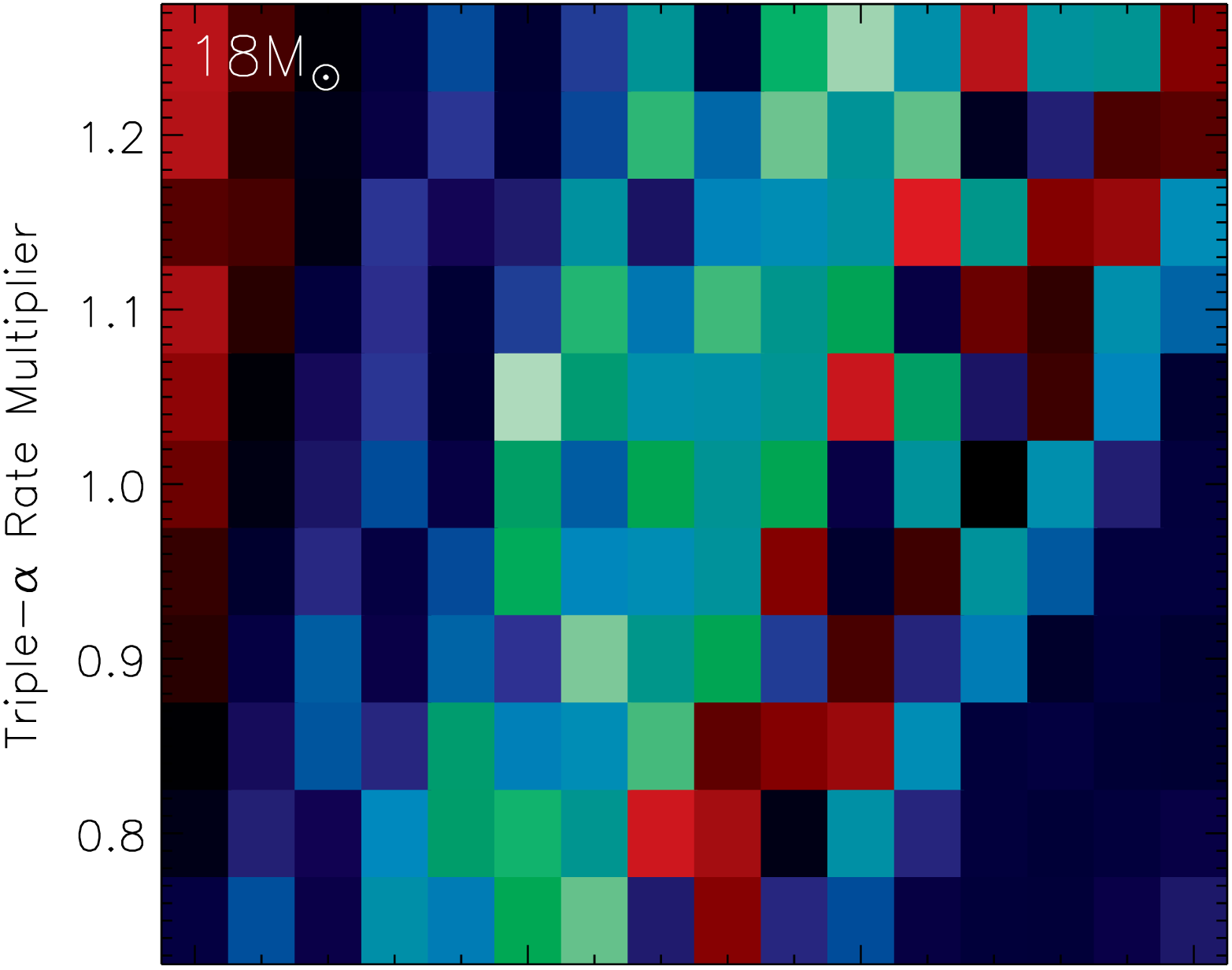}%
\end{minipage}\hspace{-1.345cm}%
\begin{minipage}[t]{0.32\columnwidth}%
\includegraphics[angle=90,scale=0.2]{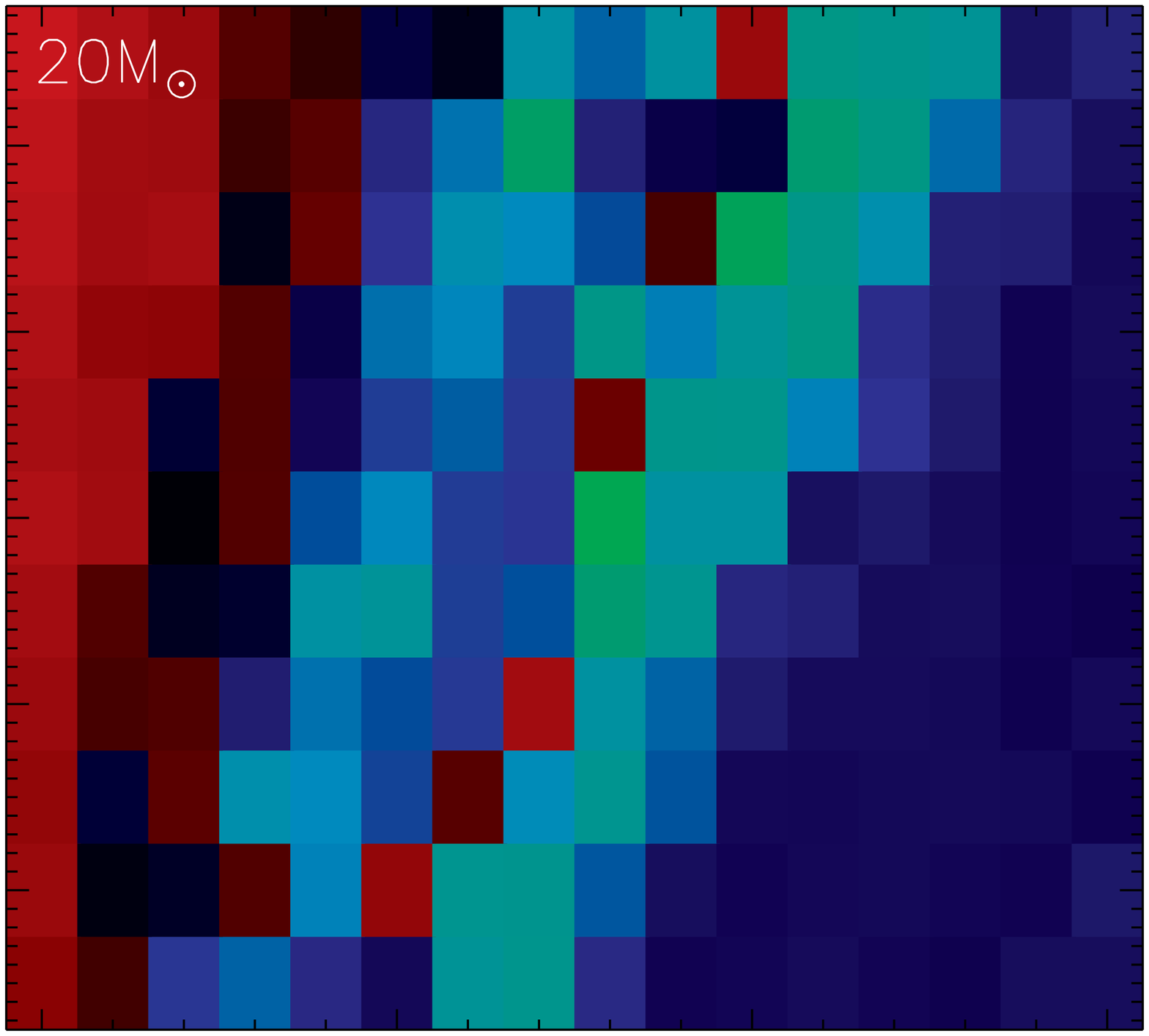}%
\end{minipage}\hspace{-1.345cm}%
\begin{minipage}[t]{0.32\columnwidth}%
\includegraphics[angle=90,scale=0.2]{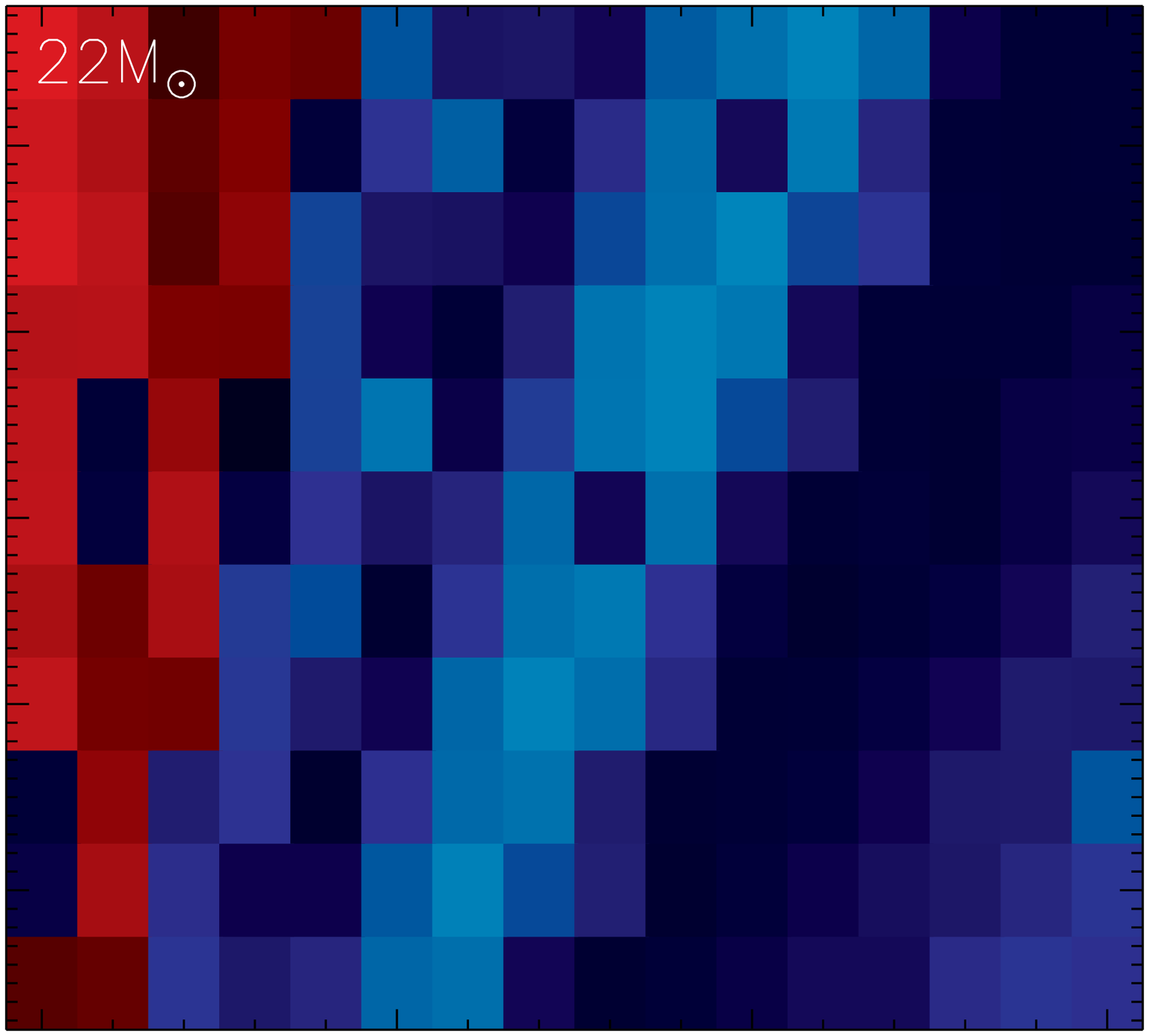}%
\end{minipage}\vspace{-0.385cm}

\begin{minipage}[t]{0.32\columnwidth}%
\includegraphics[angle=90,scale=0.2]{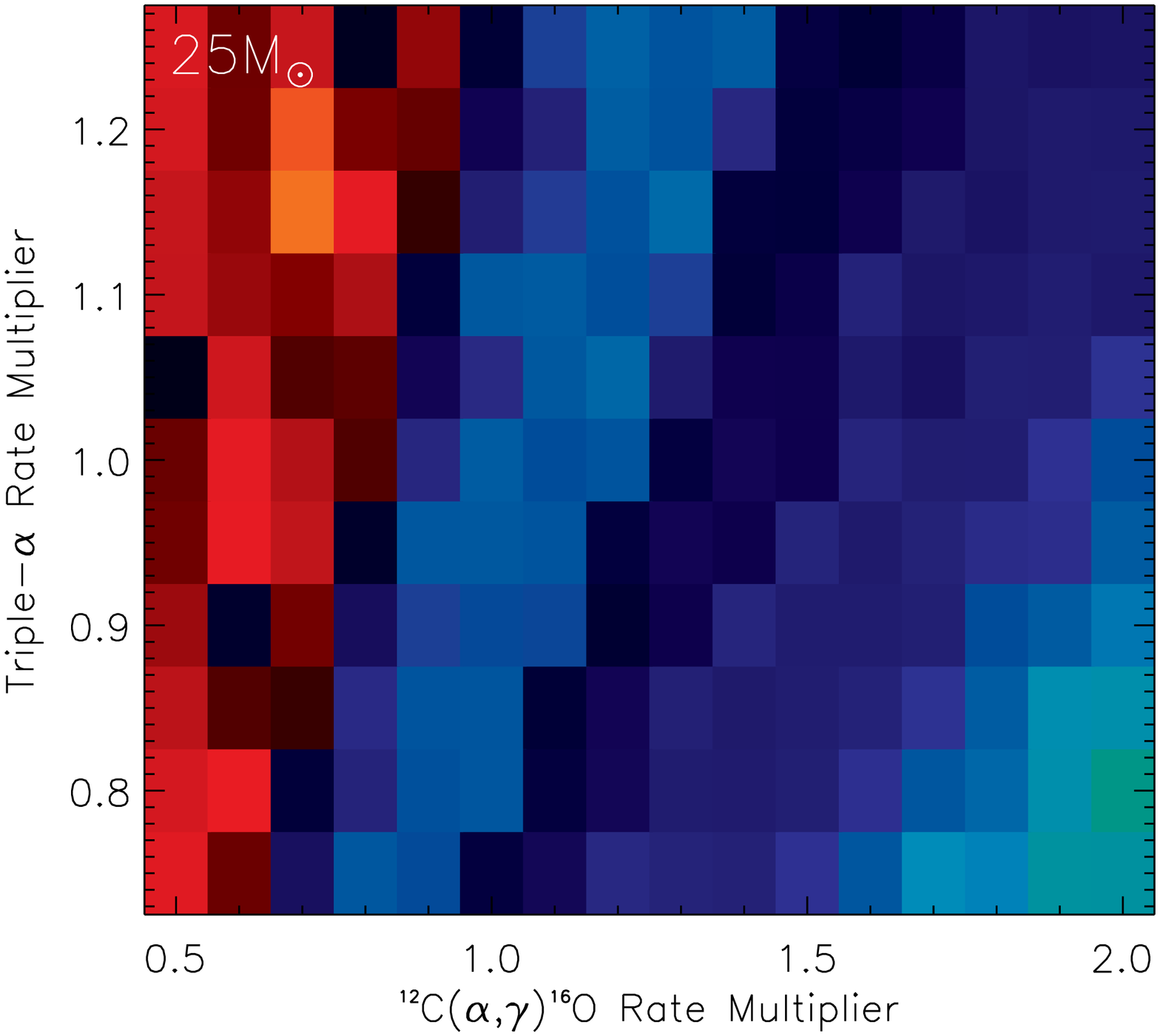}%
\end{minipage}\hspace{-1.345cm}%
\begin{minipage}[t]{0.32\columnwidth}%
\includegraphics[angle=90,scale=0.2]{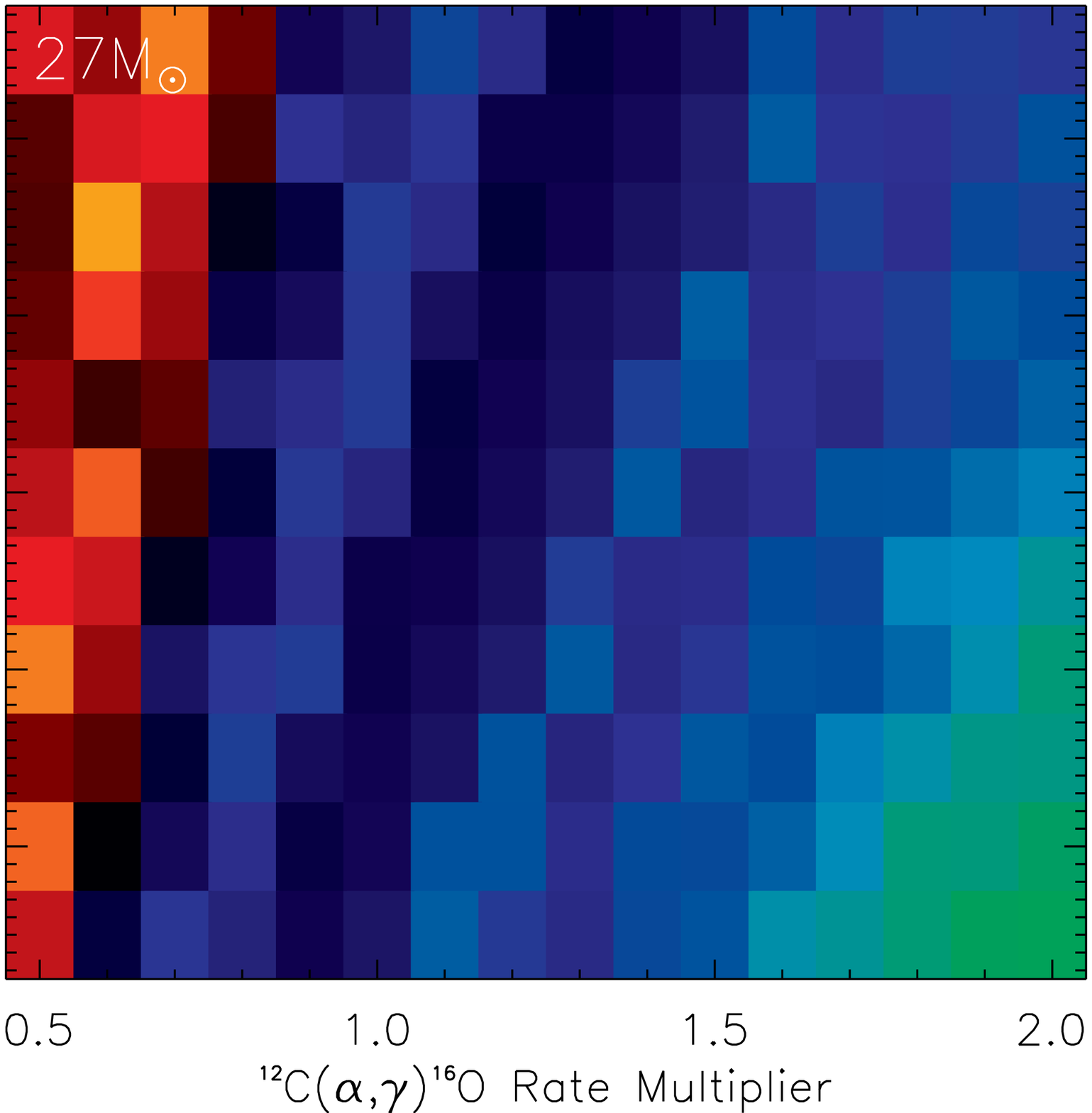}%
\end{minipage}\hspace{-1.345cm}%
\begin{minipage}[t]{0.32\columnwidth}%
\includegraphics[angle=90,scale=0.2]{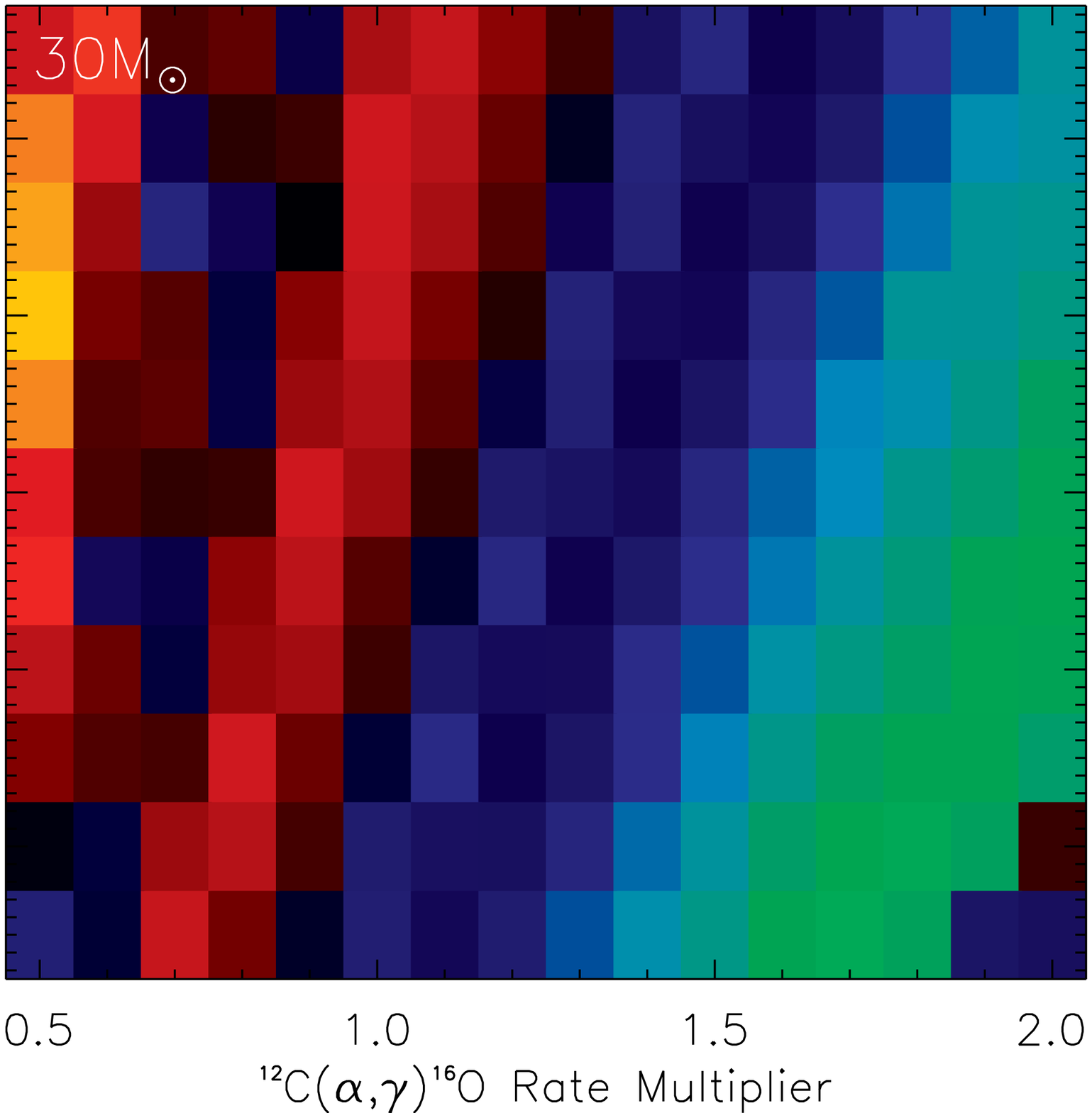}%
\end{minipage}

\caption{\label{fig:Intermediate_sigmas}Standard deviations of the production
factors of intermediate-mass isotopes for all models as a function
of the $R_{3\alpha}$ and $R_{\alpha,12}$ multipliers. The horizontal
axis gives the $R_{\alpha,12}$ multiplier, the vertical axis gives
the $R_{3\alpha}$ multiplier, and the color scale gives the logarithm
of the standard deviations.}
\end{figure}

\begin{figure}[H]
\begin{minipage}[t]{0.32\columnwidth}%
\includegraphics[angle=90,scale=0.2]{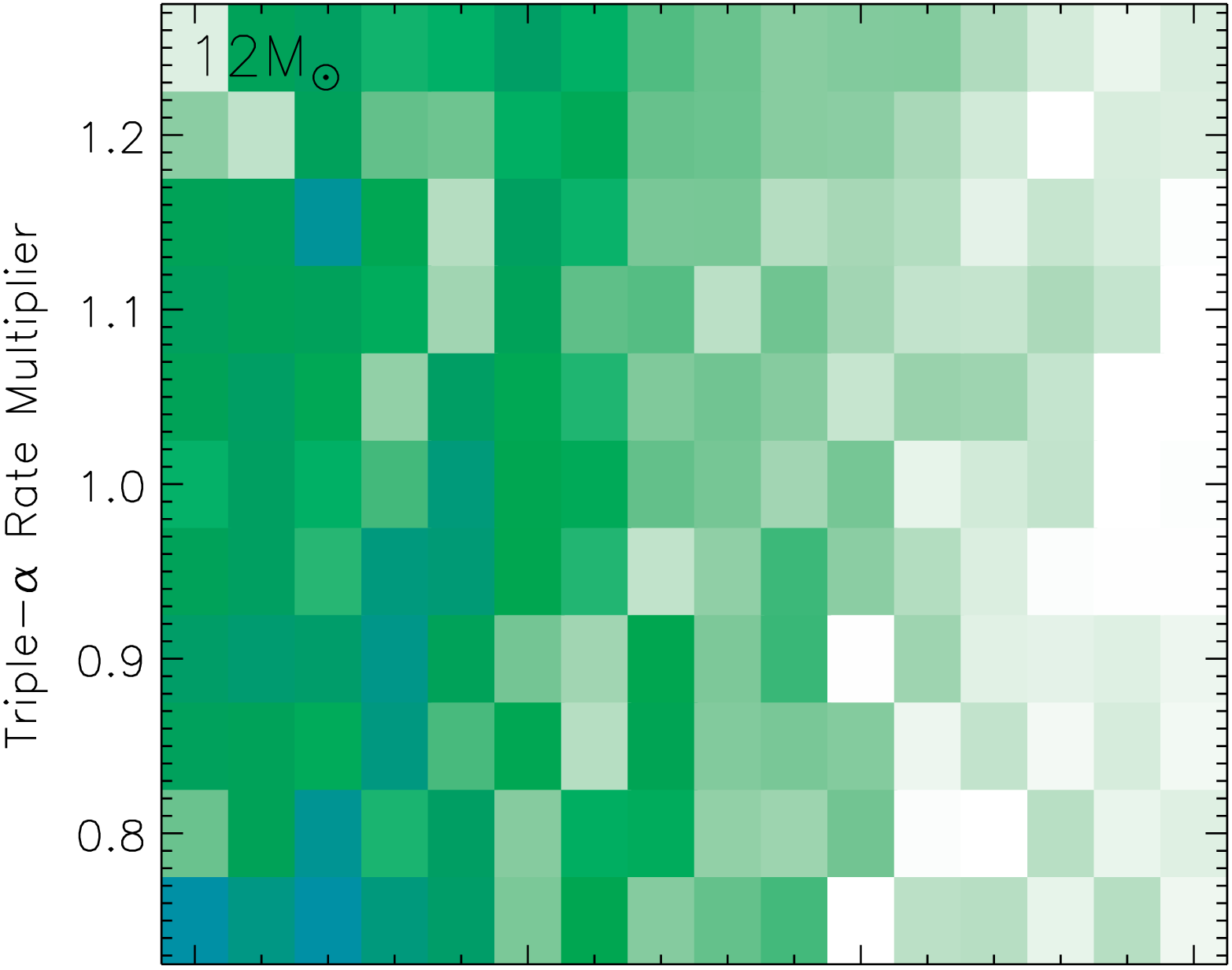}%
\end{minipage}\hspace{-1.345cm}%
\begin{minipage}[t]{0.32\columnwidth}%
\includegraphics[angle=90,scale=0.2]{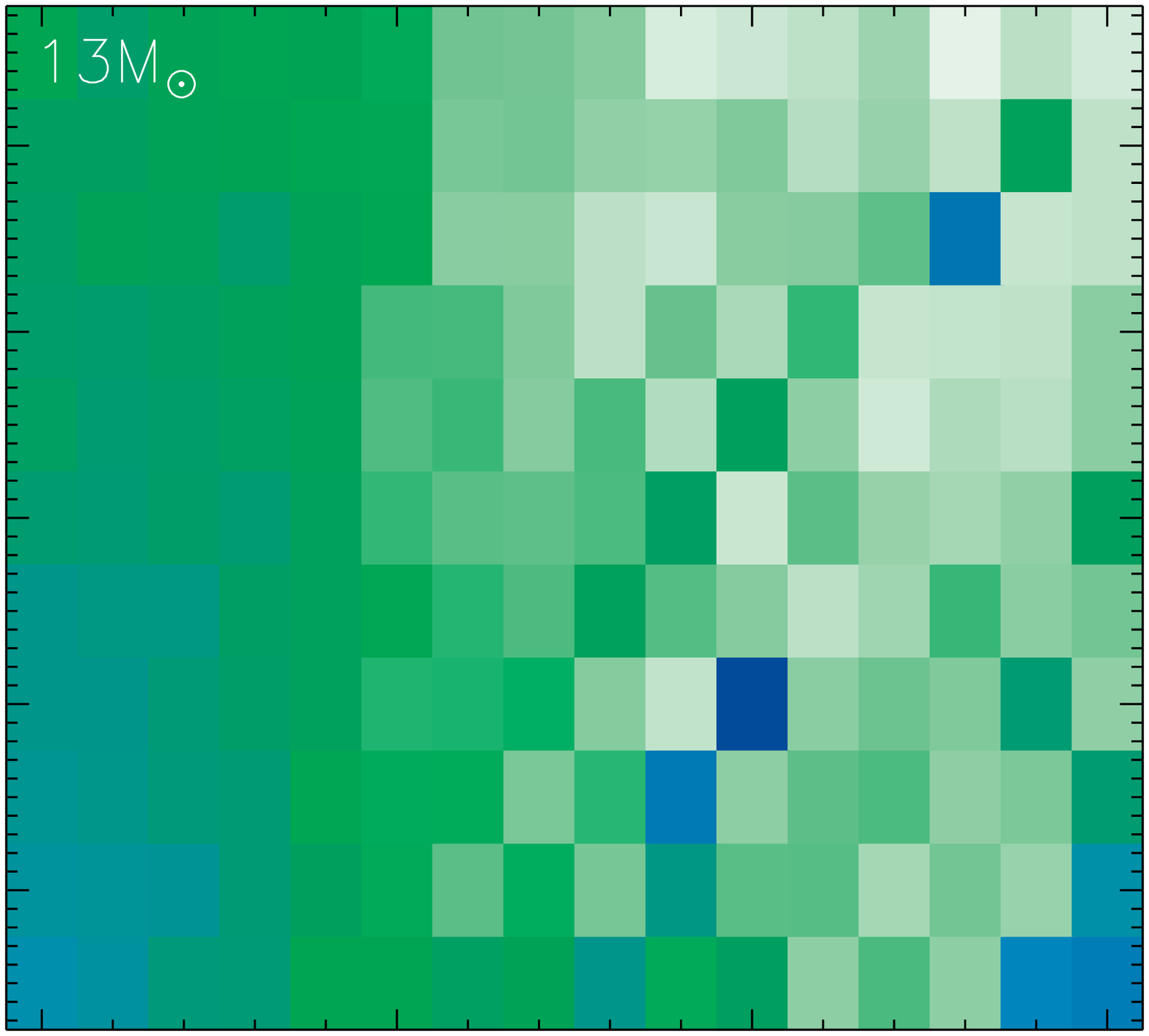}%
\end{minipage}\hspace{-1.345cm}%
\begin{minipage}[t]{0.32\columnwidth}%
\includegraphics[angle=90,scale=0.2]{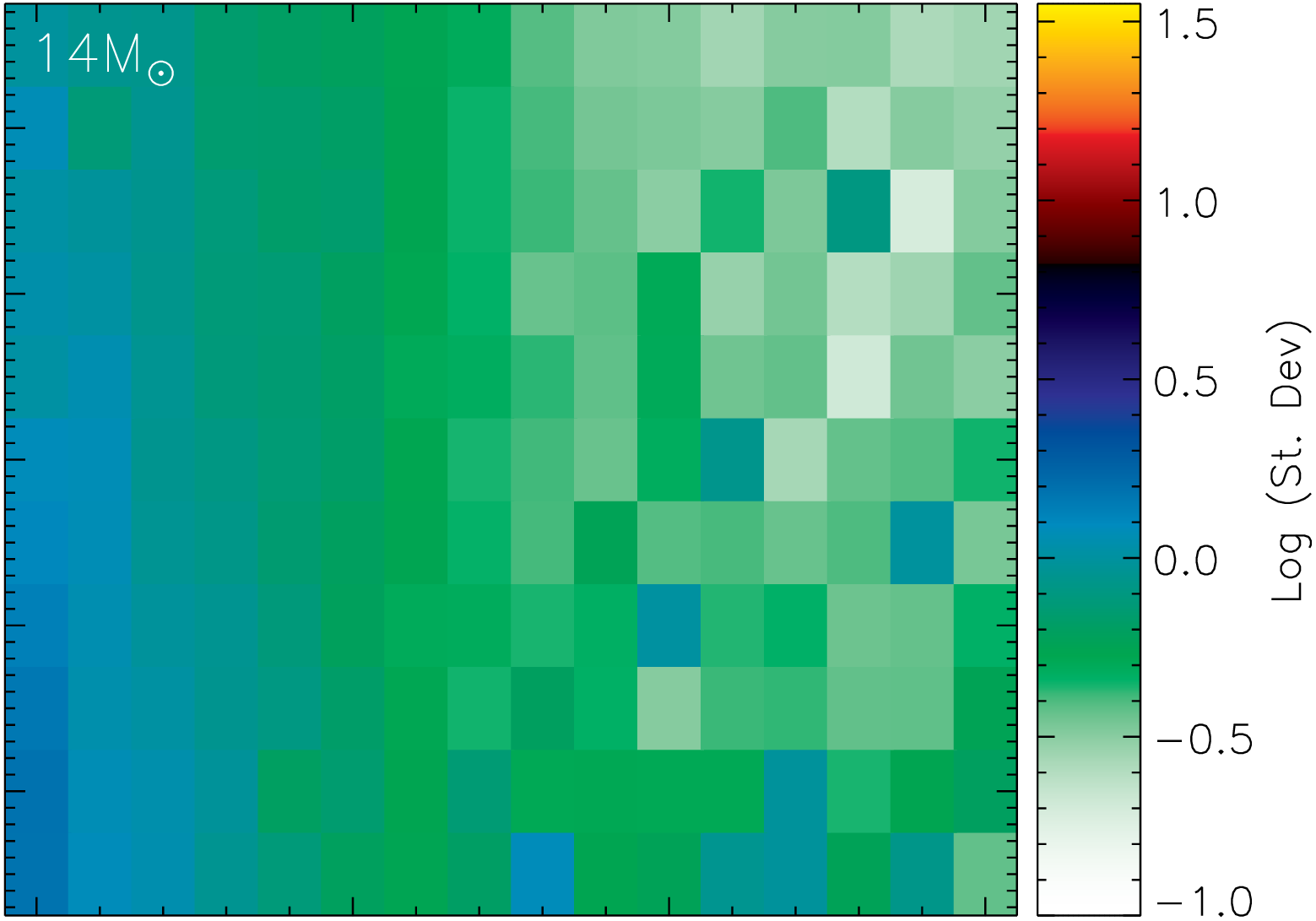}%
\end{minipage}

\vspace{-0.385cm}

\begin{minipage}[t]{0.32\columnwidth}%
\includegraphics[angle=90,scale=0.2]{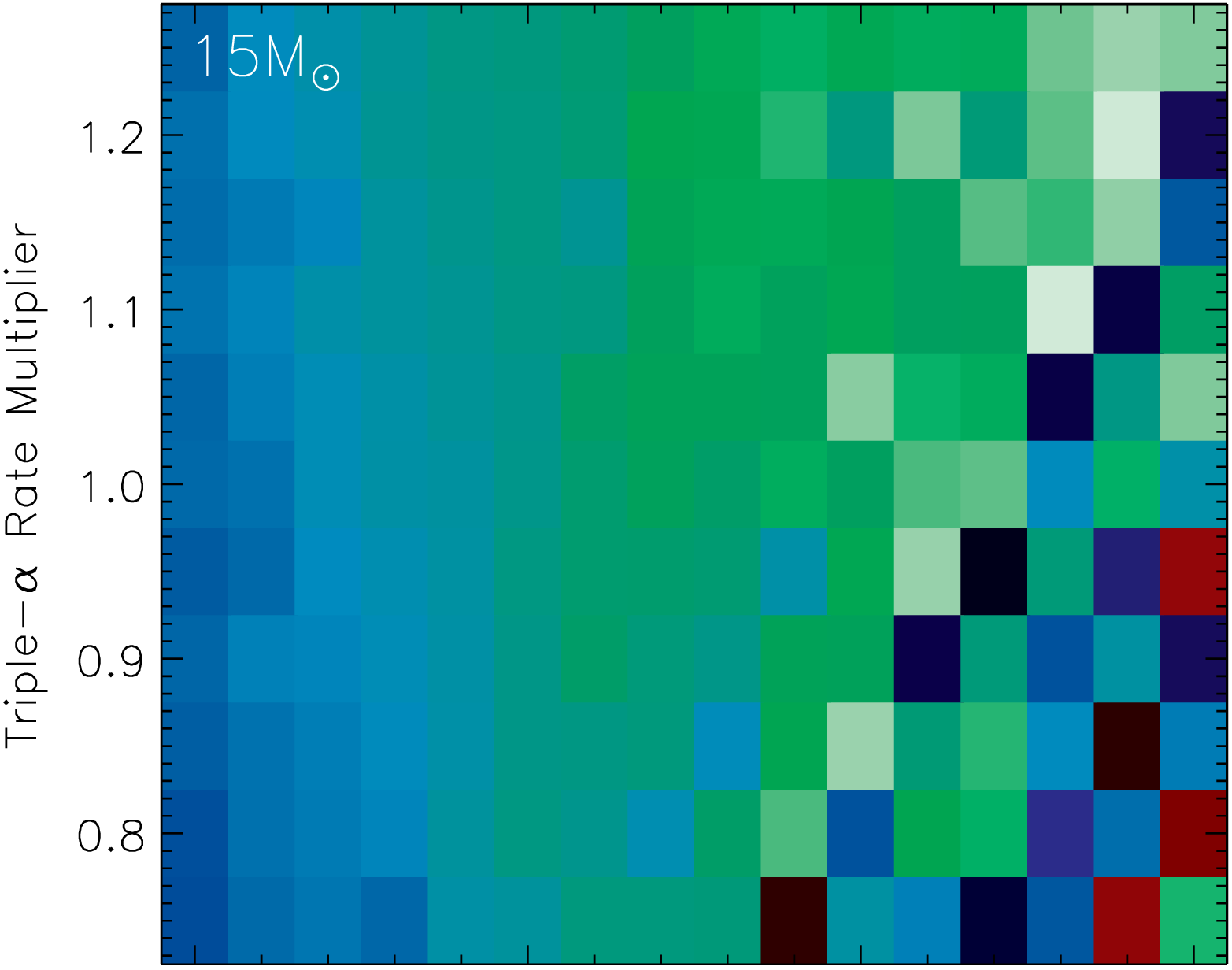}%
\end{minipage}\hspace{-1.345cm}%
\begin{minipage}[t]{0.32\columnwidth}%
\includegraphics[angle=90,scale=0.2]{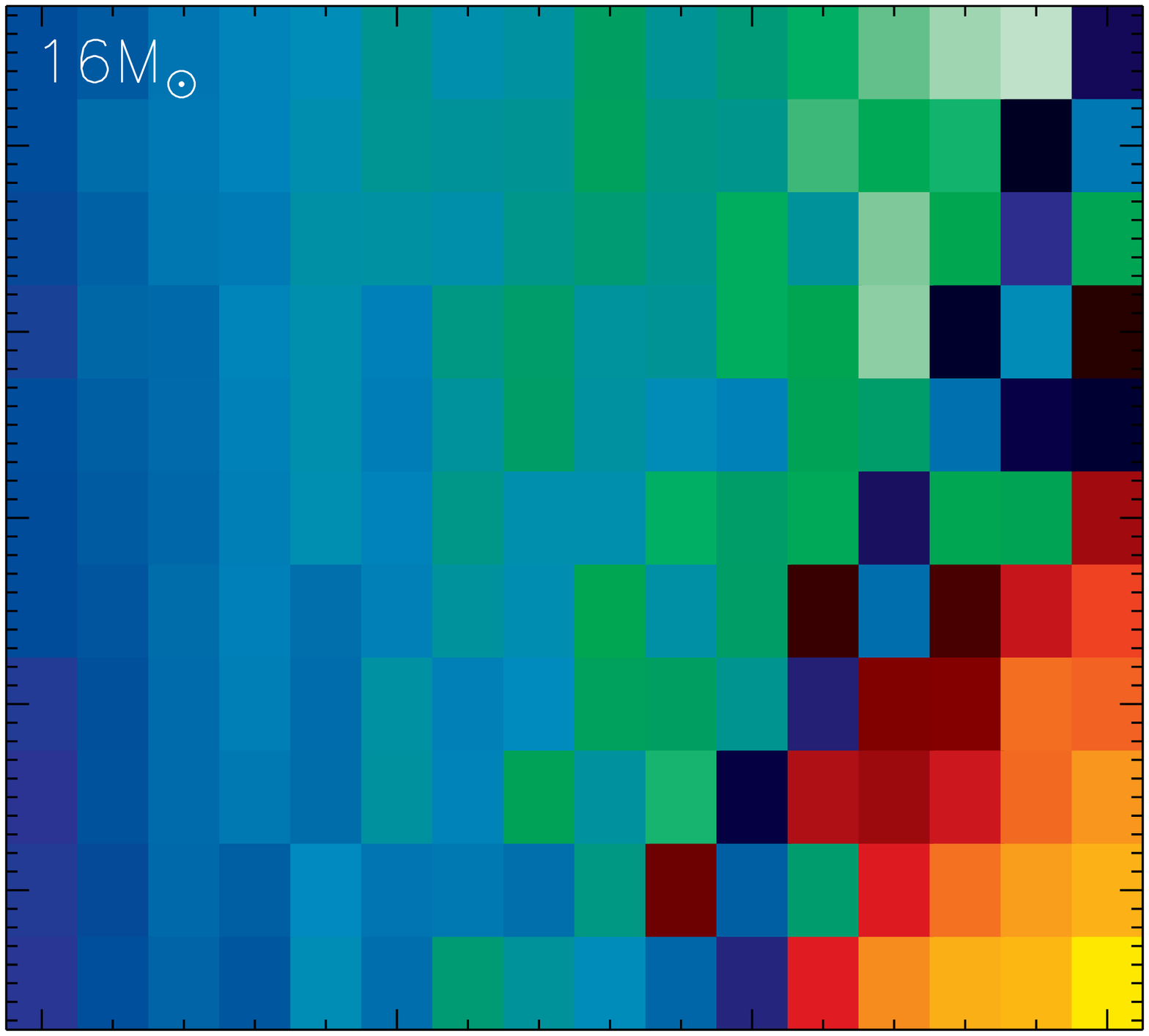}%
\end{minipage}\hspace{-1.345cm}%
\begin{minipage}[t]{0.32\columnwidth}%
\includegraphics[angle=90,scale=0.2]{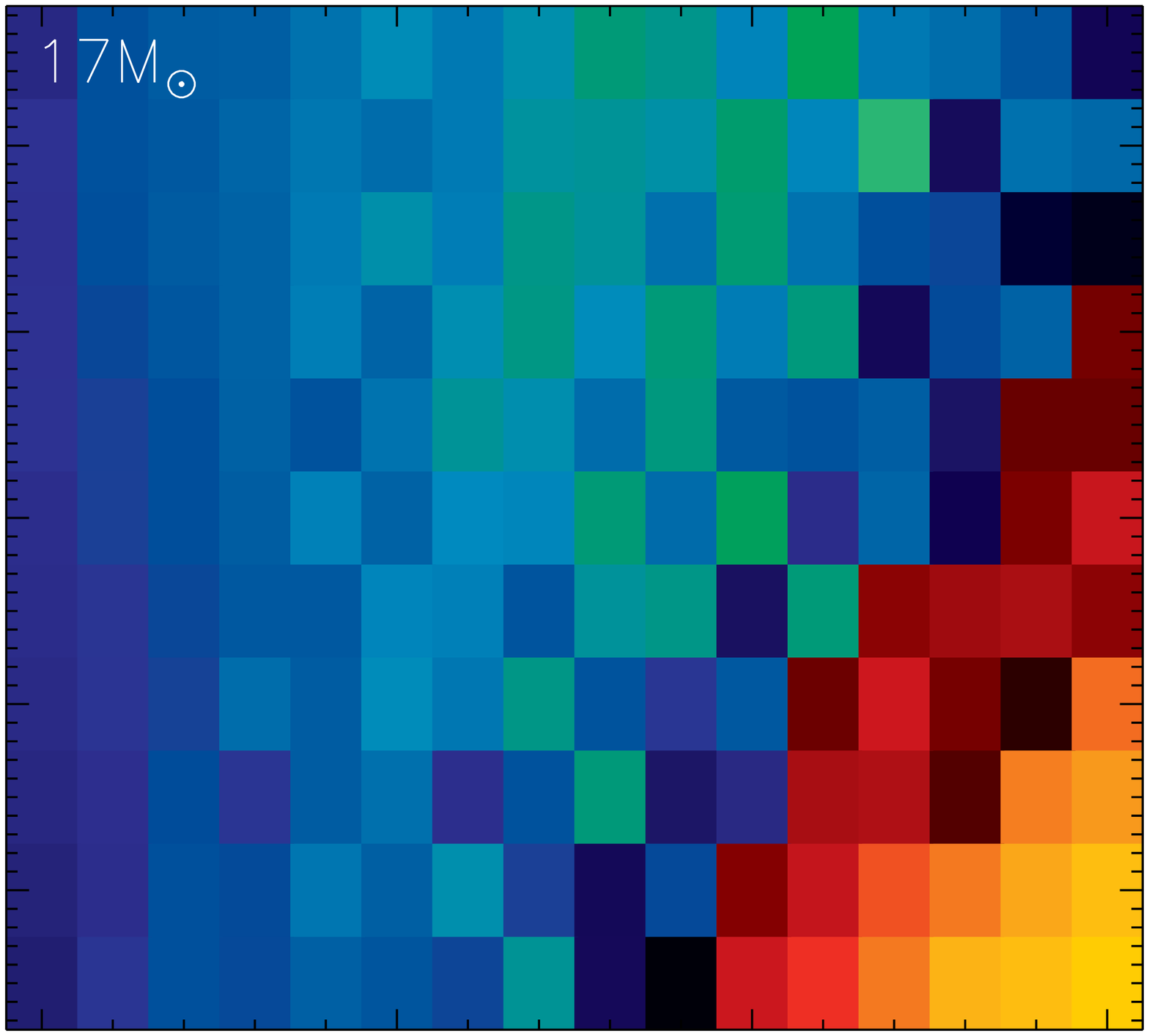}%
\end{minipage}

\vspace{-0.385cm}

\begin{minipage}[t]{0.32\columnwidth}%
\includegraphics[angle=90,scale=0.2]{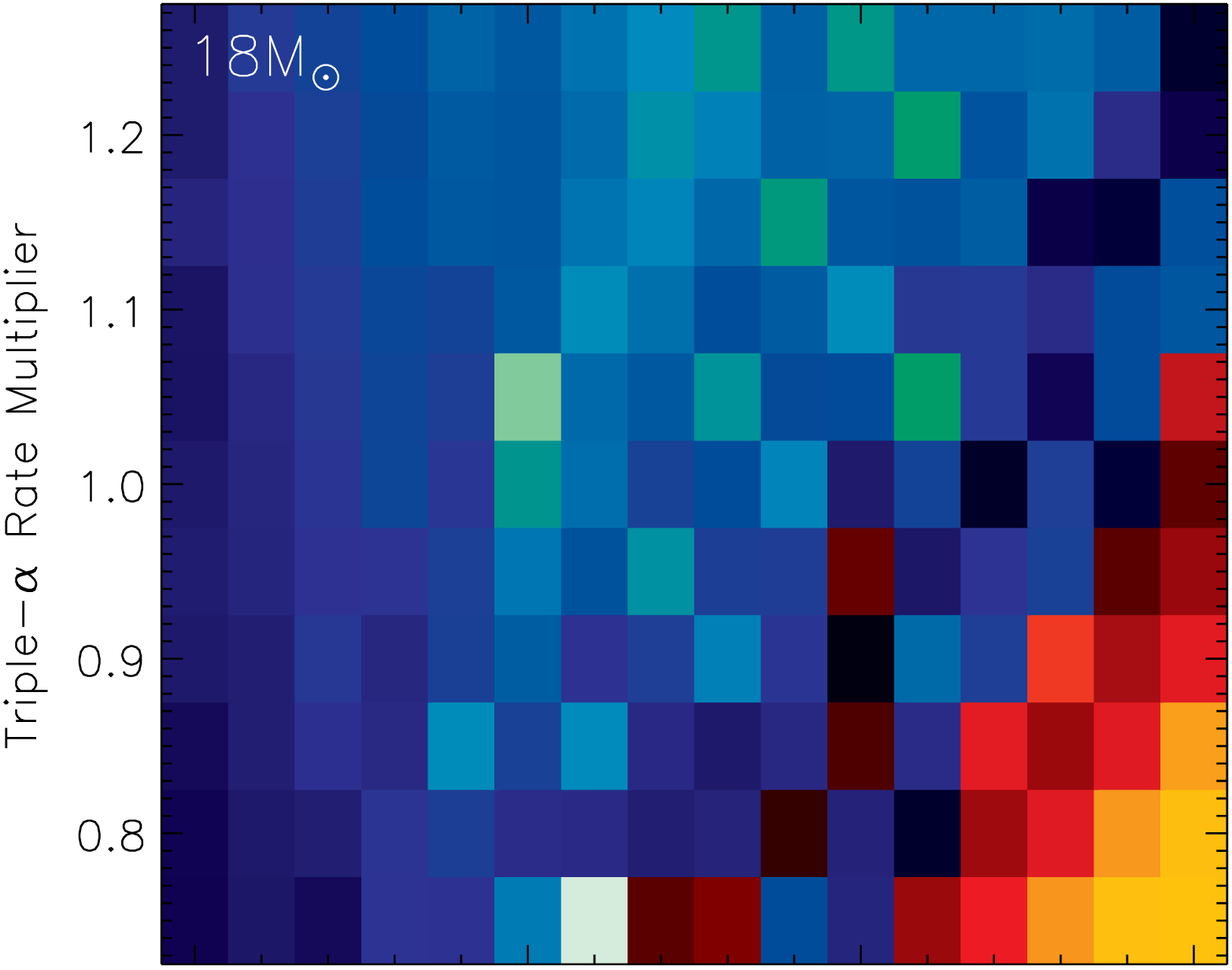}%
\end{minipage}\hspace{-1.345cm}%
\begin{minipage}[t]{0.32\columnwidth}%
\includegraphics[angle=90,scale=0.2]{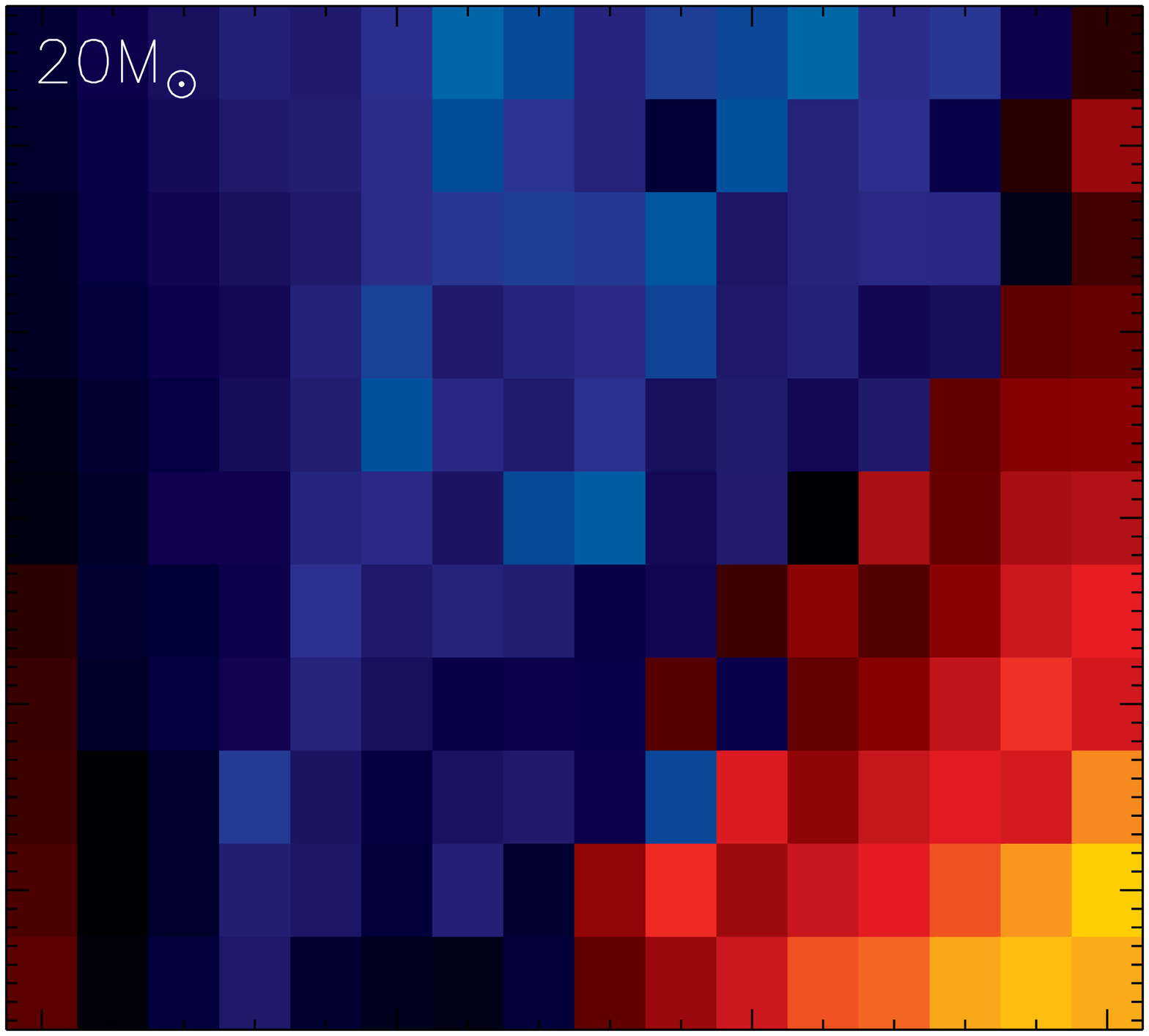}%
\end{minipage}\hspace{-1.345cm}%
\begin{minipage}[t]{0.32\columnwidth}%
\includegraphics[angle=90,scale=0.2]{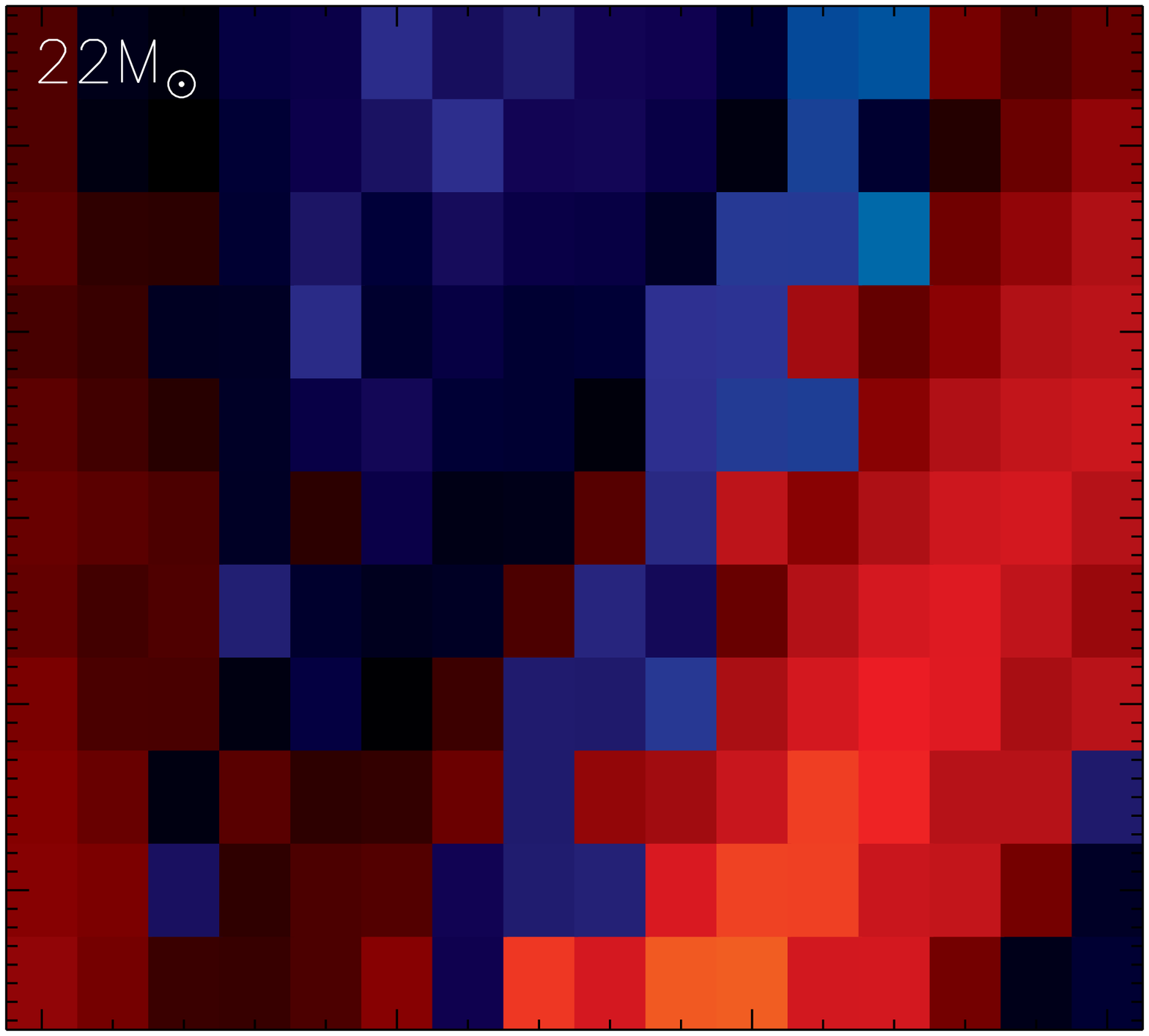}%
\end{minipage}\vspace{-0.385cm}

\begin{minipage}[t]{0.32\columnwidth}%
\includegraphics[angle=90,scale=0.2]{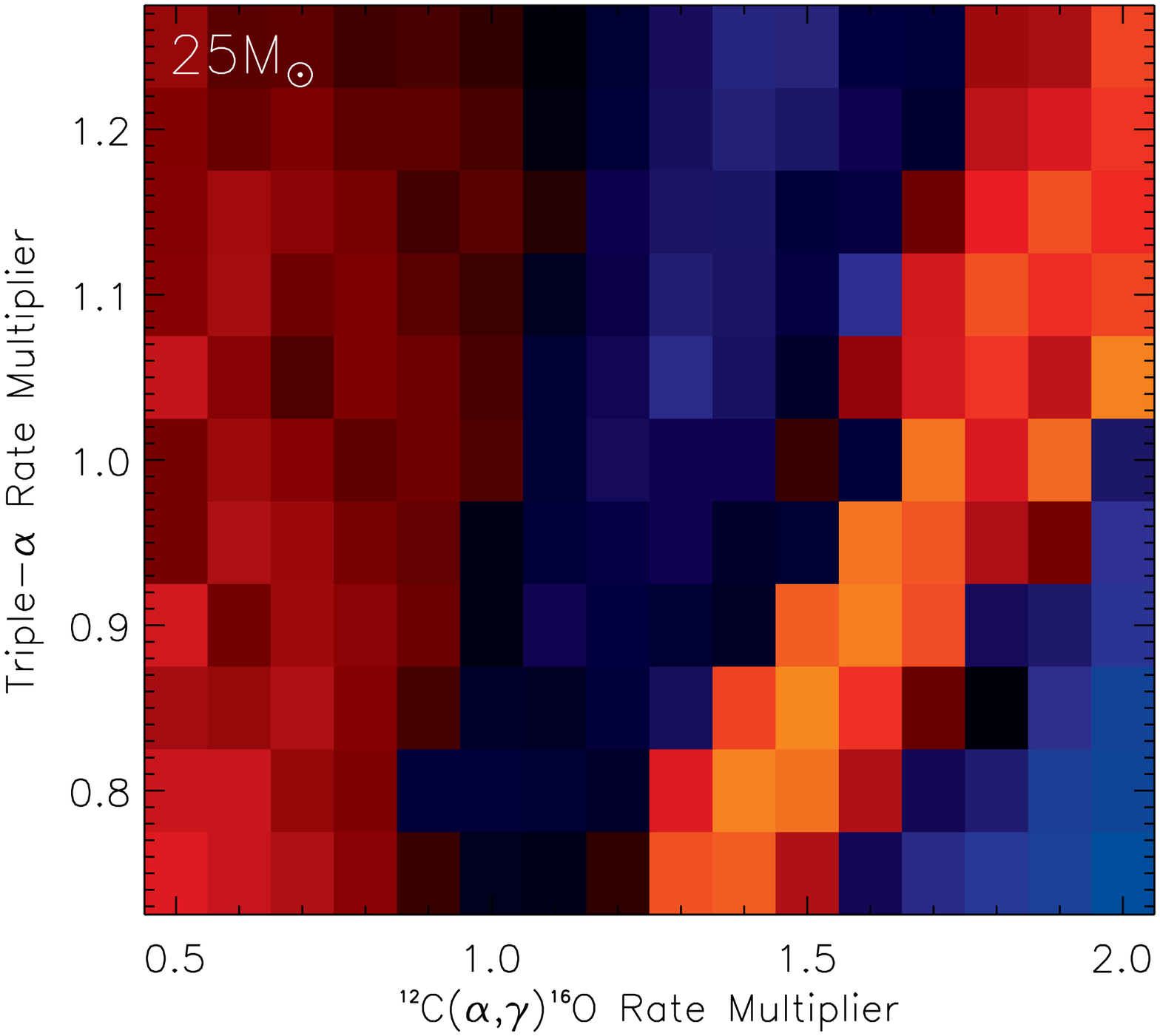}%
\end{minipage}\hspace{-1.345cm}%
\begin{minipage}[t]{0.32\columnwidth}%
\includegraphics[angle=90,scale=0.2]{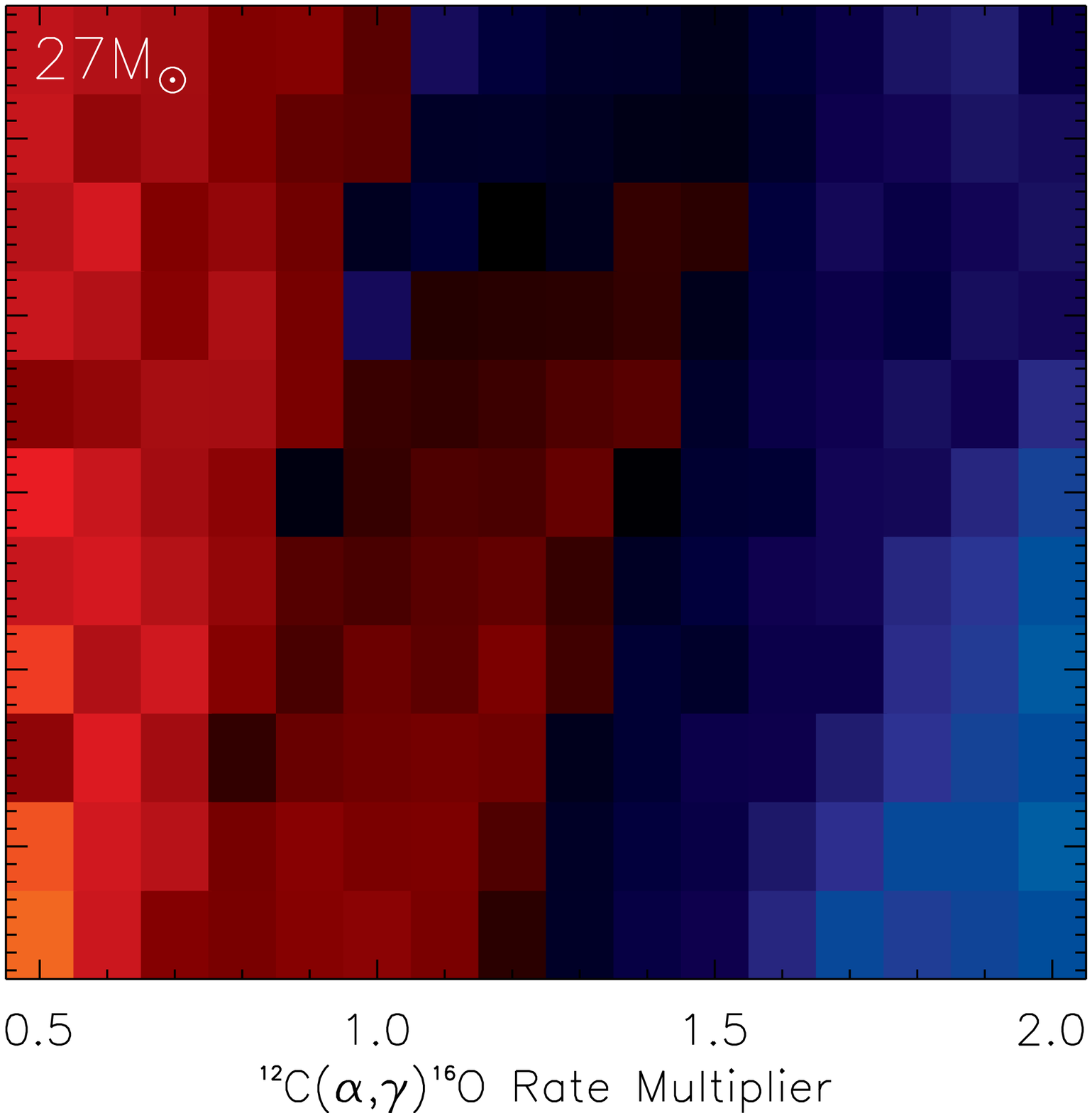}%
\end{minipage}\hspace{-1.345cm}%
\begin{minipage}[t]{0.32\columnwidth}%
\includegraphics[angle=90,scale=0.2]{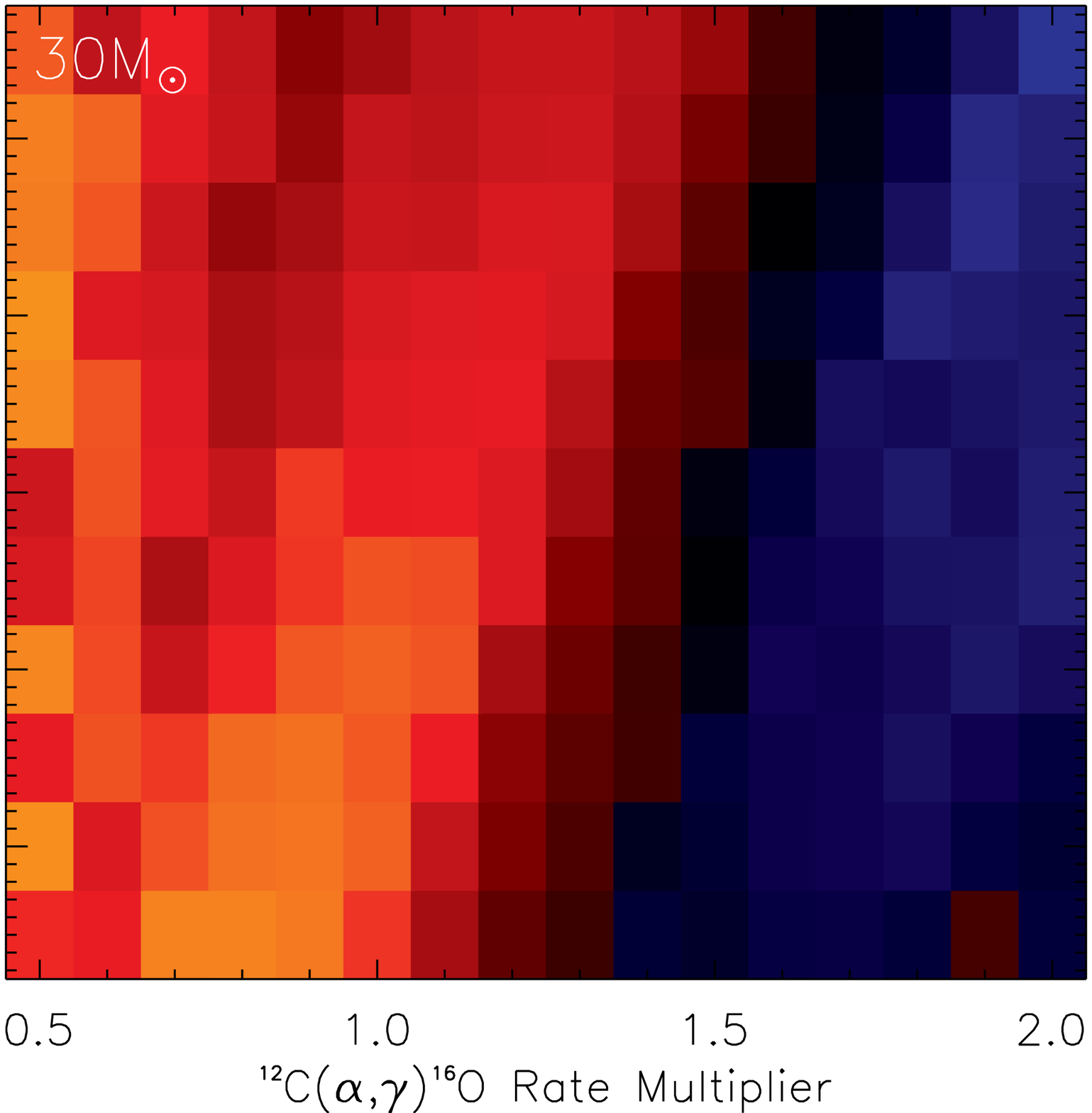}%
\end{minipage}

\caption{\label{fig:weak_sigmas}Standard deviations of the production factors
of weak \emph{s}-only isotopes for all models as a function of the
$R_{3\alpha}$ and $R_{\alpha,12}$ multipliers. The figure follows
the convention of Fig.\,\ref{fig:Intermediate_sigmas}.}
\end{figure}

\begin{figure}[H]
\begin{minipage}[t]{0.32\columnwidth}%
\includegraphics[angle=90,scale=0.2]{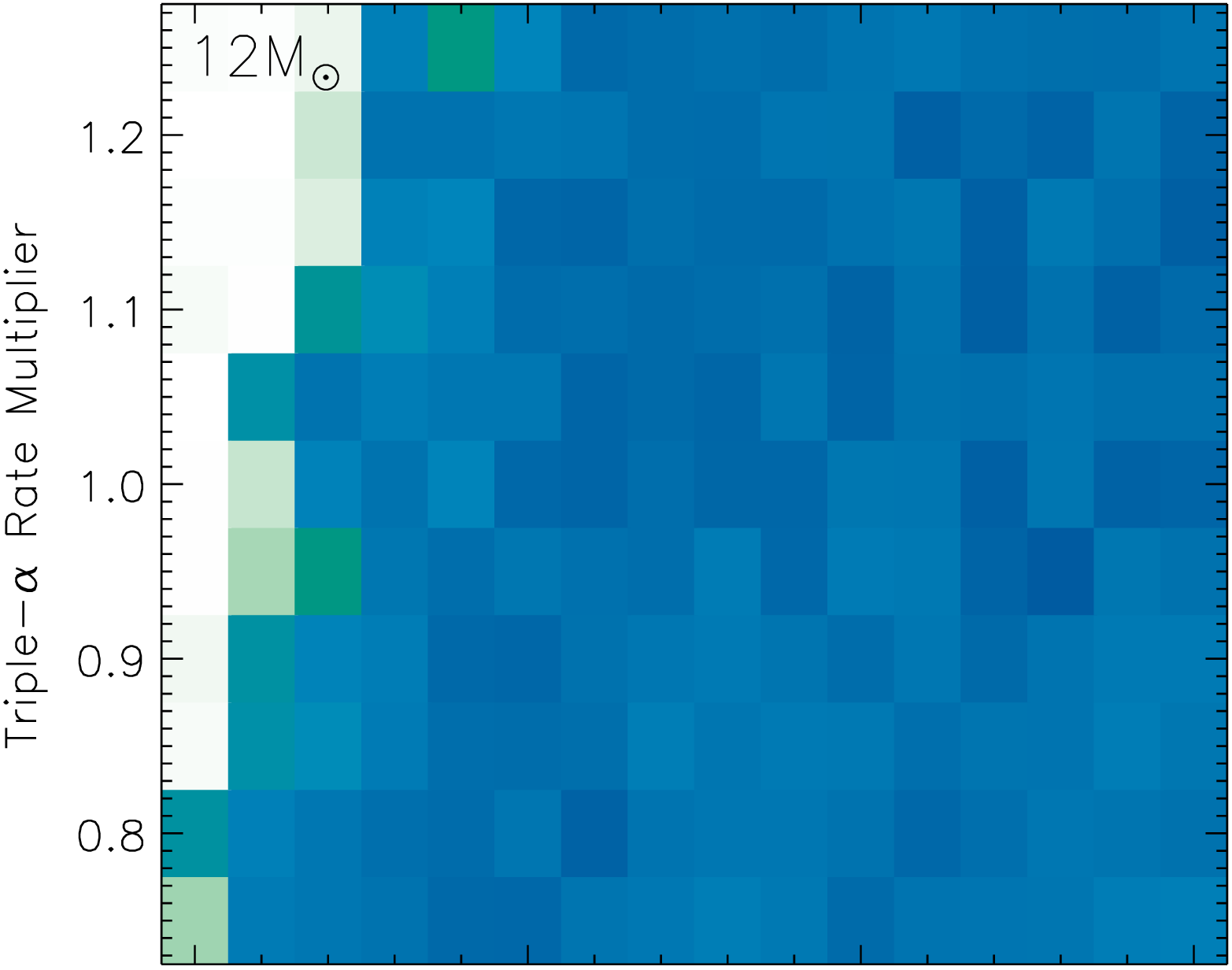}%
\end{minipage}\hspace{-1.345cm}%
\begin{minipage}[t]{0.32\columnwidth}%
\includegraphics[angle=90,scale=0.2]{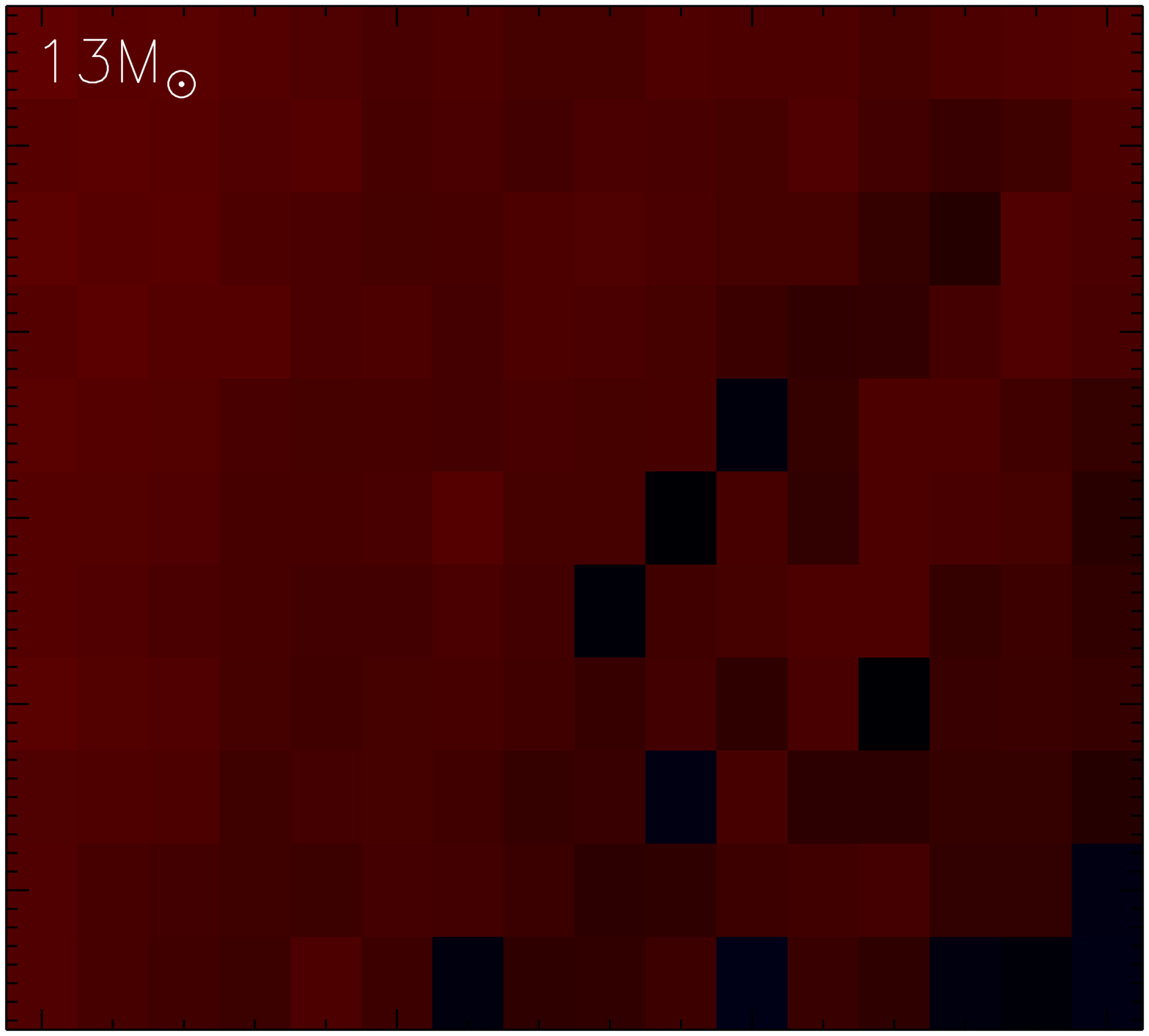}%
\end{minipage}\hspace{-1.345cm}%
\begin{minipage}[t]{0.32\columnwidth}%
\includegraphics[angle=90,scale=0.2]{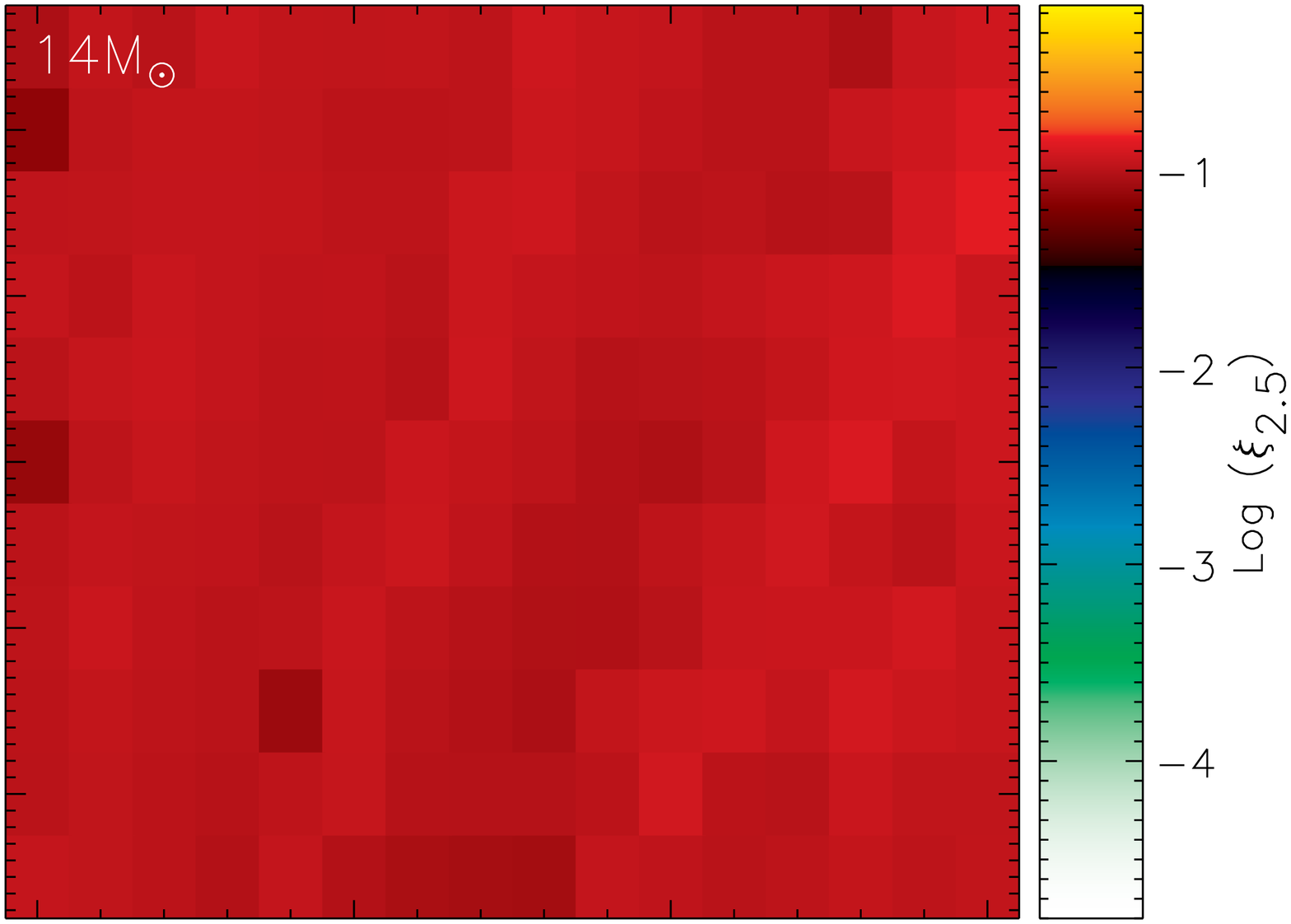}%
\end{minipage}

\vspace{-0.385cm}

\begin{minipage}[t]{0.32\columnwidth}%
\includegraphics[angle=90,scale=0.2]{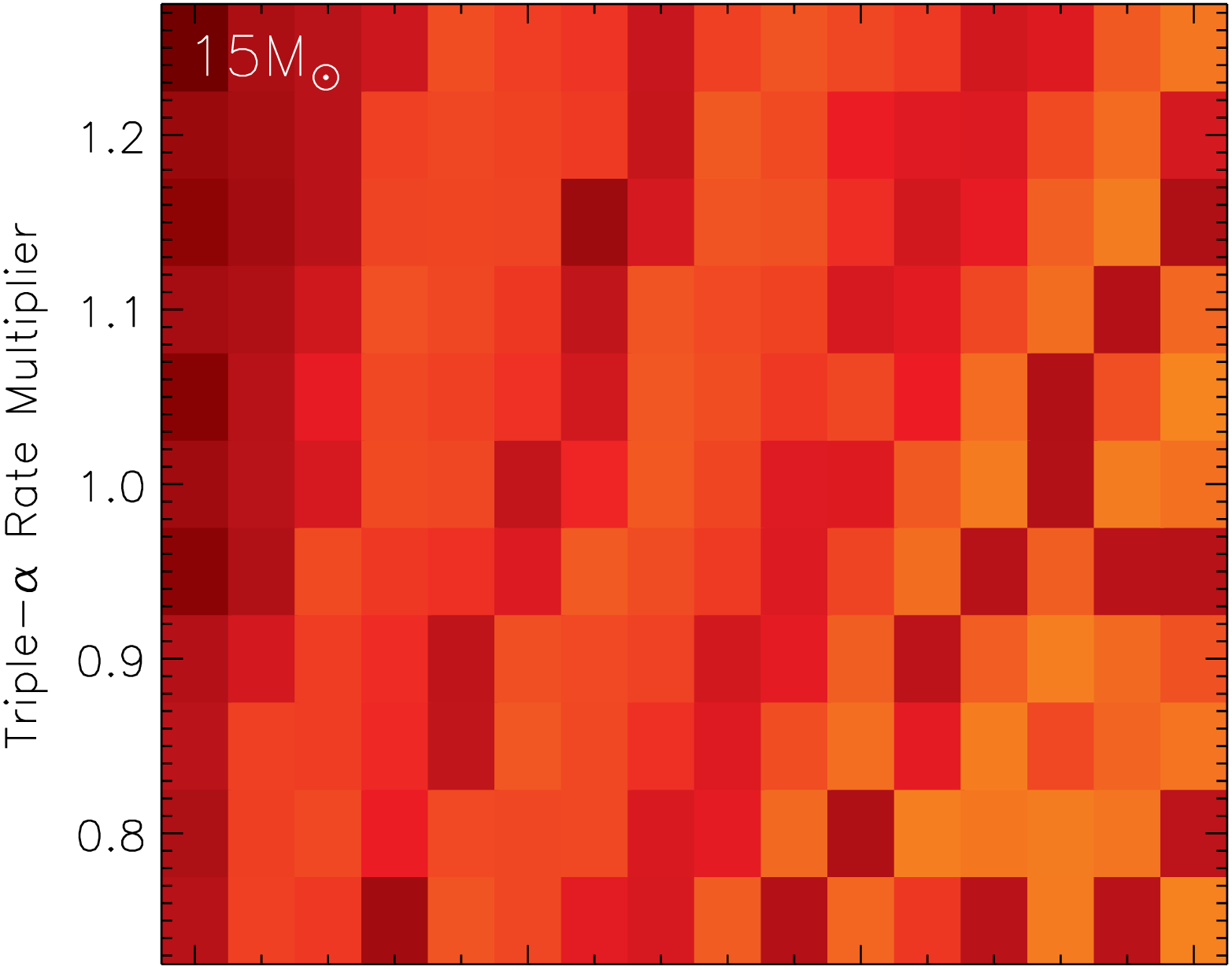}%
\end{minipage}\hspace{-1.345cm}%
\begin{minipage}[t]{0.32\columnwidth}%
\includegraphics[angle=90,scale=0.2]{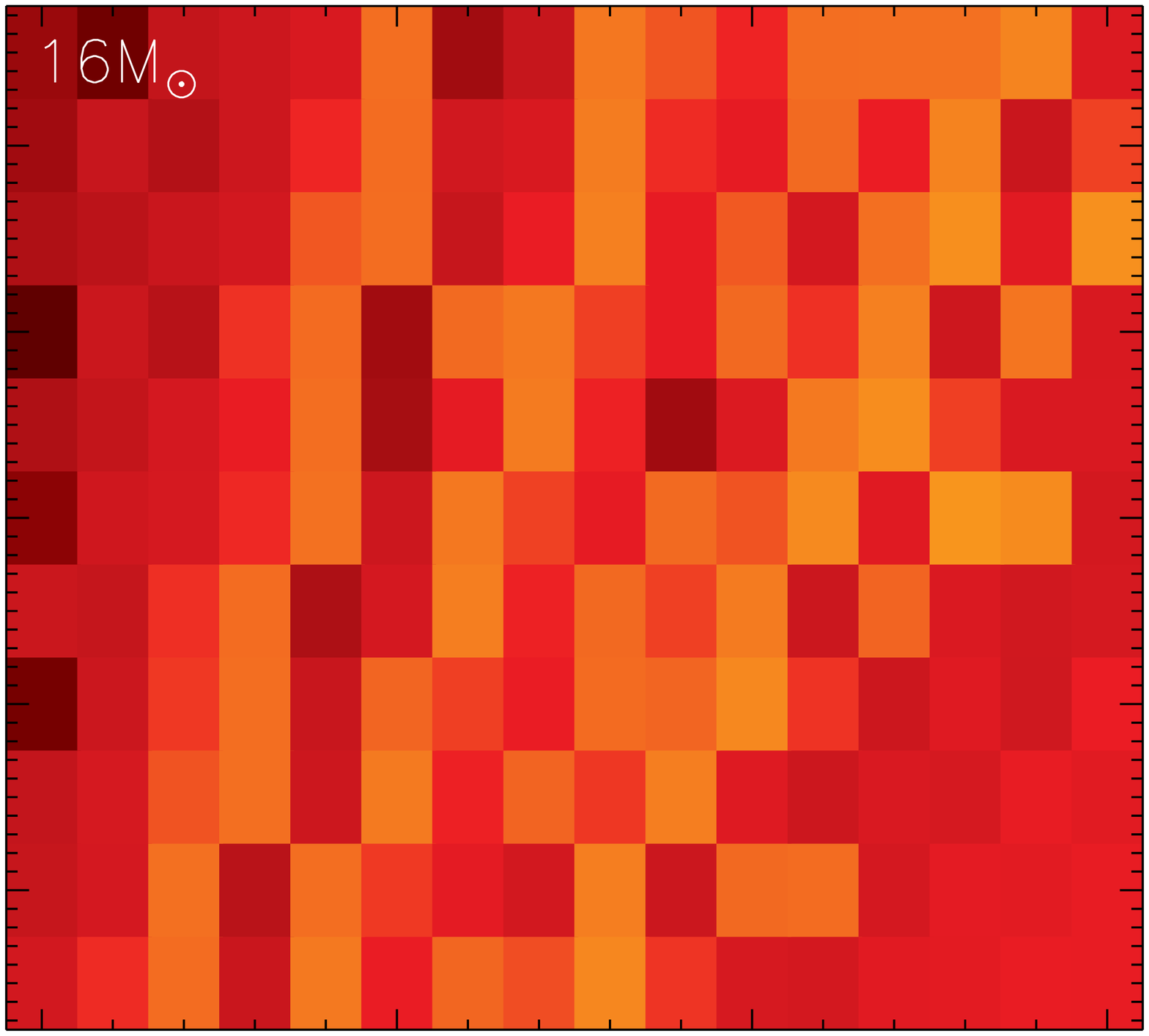}%
\end{minipage}\hspace{-1.345cm}%
\begin{minipage}[t]{0.32\columnwidth}%
\includegraphics[angle=90,scale=0.2]{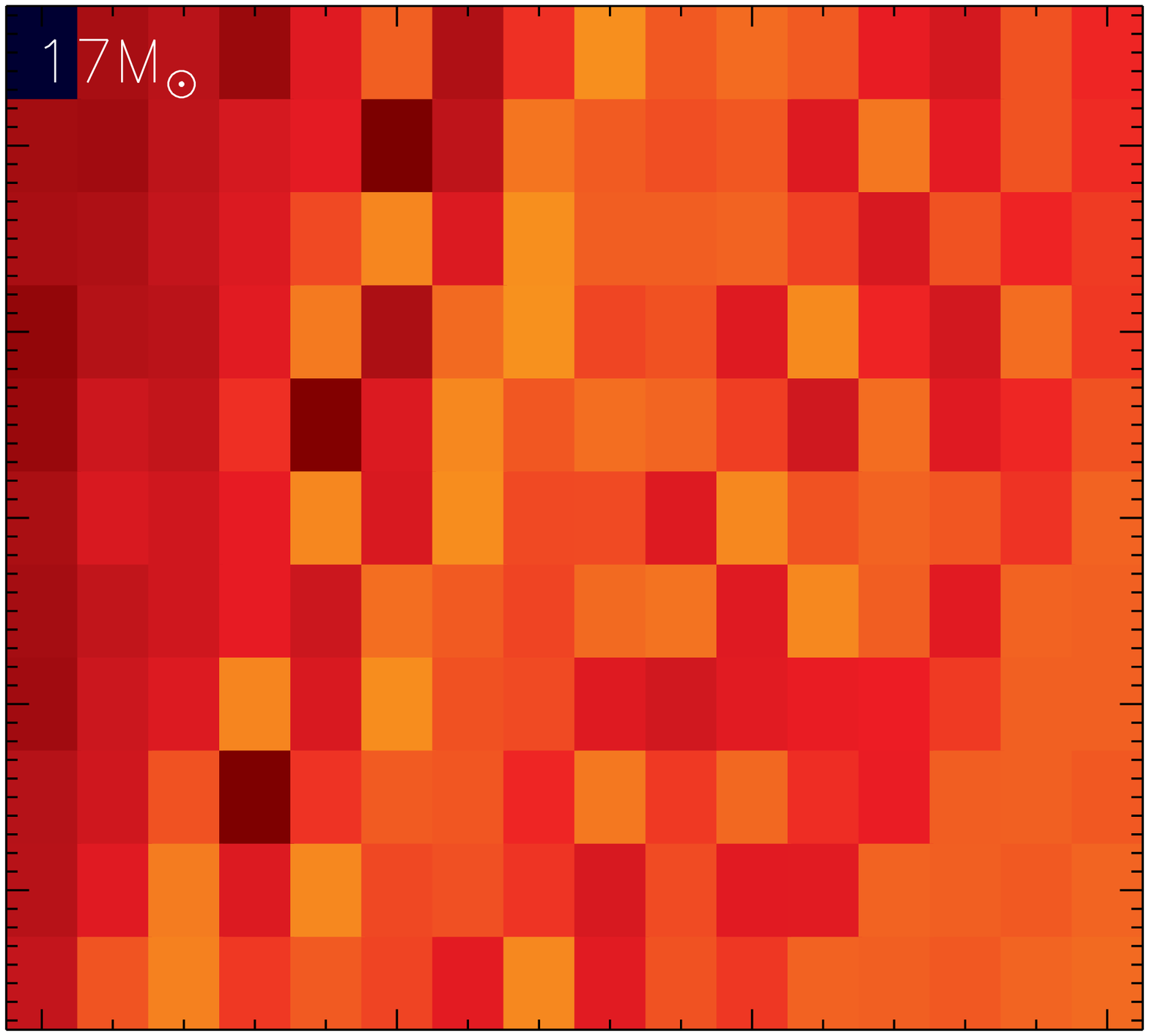}%
\end{minipage}

\vspace{-0.385cm}

\begin{minipage}[t]{0.32\columnwidth}%
\includegraphics[angle=90,scale=0.2]{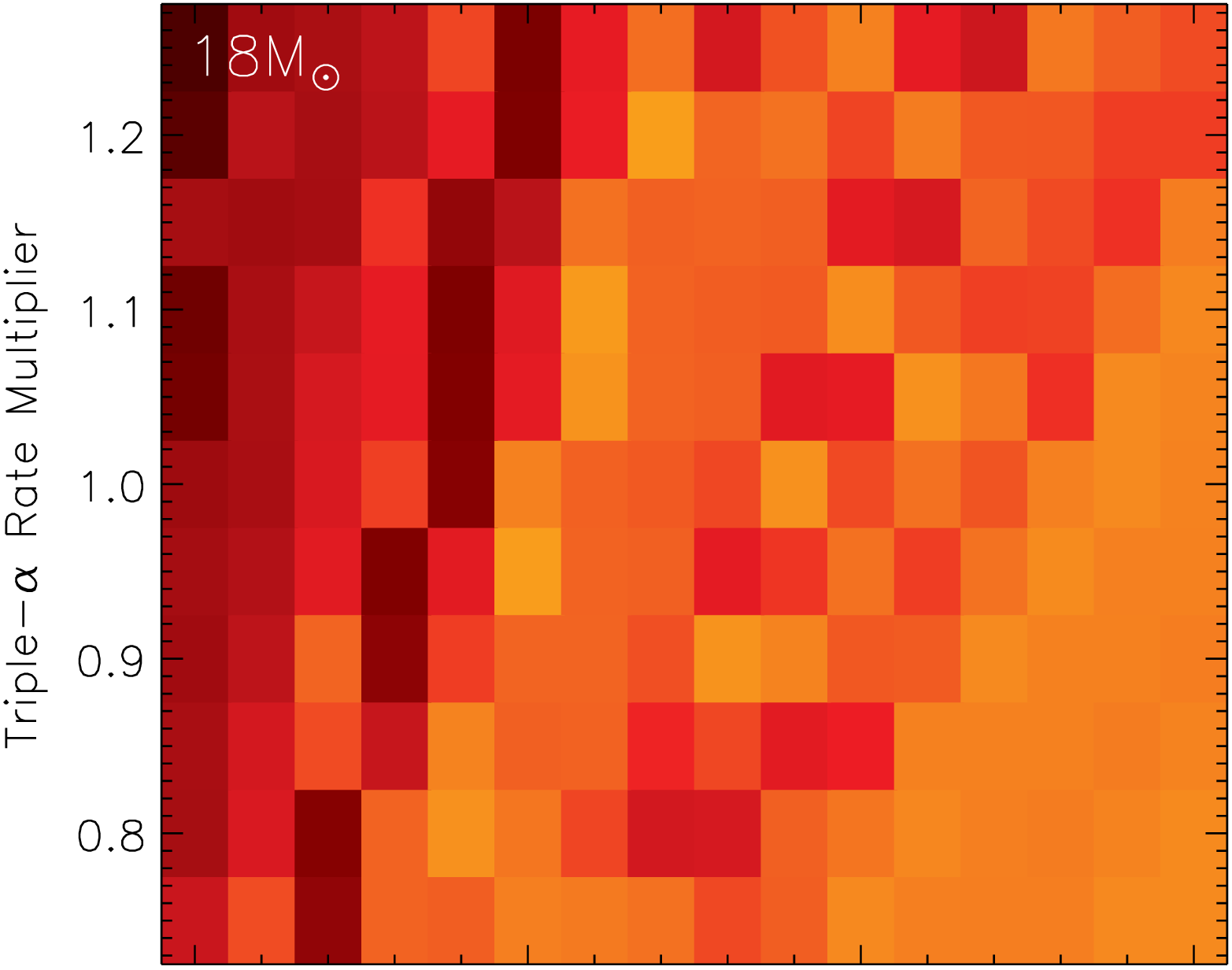}%
\end{minipage}\hspace{-1.345cm}%
\begin{minipage}[t]{0.32\columnwidth}%
\includegraphics[angle=90,scale=0.2]{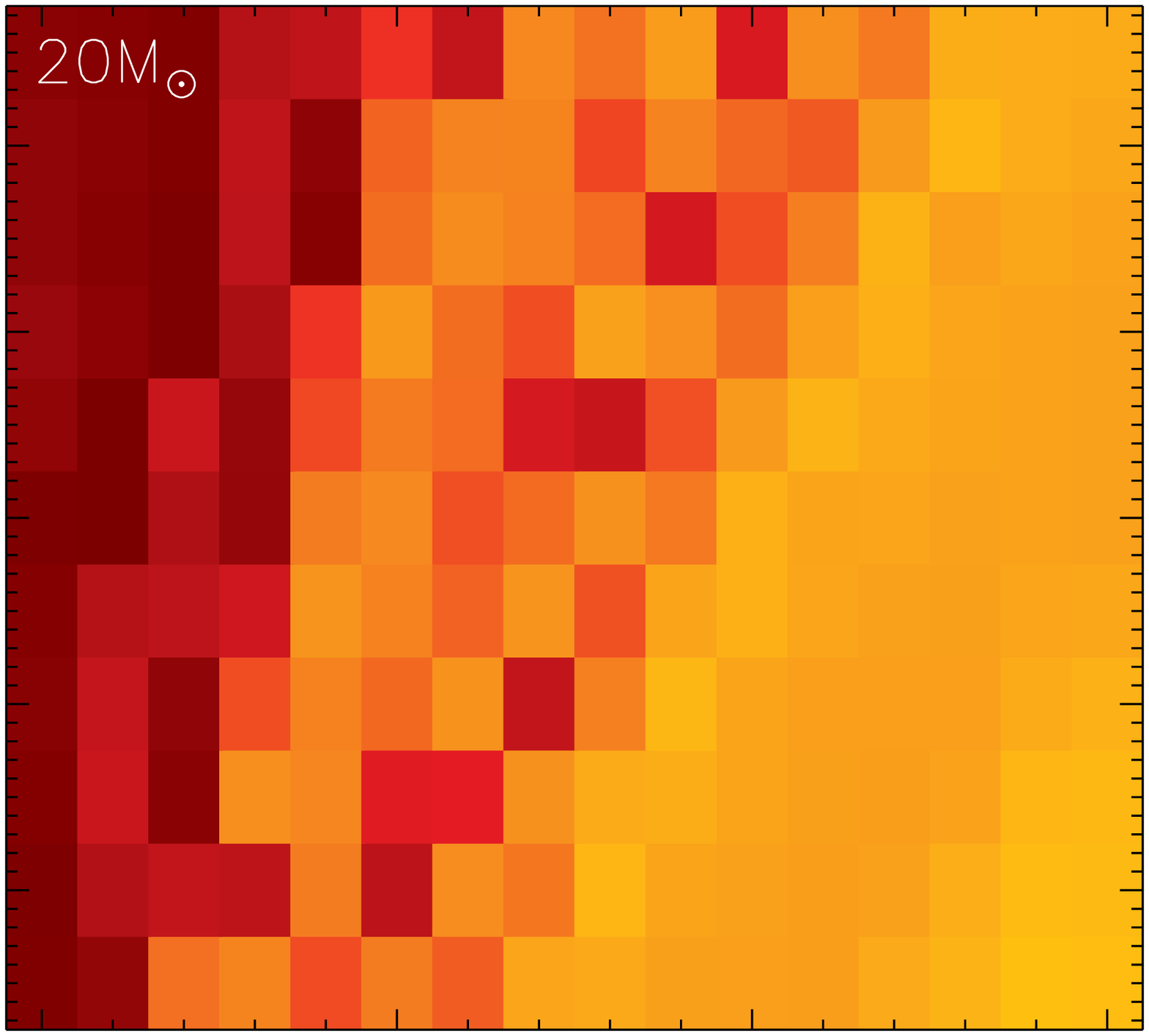}%
\end{minipage}\hspace{-1.345cm}%
\begin{minipage}[t]{0.32\columnwidth}%
\includegraphics[angle=90,scale=0.2]{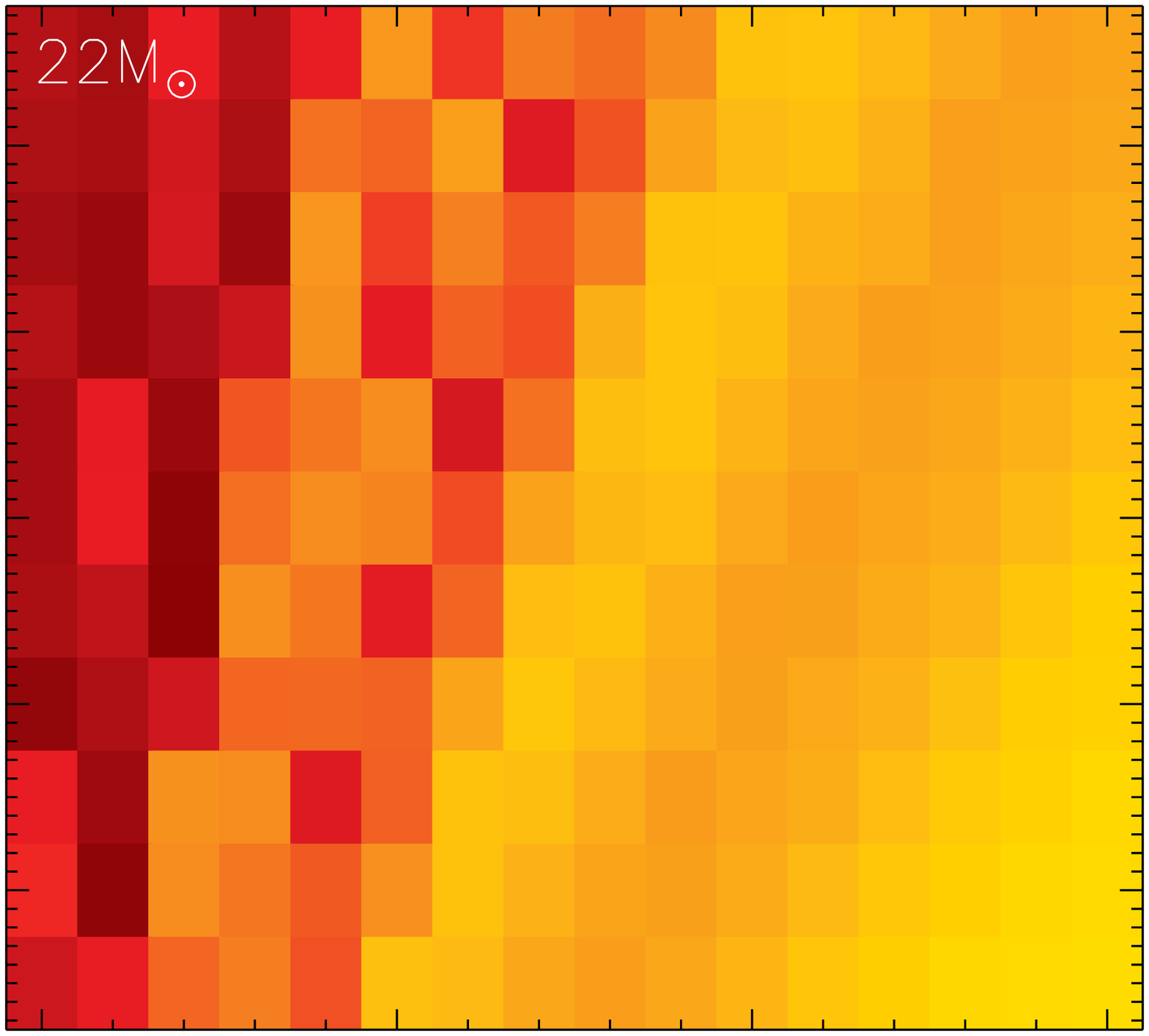}%
\end{minipage}\vspace{-0.385cm}

\begin{minipage}[t]{0.32\columnwidth}%
\includegraphics[angle=90,scale=0.2]{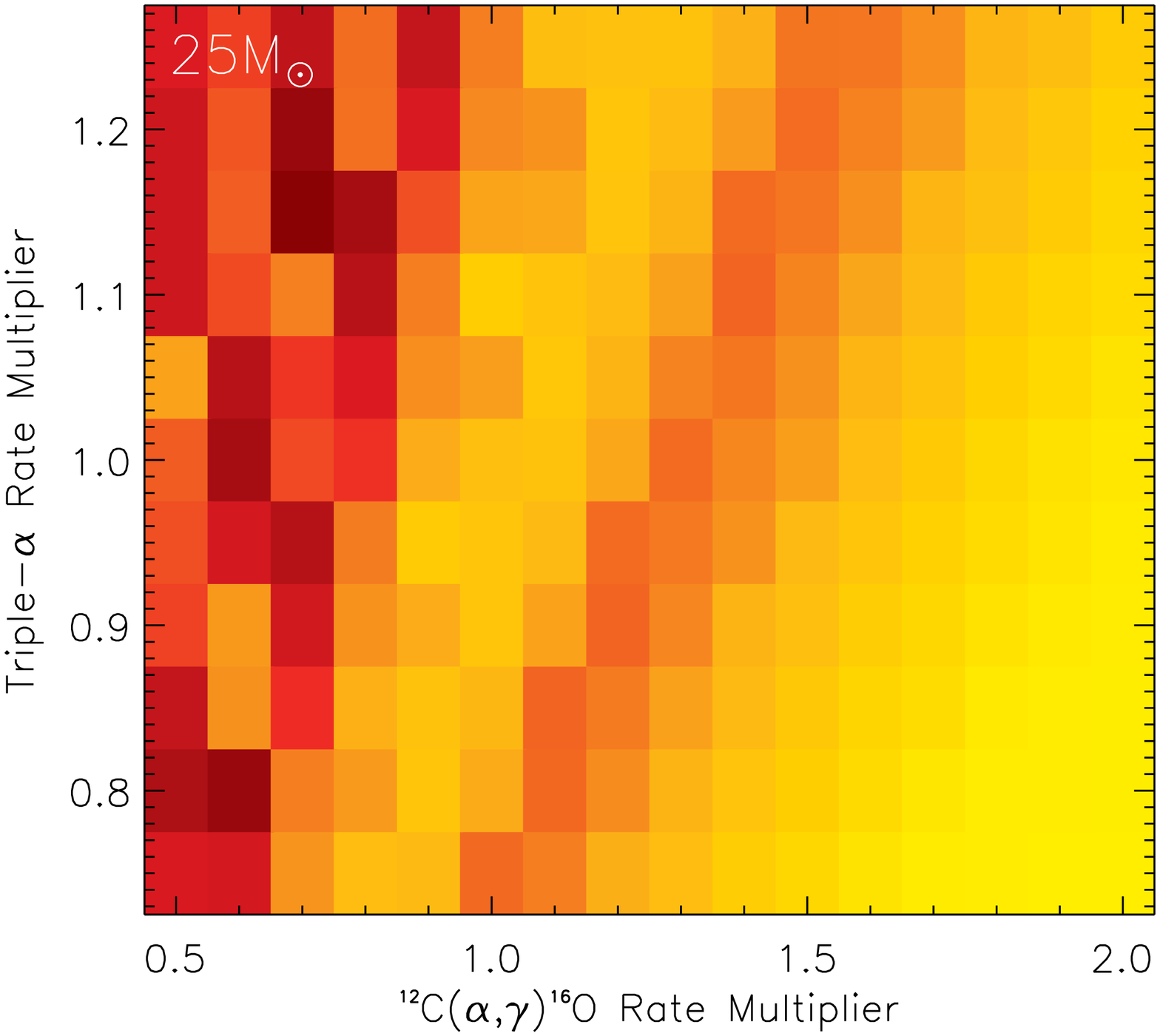}%
\end{minipage}\hspace{-1.345cm}%
\begin{minipage}[t]{0.32\columnwidth}%
\includegraphics[angle=90,scale=0.2]{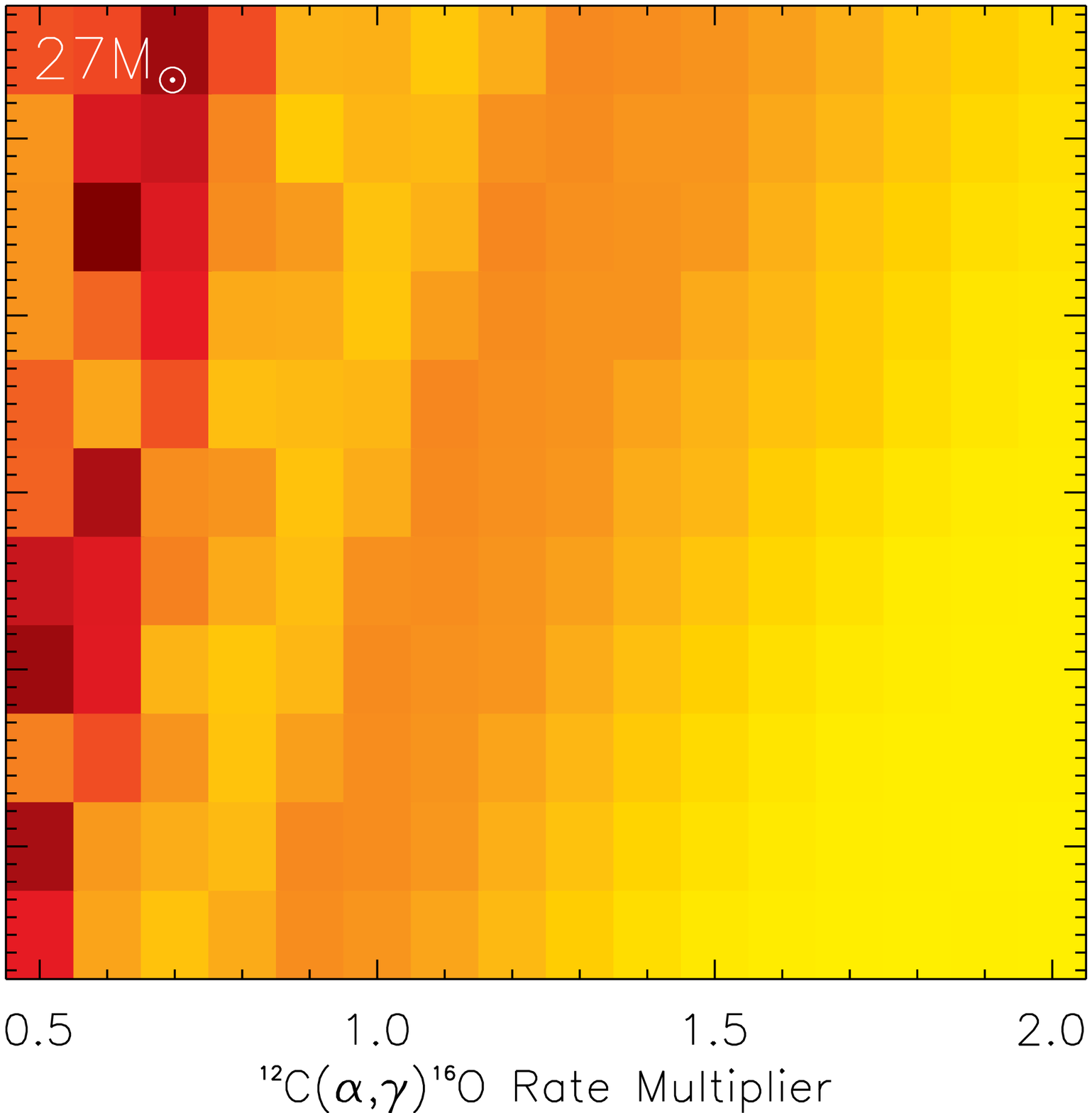}%
\end{minipage}\hspace{-1.345cm}%
\begin{minipage}[t]{0.32\columnwidth}%
\includegraphics[angle=90,scale=0.2]{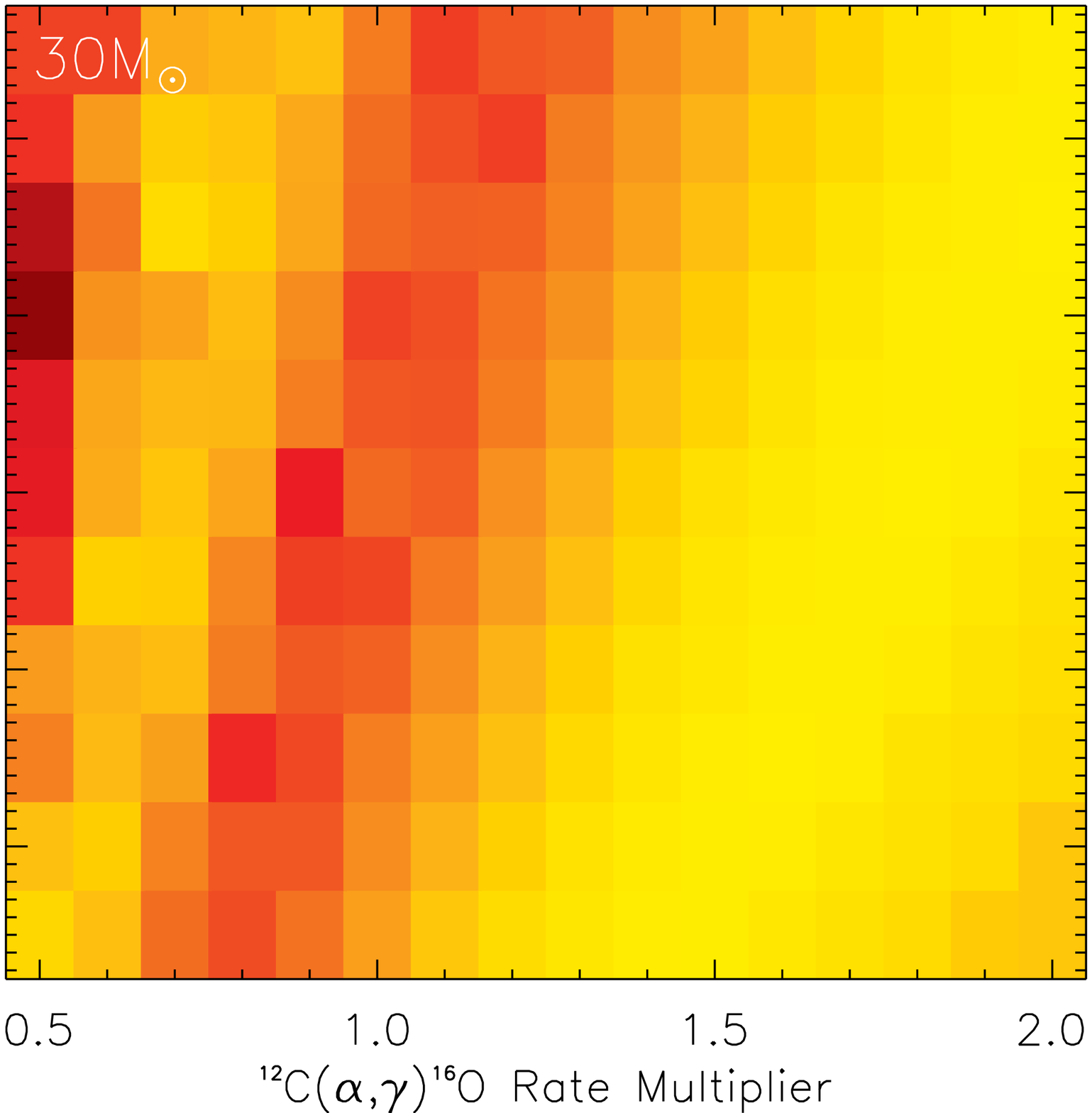}%
\end{minipage}

\caption{\label{fig:compacts}Compactness parameter values for all models as
a function of the $R_{3\alpha}$ and $R_{\alpha,12}$ multipliers.
The color scale gives the logarithm of the compactness parameter,
and the rest of the figure follows the convention of Fig.\,\ref{fig:Intermediate_sigmas}.}
\end{figure}

\begin{figure}[H]
\begin{minipage}[t]{0.32\columnwidth}%
\includegraphics[angle=90,scale=0.2]{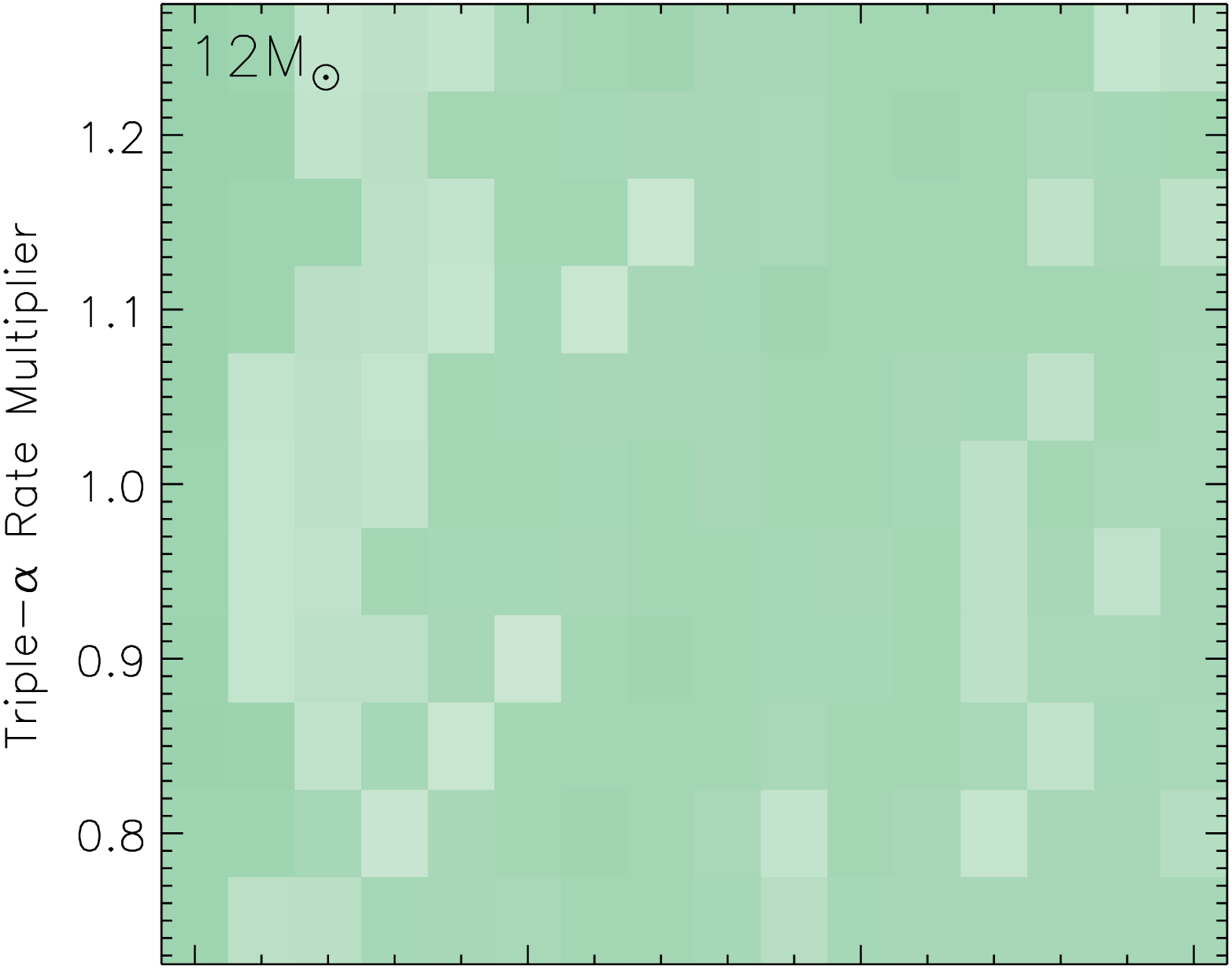}%
\end{minipage}\hspace{-1.345cm}%
\begin{minipage}[t]{0.32\columnwidth}%
\includegraphics[angle=90,scale=0.2]{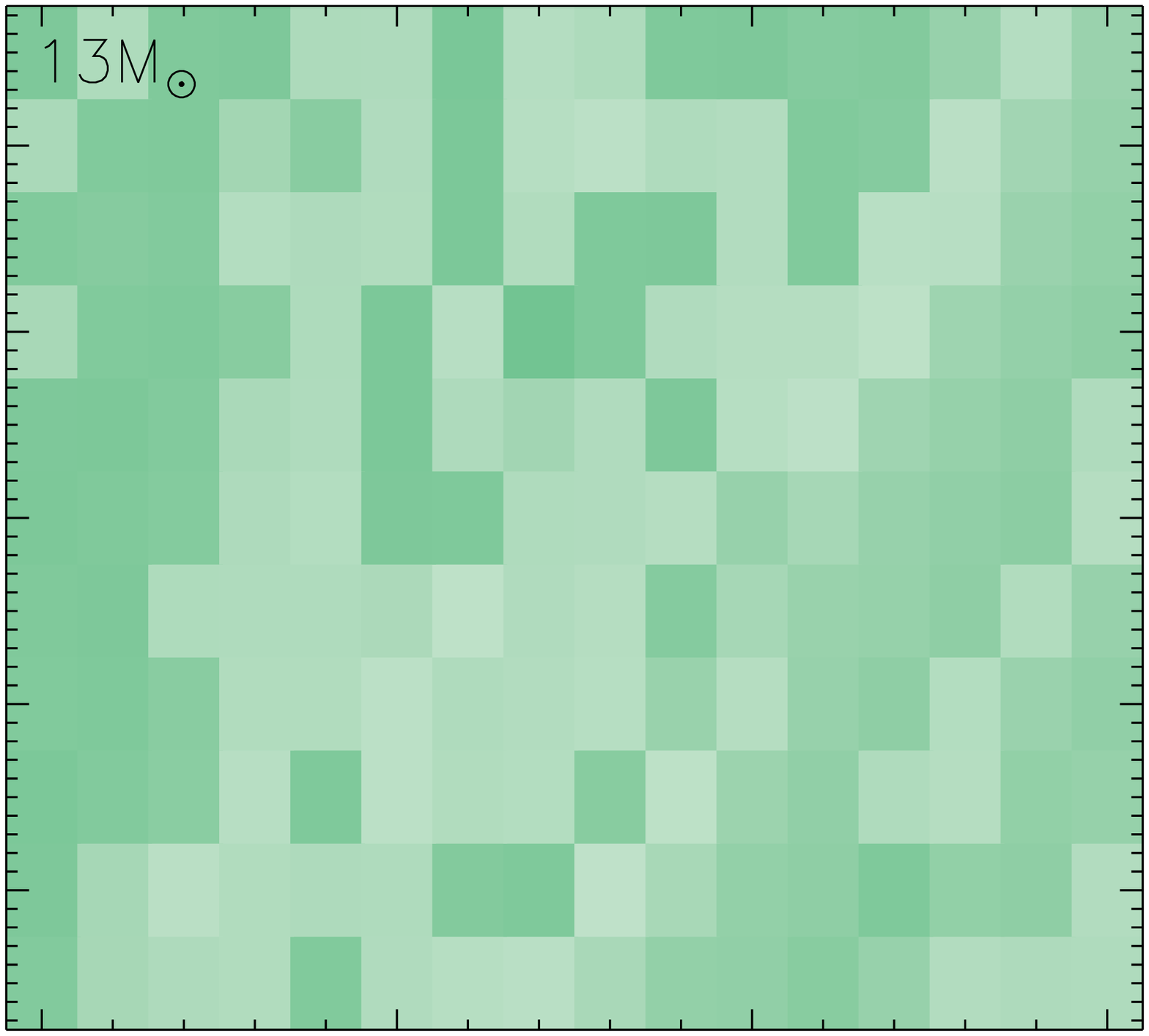}%
\end{minipage}\hspace{-1.345cm}%
\begin{minipage}[t]{0.32\columnwidth}%
\includegraphics[angle=90,scale=0.2]{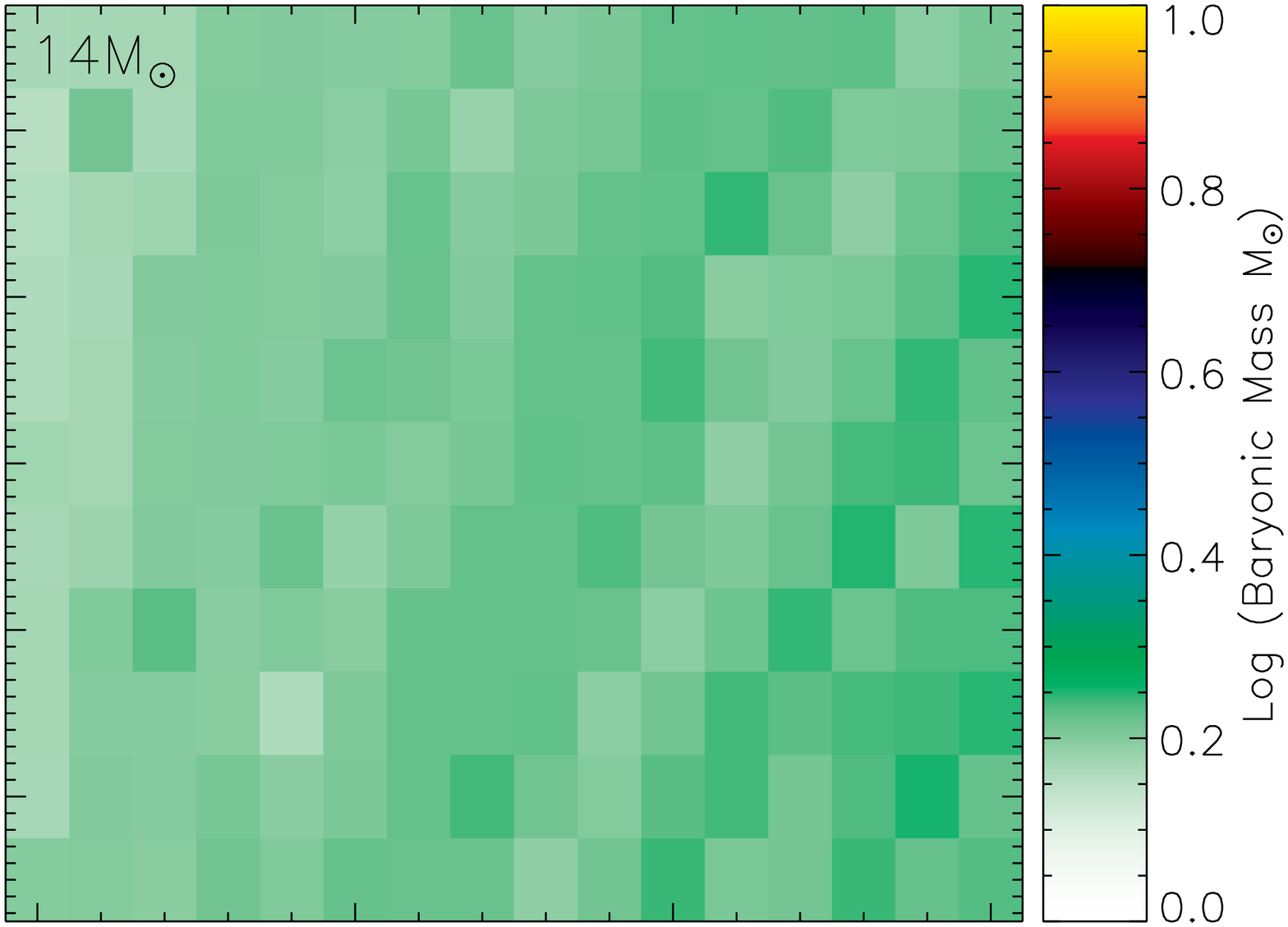}%
\end{minipage}

\vspace{-0.385cm}

\begin{minipage}[t]{0.32\columnwidth}%
\includegraphics[angle=90,scale=0.2]{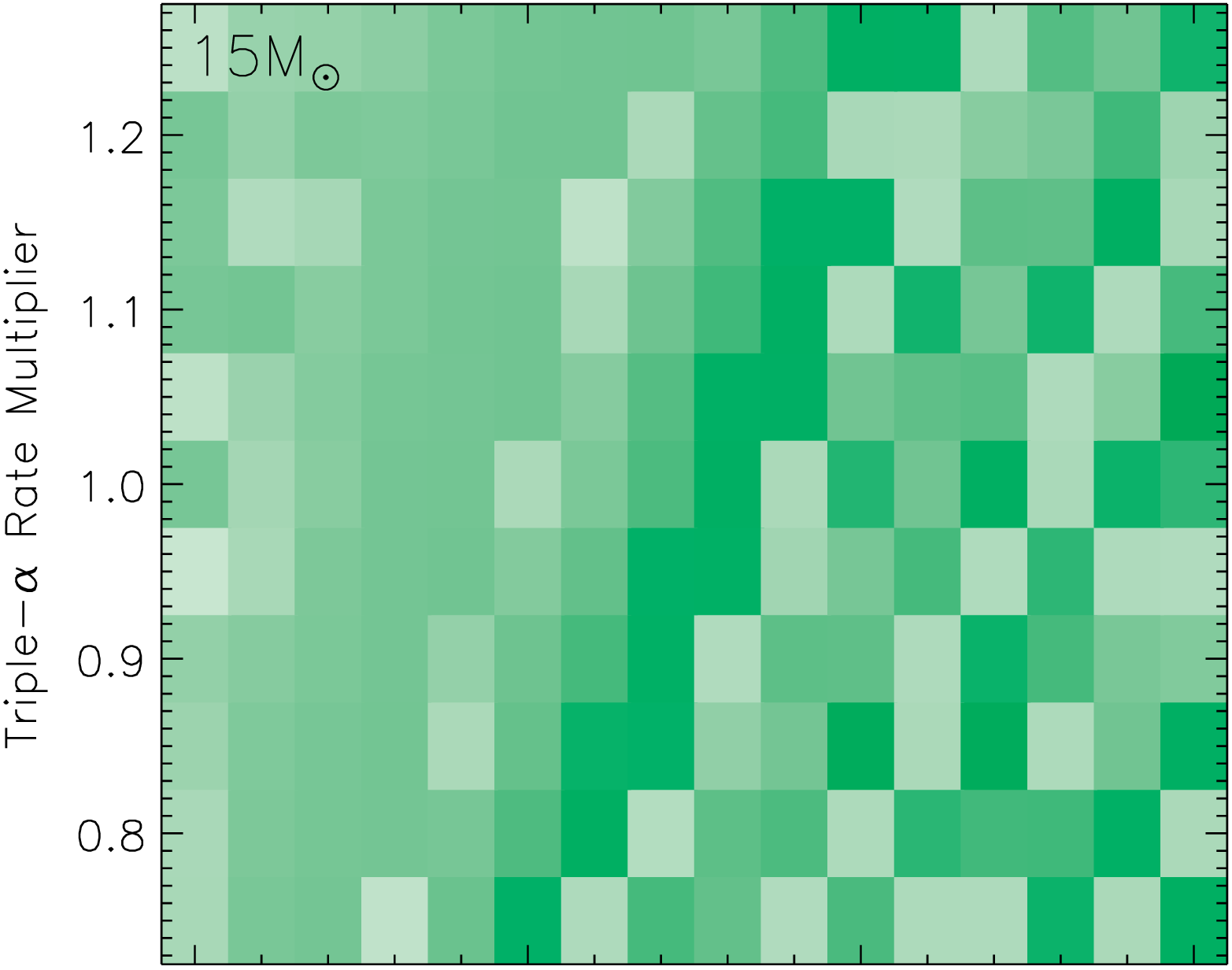}%
\end{minipage}\hspace{-1.345cm}%
\begin{minipage}[t]{0.32\columnwidth}%
\includegraphics[angle=90,scale=0.2]{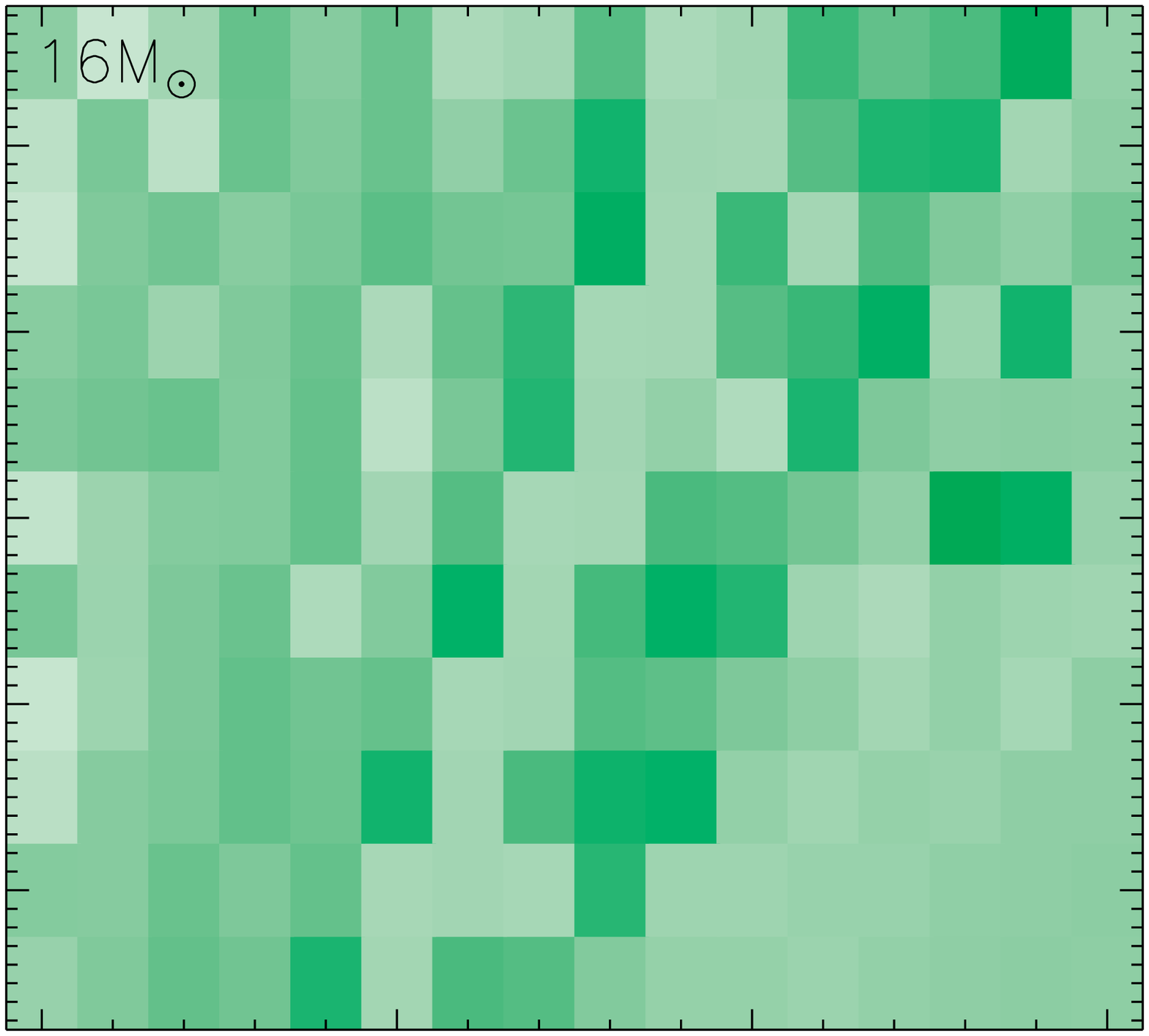}%
\end{minipage}\hspace{-1.345cm}%
\begin{minipage}[t]{0.32\columnwidth}%
\includegraphics[angle=90,scale=0.2]{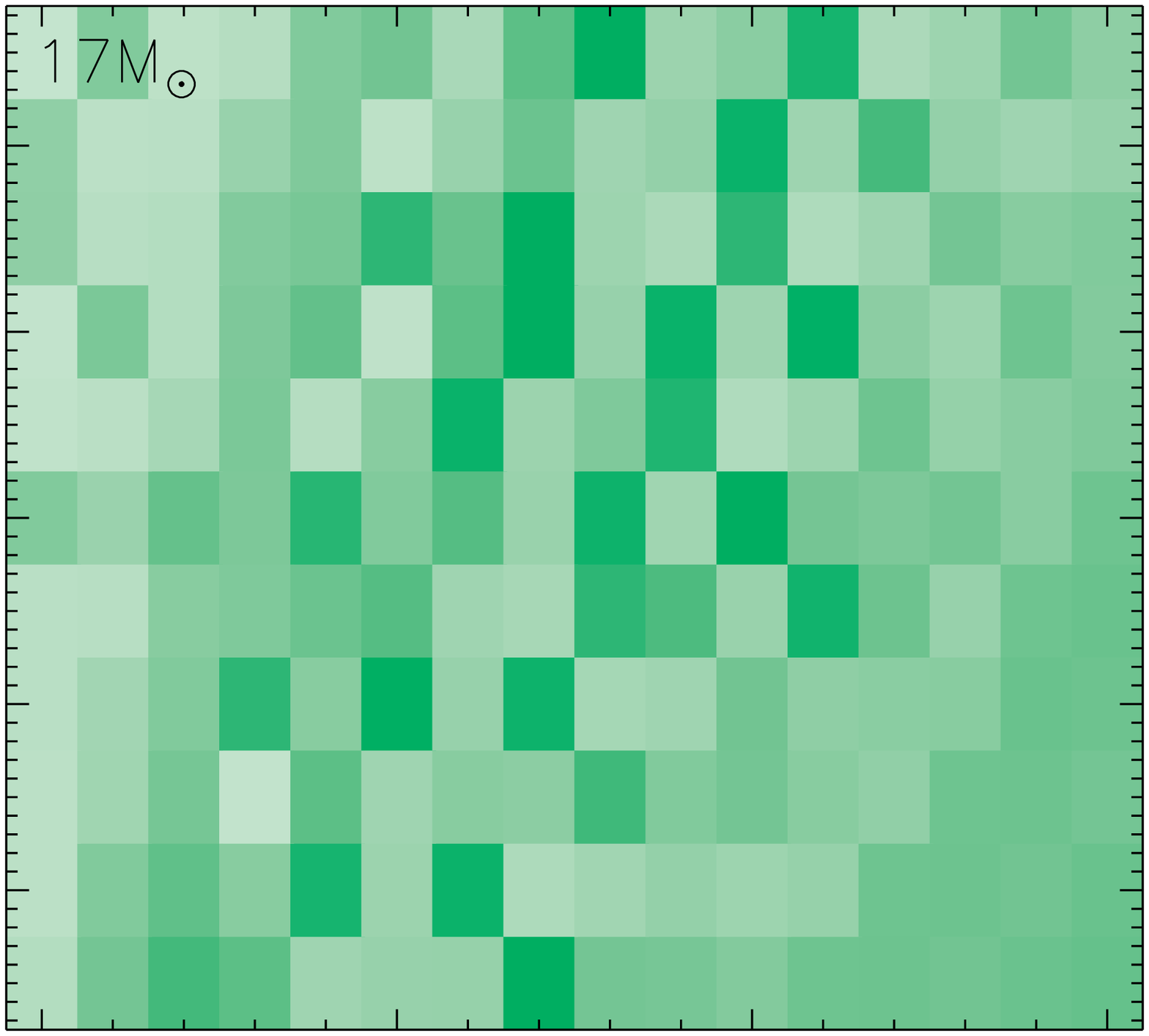}%
\end{minipage}

\vspace{-0.385cm}

\begin{minipage}[t]{0.32\columnwidth}%
\includegraphics[angle=90,scale=0.2]{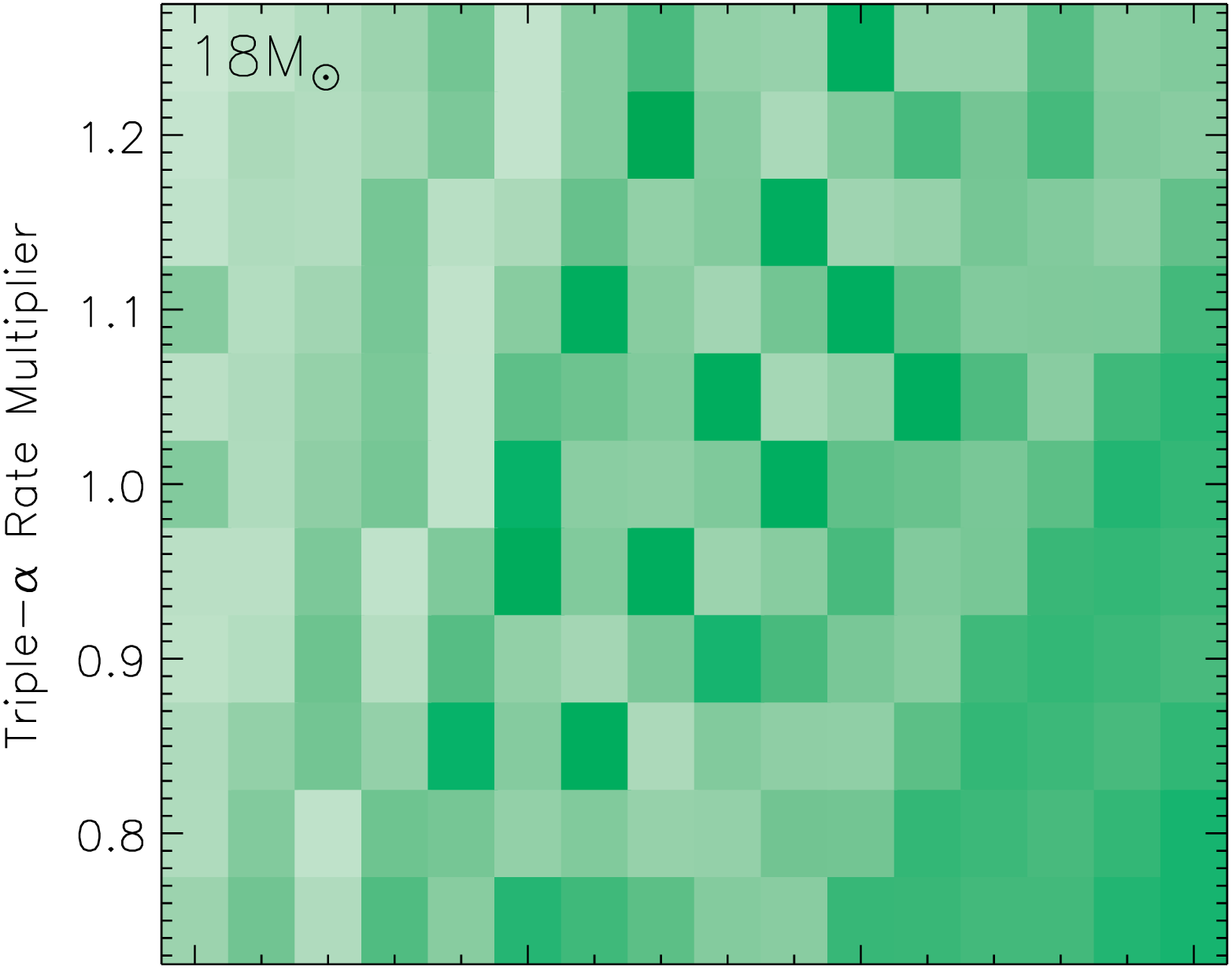}%
\end{minipage}\hspace{-1.345cm}%
\begin{minipage}[t]{0.32\columnwidth}%
\includegraphics[angle=90,scale=0.2]{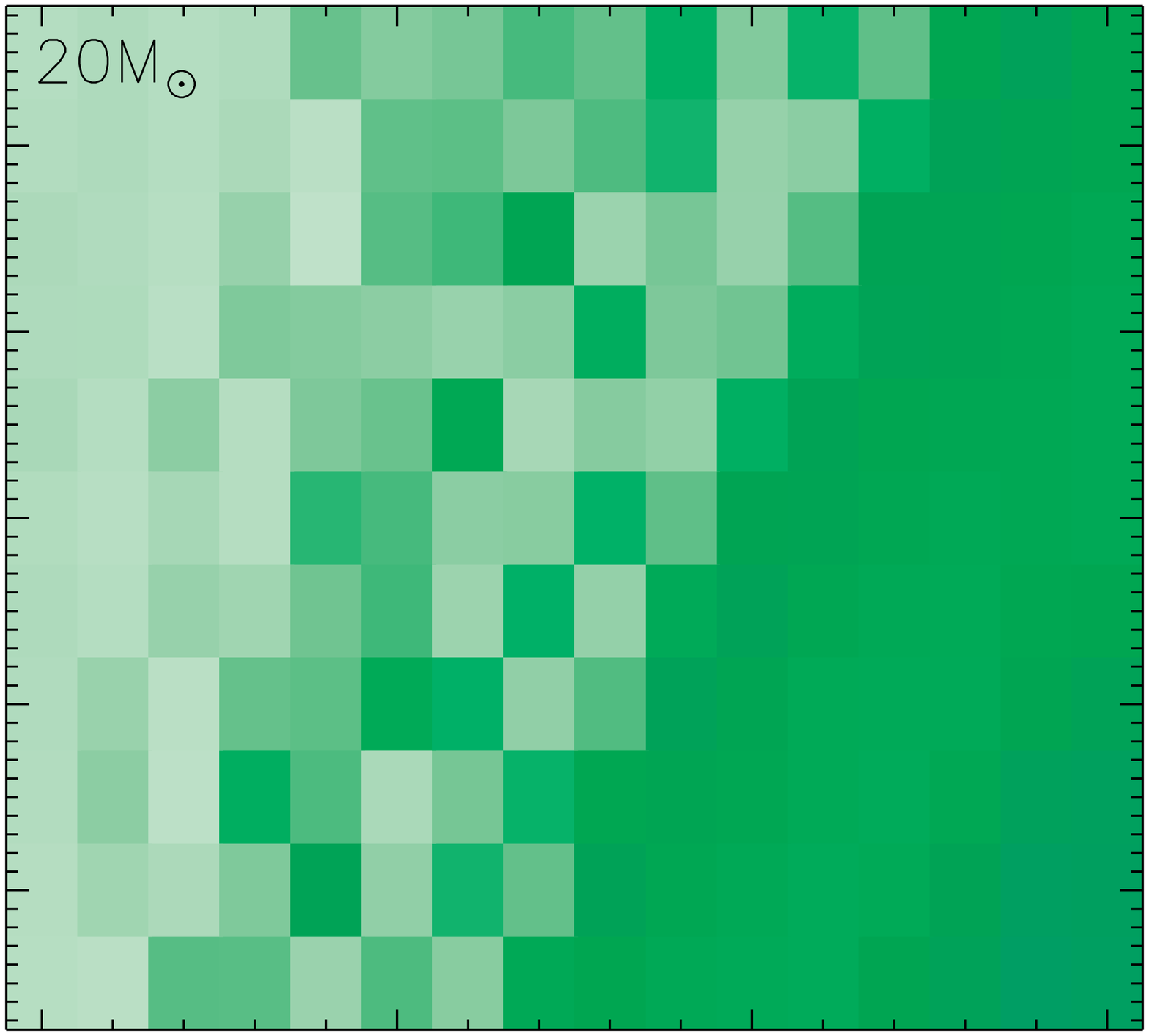}%
\end{minipage}\hspace{-1.345cm}%
\begin{minipage}[t]{0.32\columnwidth}%
\includegraphics[angle=90,scale=0.2]{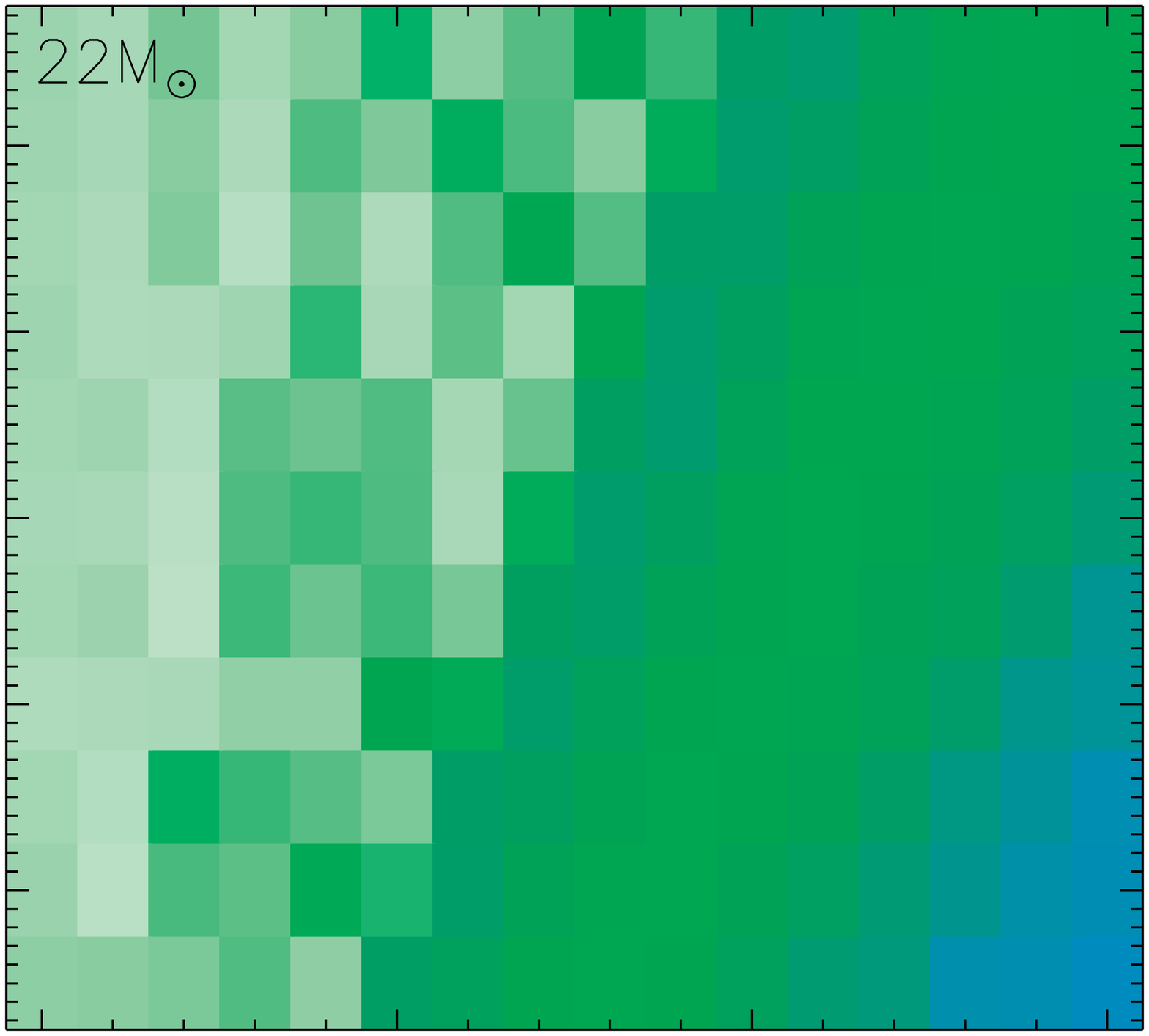}%
\end{minipage}\vspace{-0.385cm}

\begin{minipage}[t]{0.32\columnwidth}%
\includegraphics[angle=90,scale=0.2]{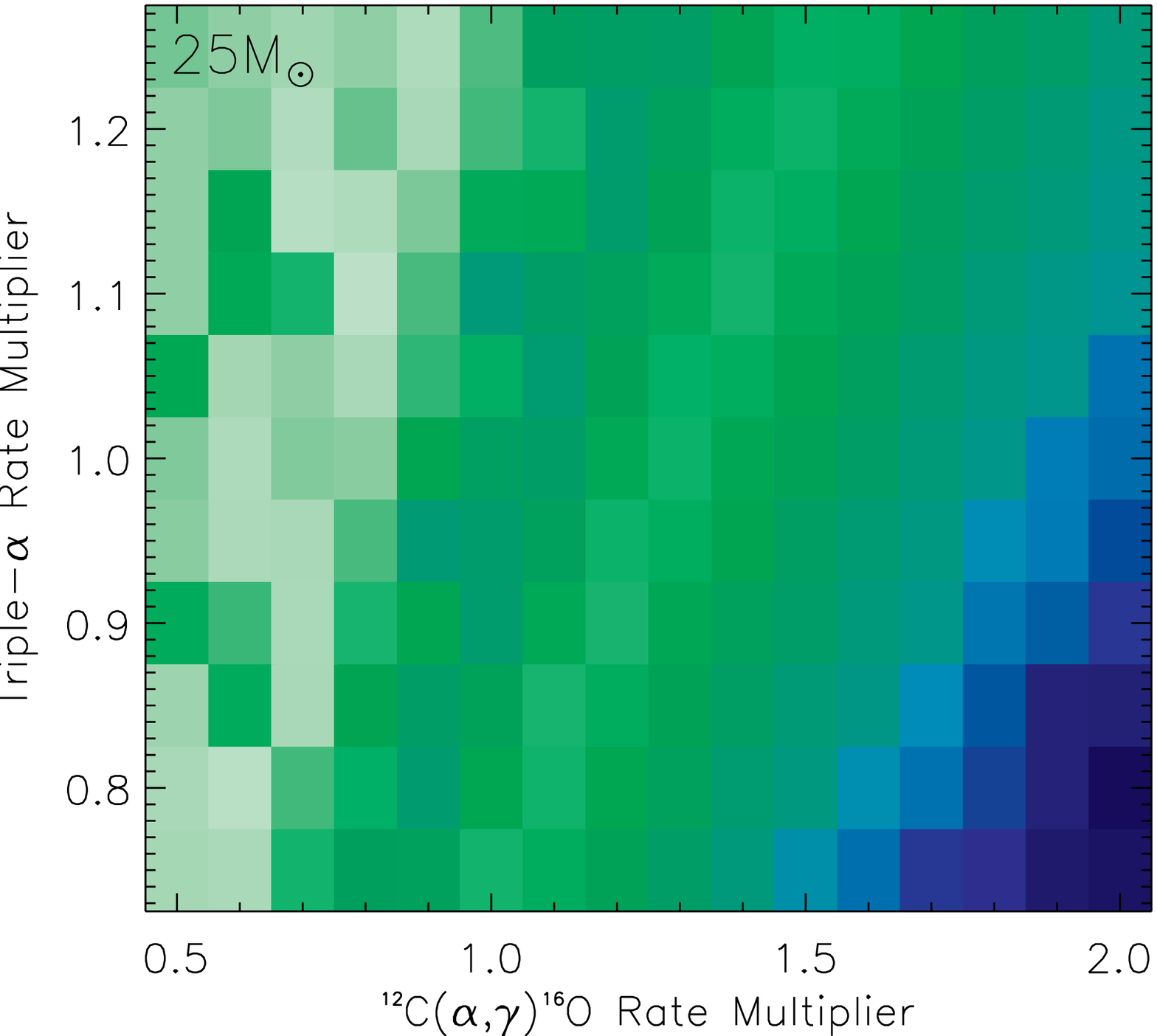}%
\end{minipage}\hspace{-1.345cm}%
\begin{minipage}[t]{0.32\columnwidth}%
\includegraphics[angle=90,scale=0.2]{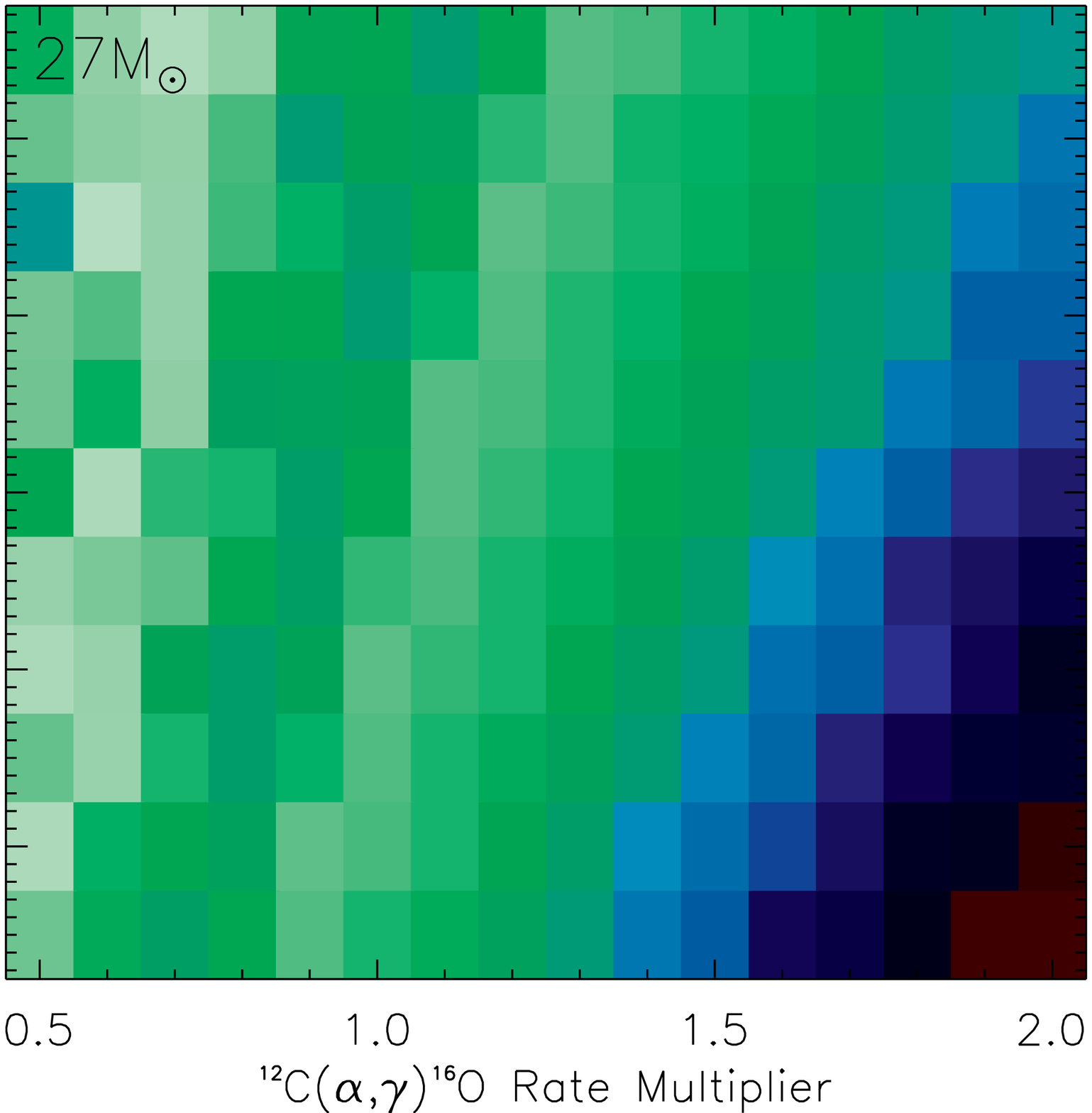}%
\end{minipage}\hspace{-1.345cm}%
\begin{minipage}[t]{0.32\columnwidth}%
\includegraphics[angle=90,scale=0.2]{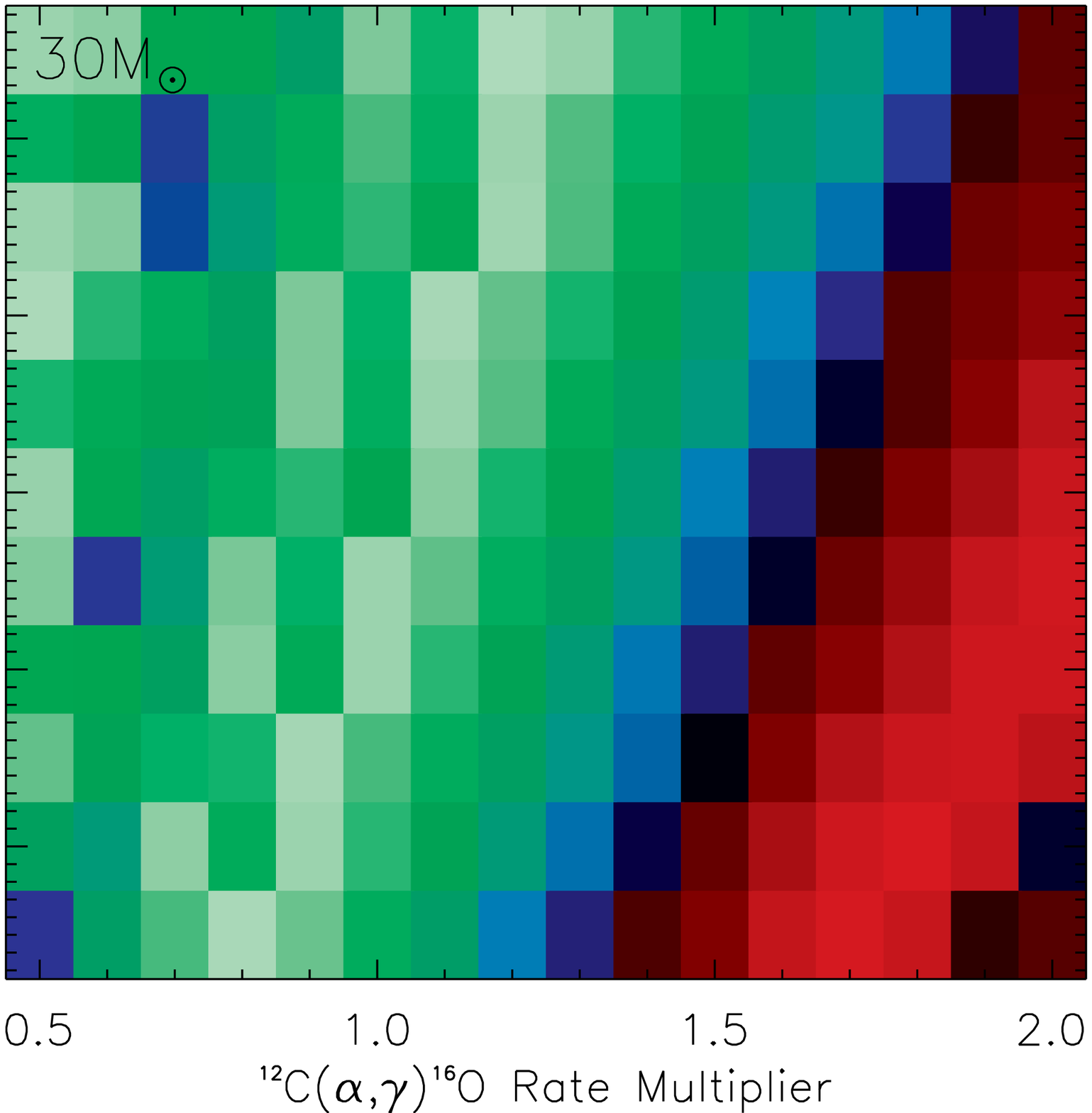}%
\end{minipage}

\caption{\label{fig:baryonics}Baryonic mass of the progenitors $\left(M_{\odot}\right)$
for all models as a function of the $R_{3\alpha}$ and $R_{\alpha,12}$
multipliers. The color scale gives the logarithm of the baryonic mass,
and the rest of the figure follows the convention of Fig.\,\ref{fig:Intermediate_sigmas}.}
\end{figure}

\begin{figure}[H]
\begin{minipage}[t]{0.32\columnwidth}%
\includegraphics[angle=90,scale=0.2]{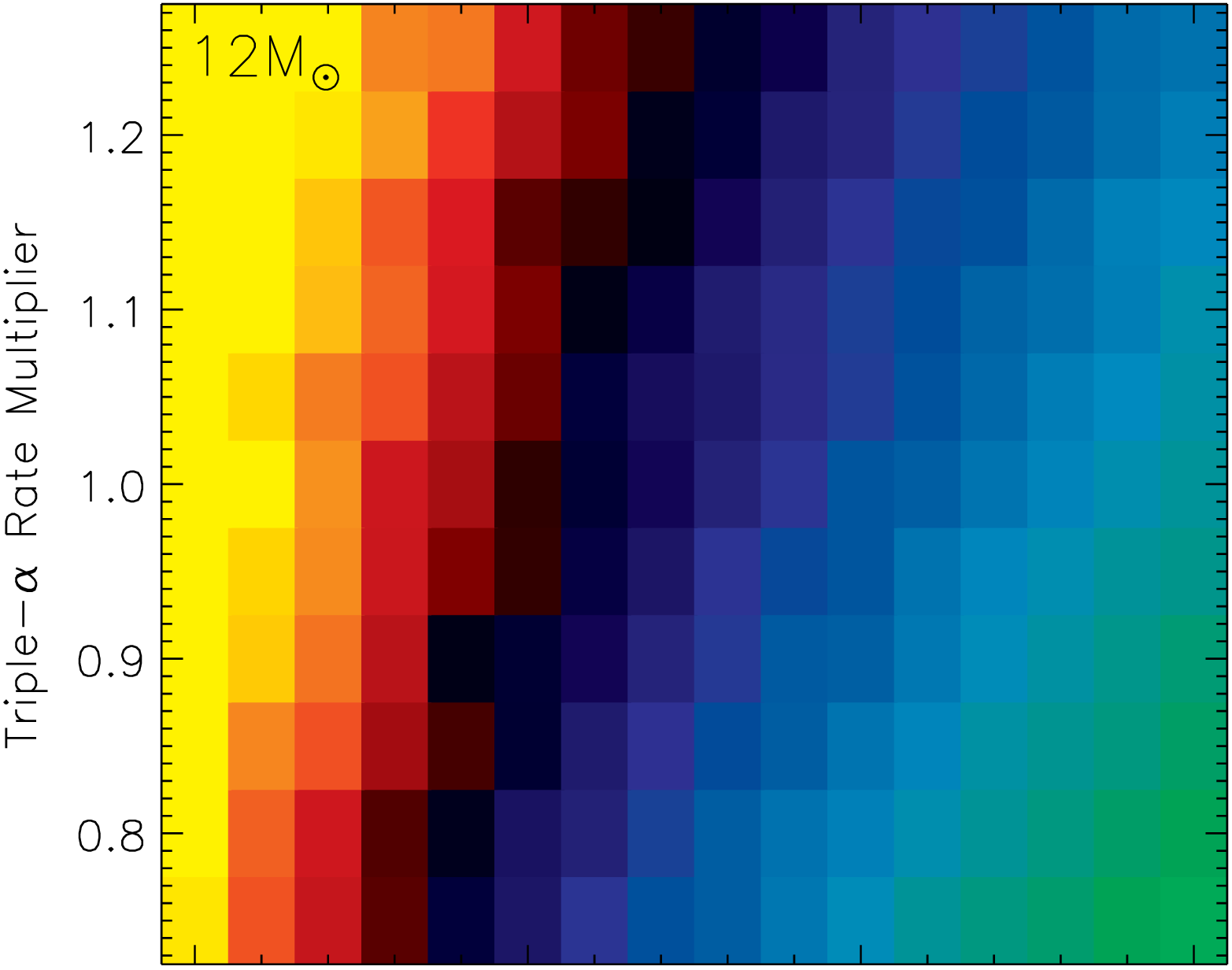}%
\end{minipage}\hspace{-1.345cm}%
\begin{minipage}[t]{0.32\columnwidth}%
\includegraphics[angle=90,scale=0.2]{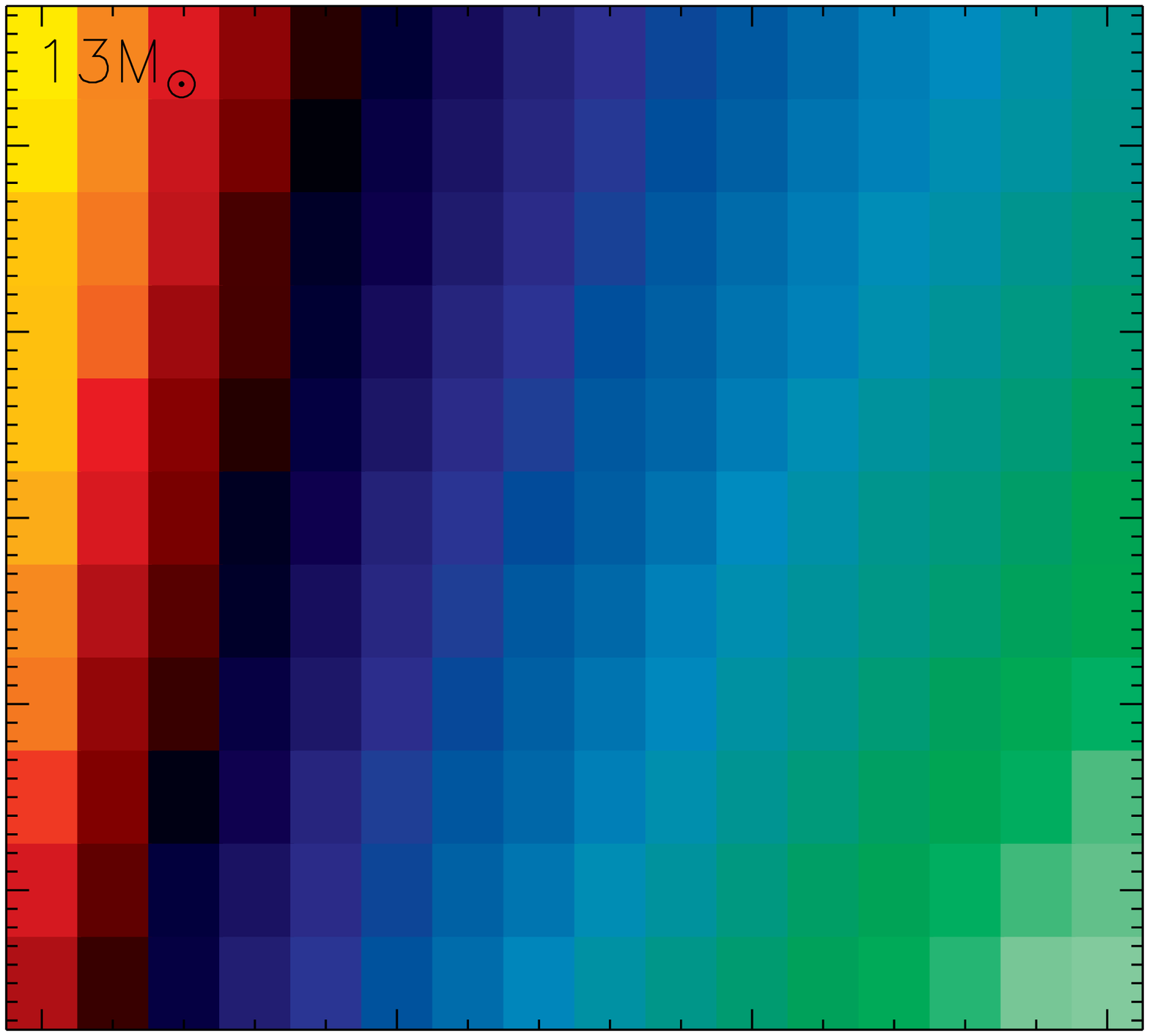}%
\end{minipage}\hspace{-1.345cm}%
\begin{minipage}[t]{0.32\columnwidth}%
\includegraphics[angle=90,scale=0.2]{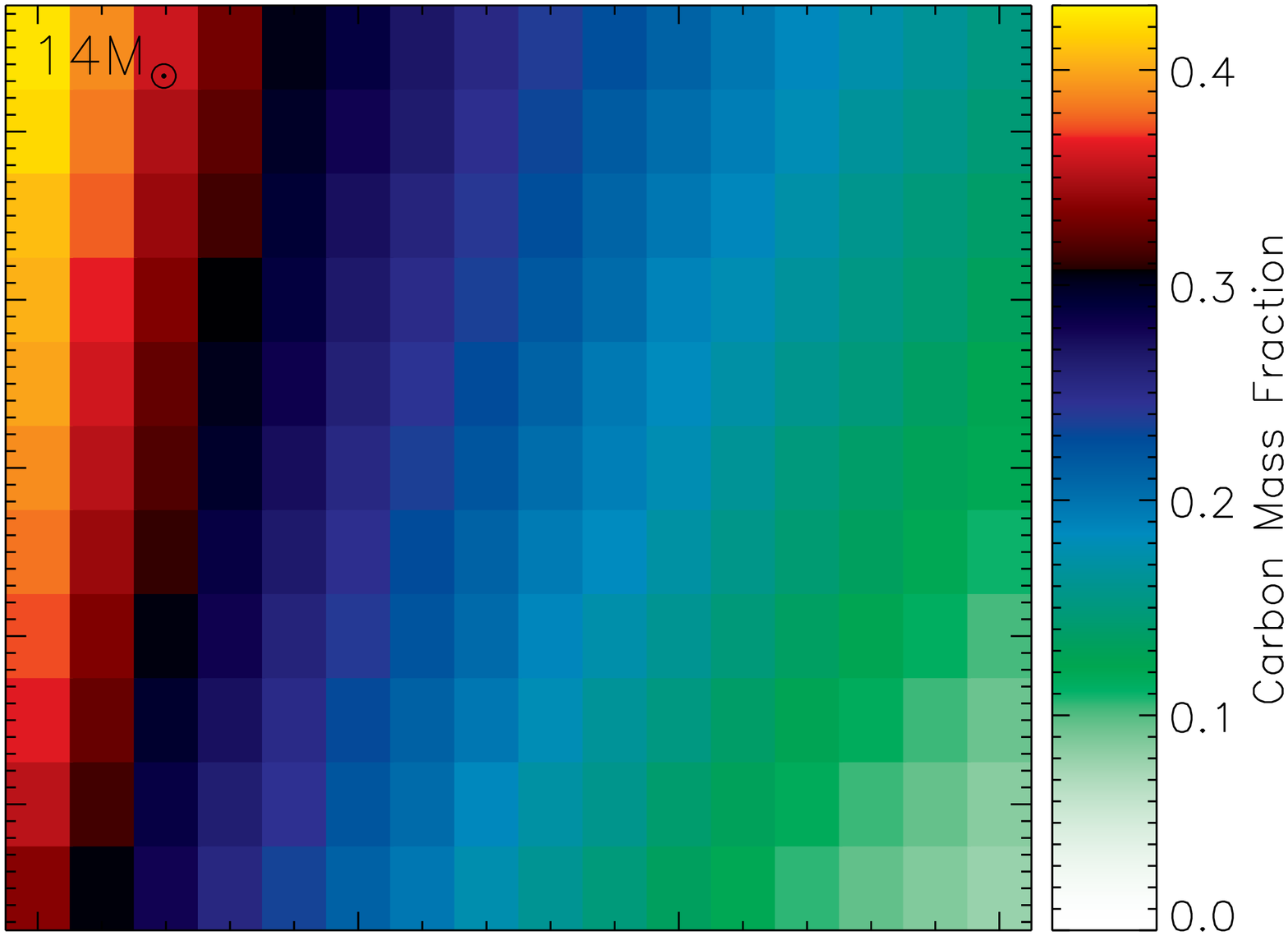}%
\end{minipage}

\vspace{-0.385cm}

\begin{minipage}[t]{0.32\columnwidth}%
\includegraphics[angle=90,scale=0.2]{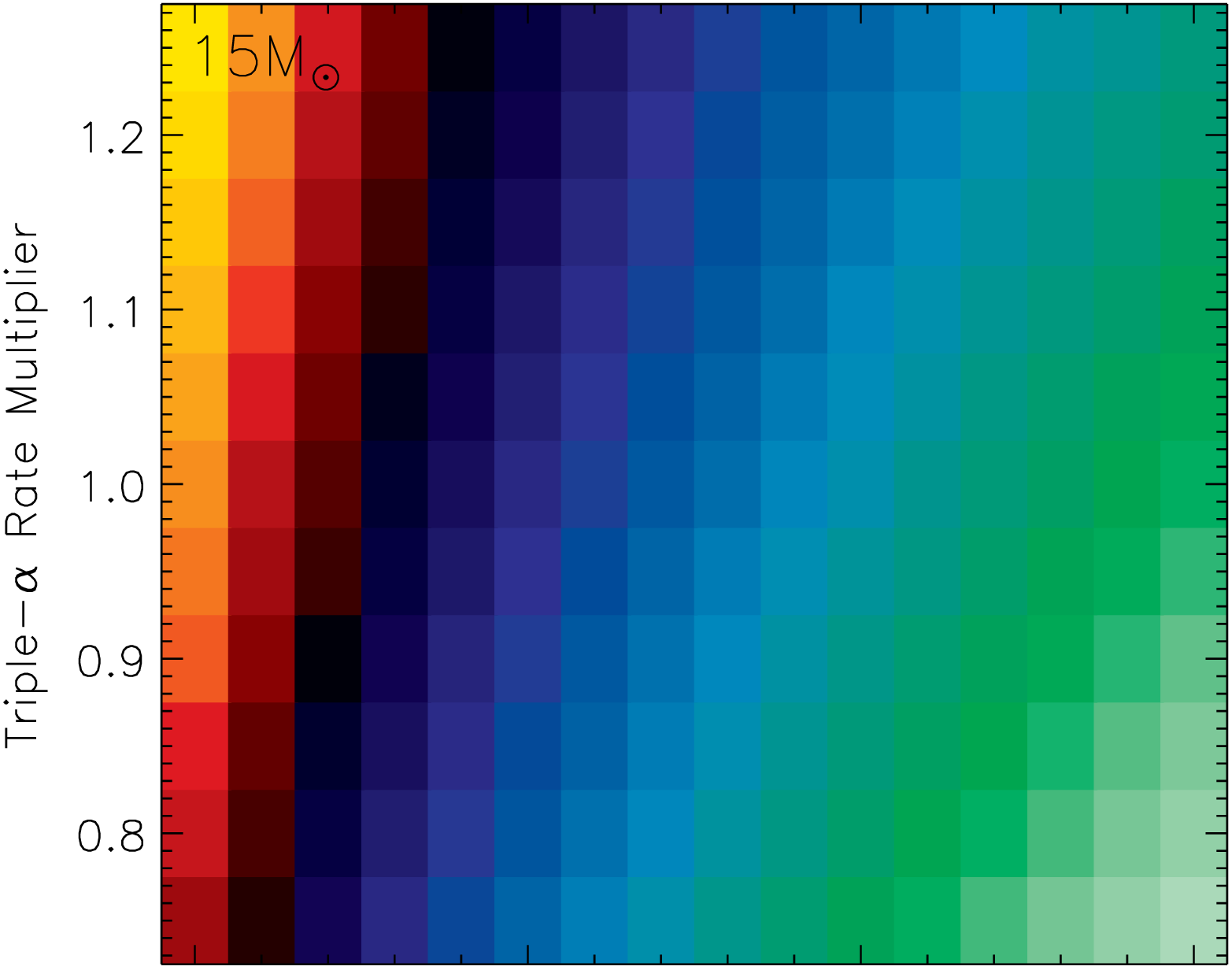}%
\end{minipage}\hspace{-1.345cm}%
\begin{minipage}[t]{0.32\columnwidth}%
\includegraphics[angle=90,scale=0.2]{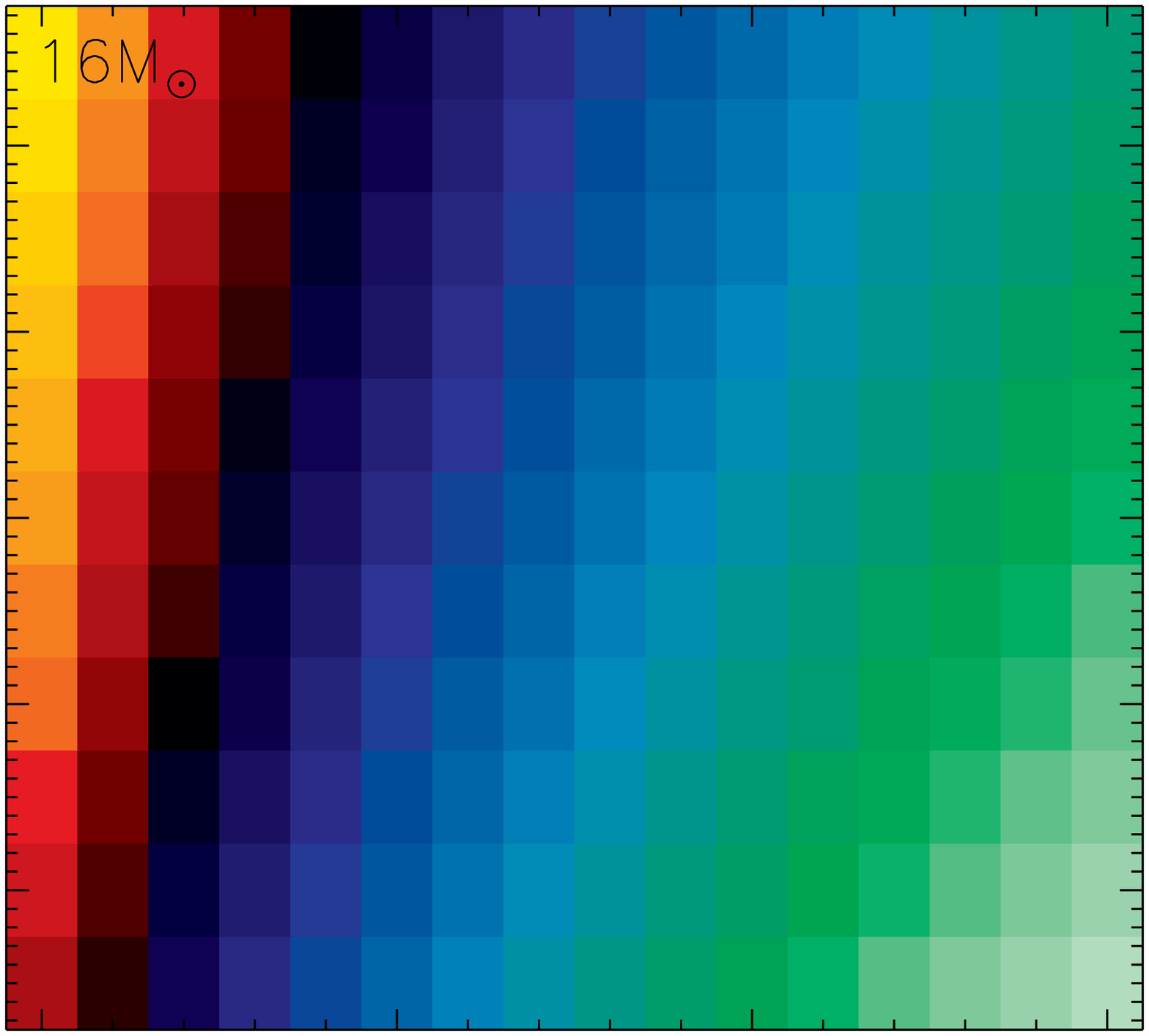}%
\end{minipage}\hspace{-1.345cm}%
\begin{minipage}[t]{0.32\columnwidth}%
\includegraphics[angle=90,scale=0.2]{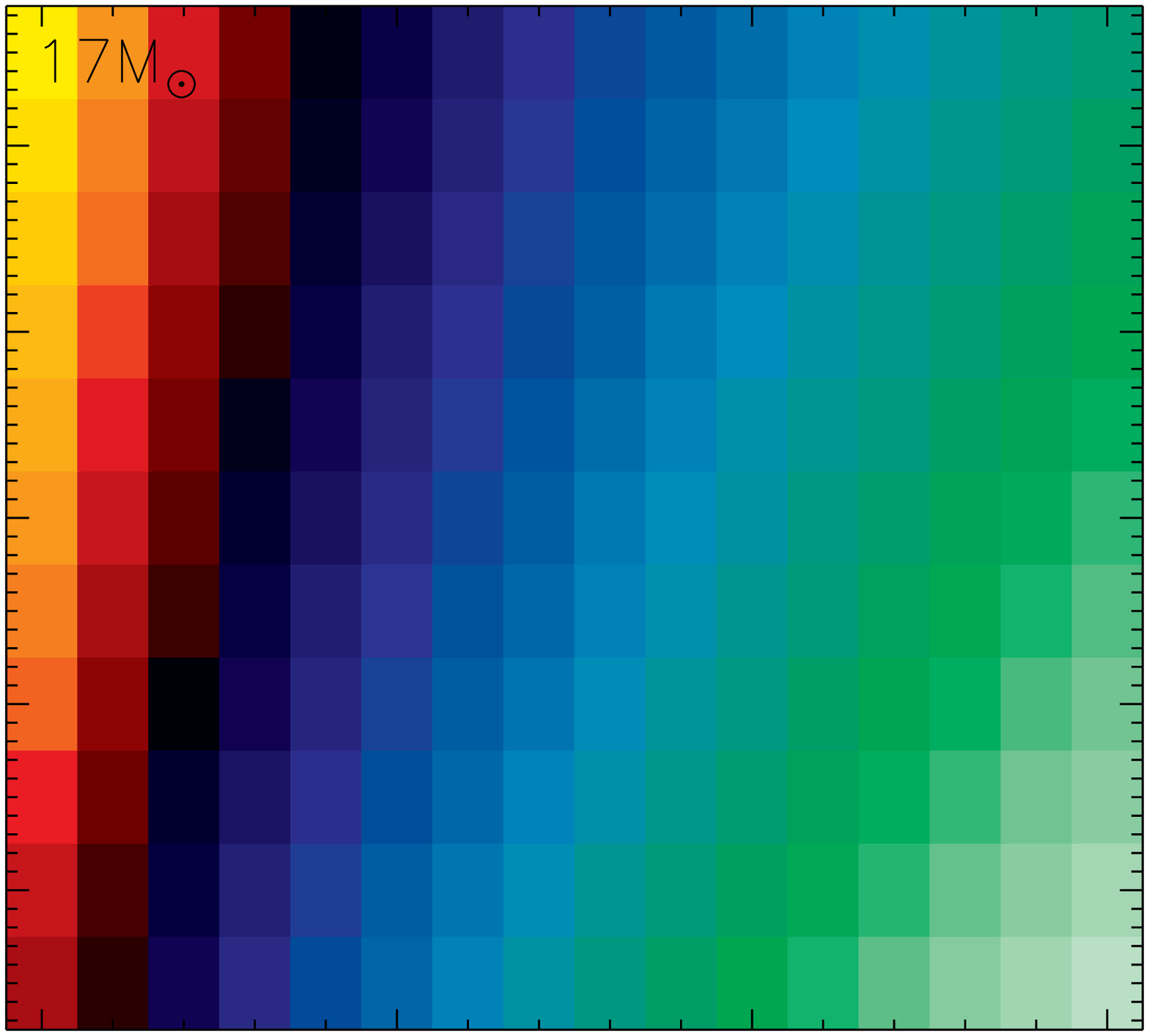}%
\end{minipage}

\vspace{-0.385cm}

\begin{minipage}[t]{0.32\columnwidth}%
\includegraphics[angle=90,scale=0.2]{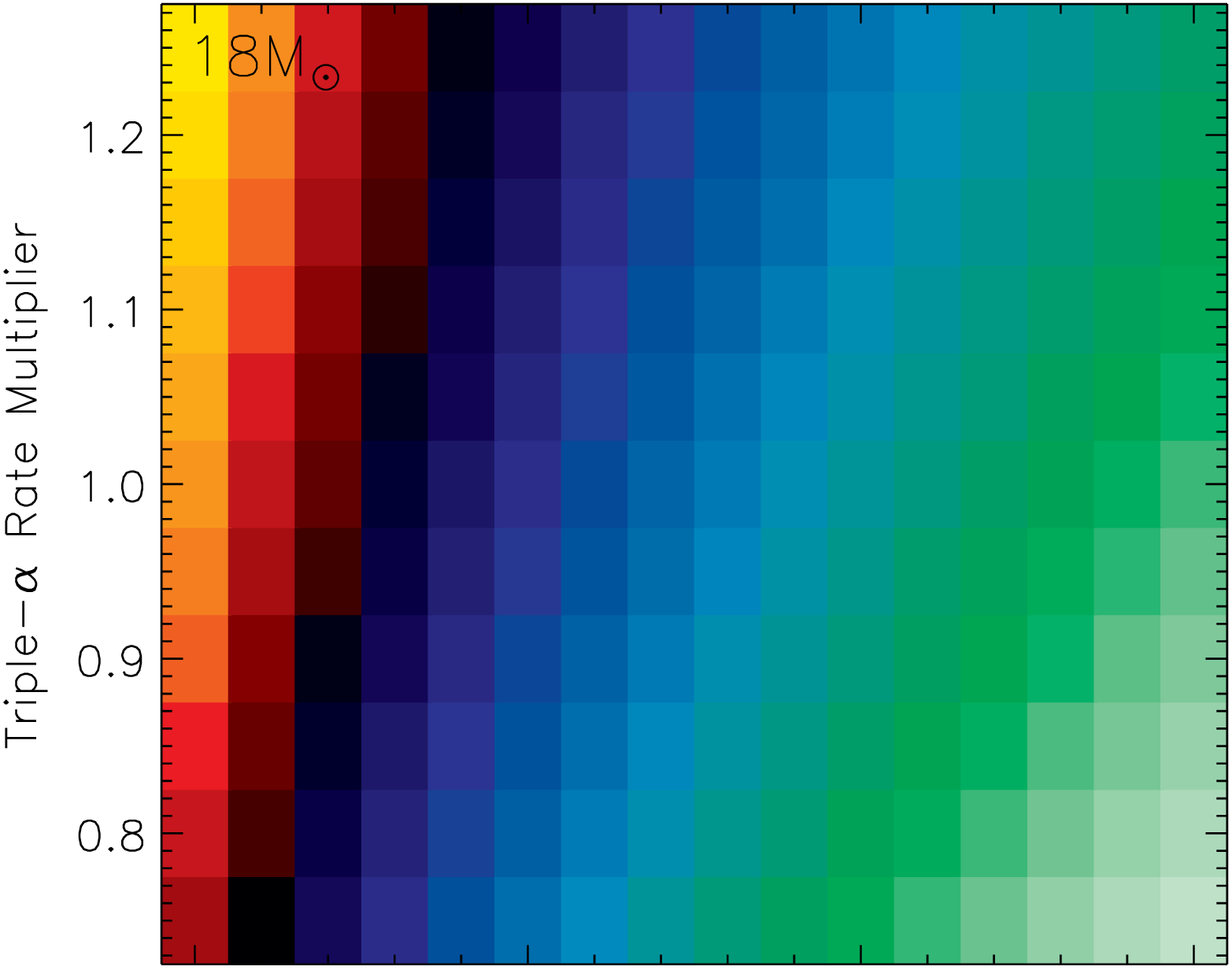}%
\end{minipage}\hspace{-1.345cm}%
\begin{minipage}[t]{0.32\columnwidth}%
\includegraphics[angle=90,scale=0.2]{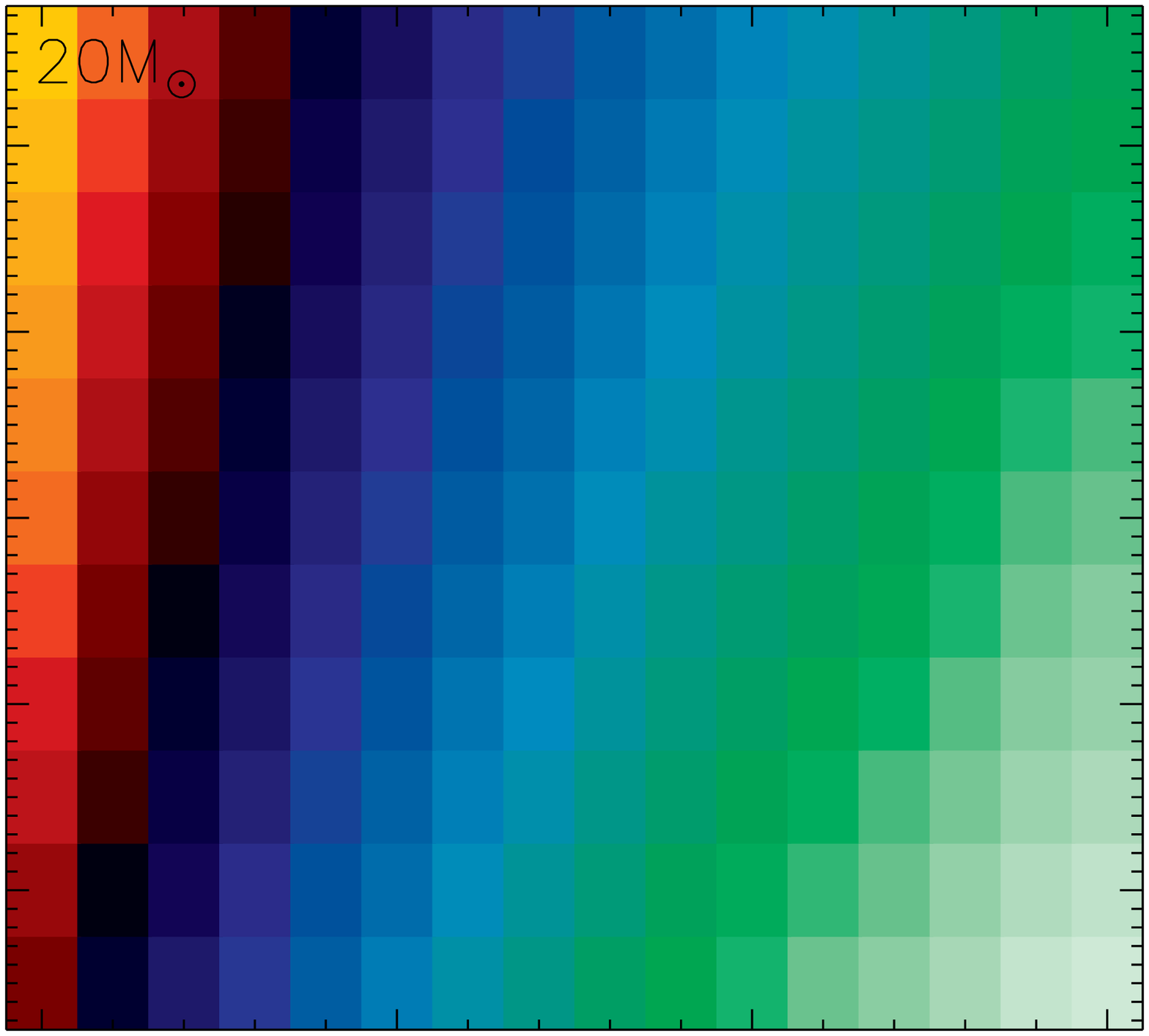}%
\end{minipage}\hspace{-1.345cm}%
\begin{minipage}[t]{0.32\columnwidth}%
\includegraphics[angle=90,scale=0.2]{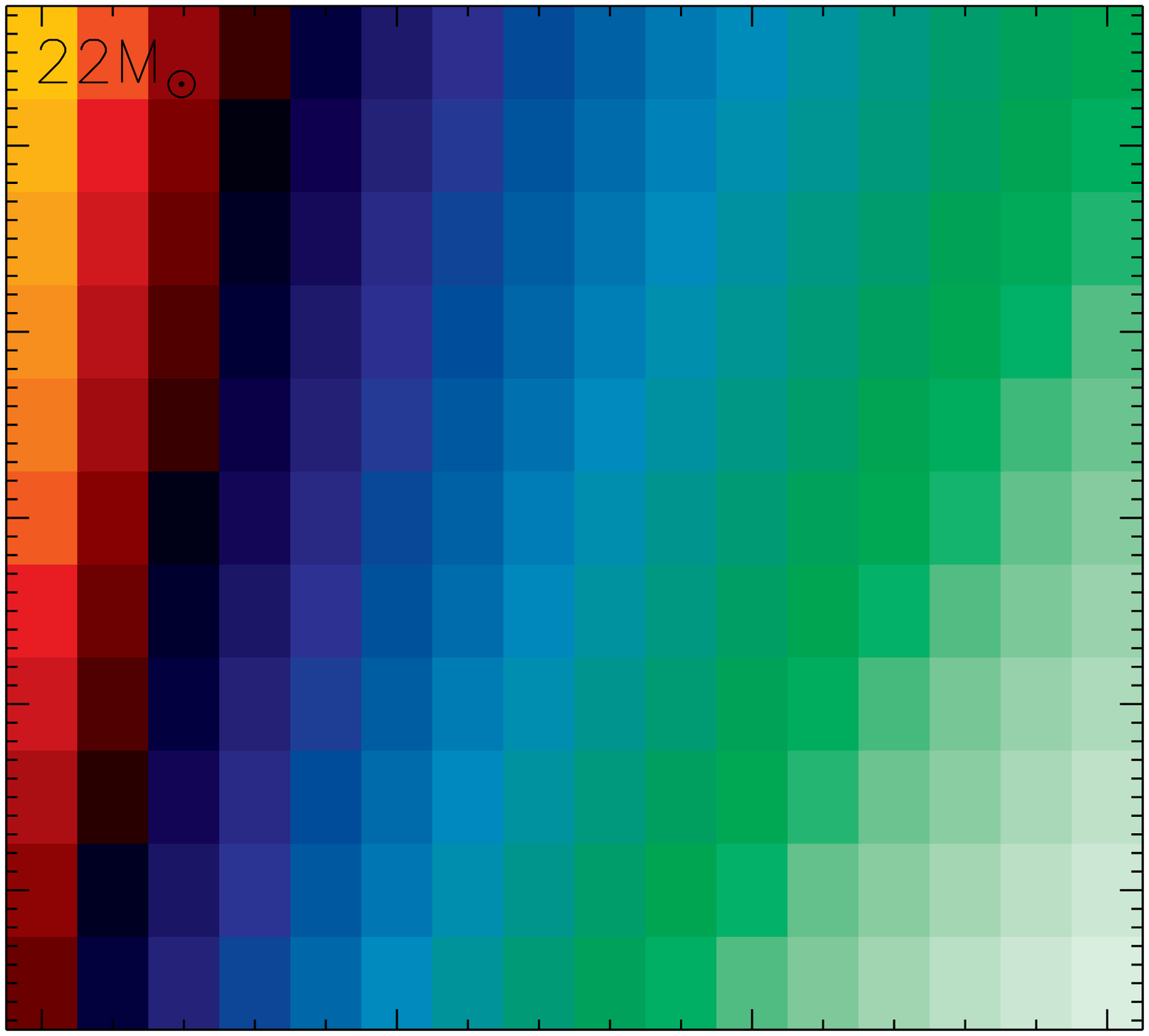}%
\end{minipage}\vspace{-0.385cm}

\begin{minipage}[t]{0.32\columnwidth}%
\includegraphics[angle=90,scale=0.2]{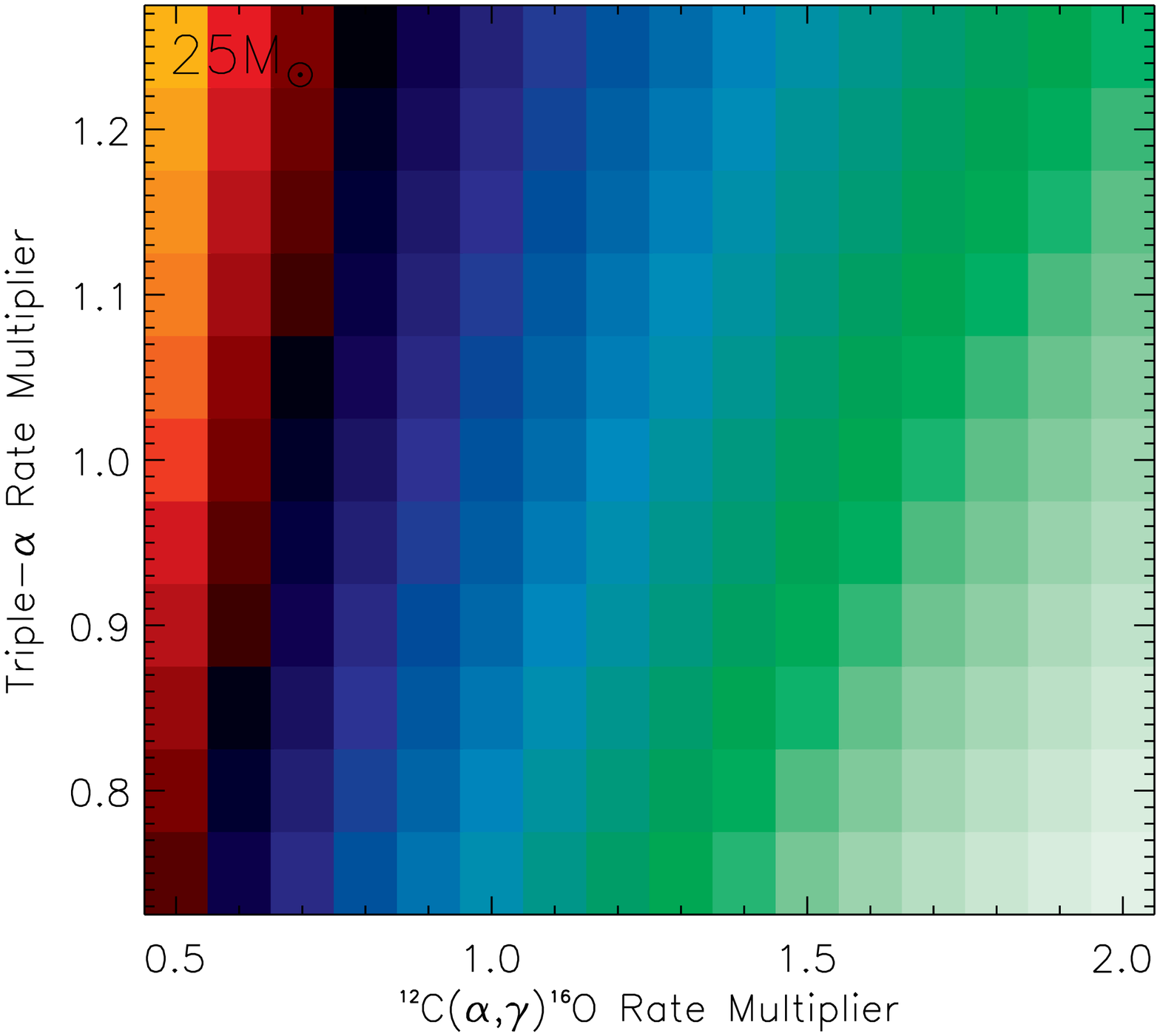}%
\end{minipage}\hspace{-1.345cm}%
\begin{minipage}[t]{0.32\columnwidth}%
\includegraphics[angle=90,scale=0.2]{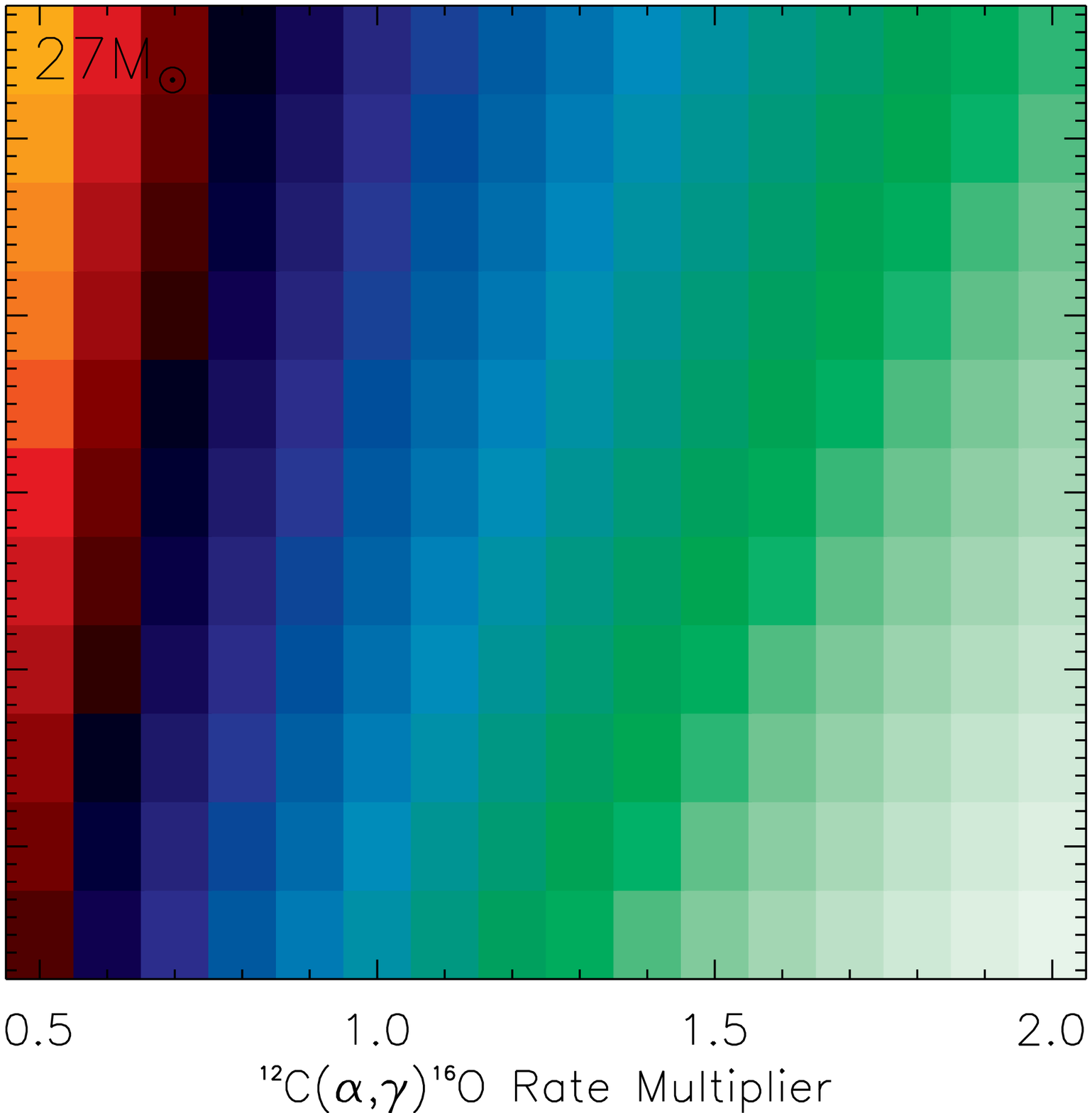}%
\end{minipage}\hspace{-1.345cm}%
\begin{minipage}[t]{0.32\columnwidth}%
\includegraphics[angle=90,scale=0.2]{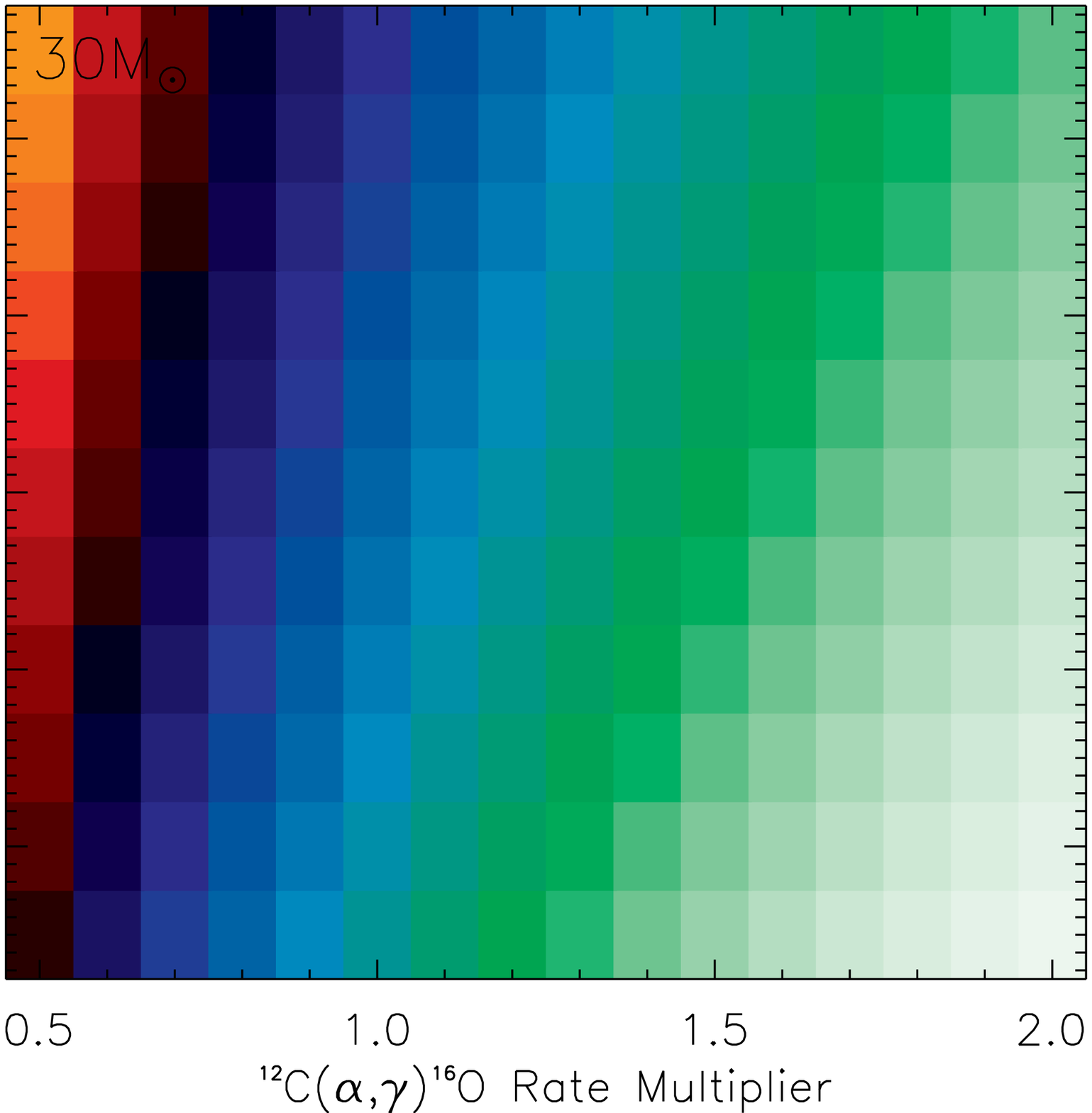}%
\end{minipage}

\caption{\label{fig:carbons}Central carbon mass fractions at the end of core-He
burning for all models as a function of the $R_{3\alpha}$ and $R_{\alpha,12}$
multipliers. The color scale gives the carbon mass fraction, and the
rest of the figure follows the convention of Fig.\,\ref{fig:Intermediate_sigmas}.}
\end{figure}

\clearpage{}
\end{document}